\documentclass[prd, twocolumn, longbibliography, amsfonts, amssymb, amsmath, showkeys, nofootinbib, superscriptaddress]{revtex4-2}
\usepackage[english]{babel}
\usepackage[utf8]{inputenc}
\usepackage{amsthm}
\usepackage{float}
\usepackage{mathtools,amsmath,bm}
\usepackage{physics}
\usepackage{xcolor}
\usepackage{graphicx}
\usepackage{adjustbox}
\usepackage{placeins}
\usepackage[T1]{fontenc}
\usepackage{lipsum}
\usepackage{csquotes}
\newcommand{\msun}{\mathrm{\, M_\odot}}
\usepackage{placeins}
\usepackage{tabularx}
\usepackage{hhline}
\usepackage[pdftex, pdftitle={Article}, pdfauthor={Author}]{hyperref} 
\usepackage{xcolor}
\usepackage{afterpage} 
\usepackage{array, booktabs, makecell}
\usepackage{multirow, bigdelim}
\usepackage{colortbl}
\usepackage{xspace}

\newcommand{\asharp}{A$^\sharp$\xspace}
\newcommand{\aplus}{A+\xspace}
\def\colwidth{.06\textwidth} 

\usepackage[acronym]{glossaries}
\newcolumntype{P}[1]{>{\centering\arraybackslash}p{#1}}

\newacronym{ce}{CE}{Cosmic Explorer}
\newacronym{et}{ET}{Einstein Telescope}
\newacronym{cehs}{CEHS}{Cosmic Explorer Horizon Study}
\newacronym{nsf}{NSF}{National Science Foundation}
\newacronym{xg}{XG}{Next Generation}

\begin{document}

\newcommand\aj{AJ}                   
\newcommand\actaa{Acta Astron.}      
\newcommand\araa{ARA\&A}             
\newcommand\apjl{ApJ}                
\newcommand\apjs{ApJS}               
\newcommand\apss{Ap\&SS}             
\newcommand\aap{A\&A}                
\newcommand\aapr{A\&A~Rev.}          
\newcommand\aaps{A\&AS}              
\newcommand\azh{AZh}                 
\newcommand\baas{BAAS}               
\newcommand\bac{Bull. astr. Inst. Czechosl.}
\newcommand\caa{Chinese Astron. Astrophys.}
\newcommand\cjaa{Chinese J. Astron. Astrophys.}
\newcommand\icarus{Icarus}           
\newcommand\jcap{J. Cosmology Astropart. Phys.}
\newcommand\jrasc{JRASC}             
\newcommand\memras{MmRAS}            
\newcommand\mnras{MNRAS}             
\newcommand\na{New A}                
\newcommand\nar{New A Rev.}          
\newcommand\pasa{PASA}               
\newcommand\pasp{PASP}               
\newcommand\pasj{PASJ}               
\newcommand\rmxaa{Rev. Mexicana Astron. Astrofis.}%
\newcommand\qjras{QJRAS}             
\newcommand\skytel{S\&T}             
\newcommand\solphys{Sol.~Phys.}      
\newcommand\sovast{Soviet~Ast.}      
\newcommand\ssr{Space~Sci.~Rev.}     
\newcommand\zap{ZAp}                 
\newcommand\iaucirc{IAU~Circ.}       
\newcommand\aplett{Astrophys.~Lett.} 
\newcommand\apspr{Astrophys.~Space~Phys.~Res.}
\newcommand\bain{Bull.~Astron.~Inst.~Netherlands} 
\newcommand\fcp{Fund.~Cosmic~Phys.}  
\newcommand\gca{Geochim.~Cosmochim.~Acta}   
\newcommand\grl{Geophys.~Res.~Lett.} 
\newcommand\jgr{J.~Geophys.~Res.}    
\newcommand\jqsrt{J.~Quant.~Spec.~Radiat.~Transf.}
\newcommand\memsai{Mem.~Soc.~Astron.~Italiana}
\newcommand\nphysa{Nucl.~Phys.~A}   
\newcommand\physrep{Phys.~Rep.}   
\newcommand\physscr{Phys.~Scr}   
\newcommand\planss{Planet.~Space~Sci.}   
\newcommand\procspie{Proc.~SPIE}   

\let\astap=\aap
\let\apjlett=\apjl
\let\apjsupp=\apjs
\let\applopt=\ao

\title{Characterizing Gravitational Wave Detector Networks: From A$^\sharp$ to Cosmic Explorer}
\author{Ish Gupta}
\affiliation{Institute for Gravitation and the Cosmos, Department of Physics, Pennsylvania State University, University Park, PA 16802, USA}

\author{Chaitanya~Afle}
\affiliation{Department of Physics, Syracuse University, Syracuse, NY 13244, USA}

\author{K.G. Arun}
\affiliation{Chennai Mathematical Institute, Chennai, India}
\affiliation{Institute for Gravitation and the Cosmos, Department of Physics, Pennsylvania State University, University Park, PA 16802, USA}

\author{Ananya~Bandopadhyay}
\affiliation{Department of Physics, Syracuse University, Syracuse, NY 13244, USA}

\author{Masha~Baryakhtar}
\affiliation{Department of Physics, University of Washington, Seattle, WA 98195, USA}

\author{Sylvia~Biscoveanu}
\affiliation{LIGO Laboratory, Massachusetts Institute of Technology, Cambridge, MA 02139, USA}

\author{Ssohrab~Borhanian}
\affiliation{Theoretisch-Physikalisches Institut, Friedrich-Schiller-Universit\"at Jena, Jena 07743, Germany}

\author{Floor~Broekgaarden}
\affiliation{Center for Astrophysics, Harvard \& Smithsonian, Cambridge, MA 02138, USA}	

\author{Alessandra Corsi}
\affiliation{Department of Physics and Astronomy, Texas Tech University, Lubbock, TX 79409, USA}

\author{Arnab Dhani}
\affiliation{Max Planck Institute for Gravitational Physics (Albert Einstein Institute), Potsdam 14476, Germany}

\author{Matthew~Evans}
\affiliation{LIGO Laboratory, Massachusetts Institute of Technology, Cambridge, MA 02139, USA}

\author{Evan D. Hall}
\affiliation{LIGO Laboratory, Massachusetts Institute of Technology, Cambridge, MA 02139, USA}

\author{Otto A. Hannuksela}
\affiliation{Department of Physics, The Chinese University of Hong Kong, Shatin, New Territories, Hong Kong}

\author{Keisi Kacanja}
\affiliation{Department of Physics, Syracuse University, Syracuse, NY 13244, USA}

\author{Rahul Kashyap}
\affiliation{Institute for Gravitation and the Cosmos, Department of Physics, Pennsylvania State University, University Park, PA 16802, USA}

\author{Sanika Khadkikar}
\affiliation{Institute for Gravitation and the Cosmos, Department of Physics, Pennsylvania State University, University Park, PA 16802, USA}

\author{Kevin Kuns}
\affiliation{LIGO Laboratory, Massachusetts Institute of Technology, Cambridge, MA 02139, USA}

\author{Tjonnie G. F. Li}
\affiliation{Institute for Theoretical Physics, Department of Physics and Astronomy, KU Leuven, B-3001 Leuven, Belgium}
\affiliation{STADIUS, Department of Electrical Engineering (ESAT), KU Leuven, B-3001 Leuven, Belgium}

\author{Andrew L. Miller}
\affiliation{Nikhef -- National Institute for Subatomic Physics, 1098 XG Amsterdam, The Netherlands}
\affiliation{Institute for Gravitational and Subatomic Physics,
Utrecht University, 3584 CC Utrecht, The Netherlands}

\author{Alexander Harvey Nitz}
\affiliation{Department of Physics, Syracuse University, Syracuse, NY 13244, USA}

\author{Benjamin J. Owen}
\affiliation{Department of Physics and Astronomy, Texas Tech University, Lubbock, TX 79409, USA}

\author{Cristiano Palomba}
\affiliation{INFN, Sezione di Roma, I-00185 Roma, Italy}

\author{Anthony Pearce}
\affiliation{Department of Physics and Astronomy, Texas Tech University, Lubbock, TX 79409, USA}

\author{Hemantakumar Phurailatpam}
\affiliation{Department of Physics, The Chinese University of Hong Kong, Shatin, New Territories, Hong Kong}

\author{Binod Rajbhandari}
\affiliation{Department of Physics and Astronomy, Texas Tech University, Lubbock, TX 79409, USA}

\author{Jocelyn Read}
\affiliation{Nicholas and Lee Begovich Center for Gravitational Wave Physics and Astronomy, California State University Fullerton, Fullerton CA 92831, USA}

\author{Joseph D. Romano}
\affiliation{Department of Physics and Astronomy, Texas Tech University, Lubbock, TX 79409, USA}

\author{Bangalore S. Sathyaprakash}
\affiliation{Institute for Gravitation and the Cosmos, Department of Physics, Pennsylvania State University, University Park, PA 16802, USA}
\affiliation{Gravity Exploration Institute, School of Physics and Astronomy, Cardiff University, Cardiff, CF24 3AA, United Kingdom}

\author{David H. Shoemaker}
\affiliation{LIGO Laboratory, Massachusetts Institute of Technology, Cambridge, MA 02139, USA}

\author{Divya Singh}
\affiliation{Institute for Gravitation and the Cosmos, Department of Physics, Pennsylvania State University, University Park, PA 16802, USA}

\author{Salvatore Vitale}
\affiliation{LIGO Laboratory, Massachusetts Institute of Technology, Cambridge, MA 02139, USA}



\author{Lisa~Barsotti}
\affiliation{LIGO Laboratory, Massachusetts Institute of Technology, Cambridge, MA 02139, USA}


\author{Emanuele~Berti}
\affiliation{Department of Physics and Astronomy, Johns Hopkins University, Baltimore, MD 21218, USA}


\author{Craig Cahillane}
\affiliation{Department of Physics, Syracuse University, Syracuse, NY 13244, USA}


\author{Hsin-Yu Chen}
\affiliation{Department of Physics, The University of Texas at Austin, Austin, TX 78712, USA}


%
%
%
%
%
\author{Peter Fritschel}
\affiliation{LIGO Laboratory, Massachusetts Institute of Technology, Cambridge, MA 02139, USA}

%
%

\author{Carl-Johan Haster}
\affiliation{Department of Physics and Astronomy, University of Nevada Las Vegas, Las Vegas, NV 89154, USA}

%

%
%

\author{Philippe Landry}
\affiliation{Canadian Institute for Theoretical Astrophysics, University of Toronto, Toronto, ON M5S 3H8, Canada}

%
\author{Geoffrey Lovelace}
\affiliation{Nicholas and Lee Begovich Center for Gravitational Wave Physics and Astronomy, California State University Fullerton, Fullerton CA 92831, USA}	

\author{David McClelland}
\affiliation{OzGrav-ANU, Centre for Gravitational Astrophysics, College of Science,The Australian National University, ACT 2601, Australia}

\author{Bram J J Slagmolen}
\affiliation{OzGrav-ANU, Centre for Gravitational Astrophysics, College of Science,The Australian National University, ACT 2601, Australia}

%
%
%
%
%
%




\author{Joshua R Smith}
\affiliation{Nicholas and Lee Begovich Center for Gravitational Wave Physics and Astronomy, California State University Fullerton, Fullerton CA 92831, USA}

\author{Marcelle Soares-Santos}
\affiliation{Department of Physics, University of Michigan, Ann Arbor, MI 48109, USA}

\author{Ling Sun}
\affiliation{OzGrav-ANU, Centre for Gravitational Astrophysics, College of Science,The Australian National University, ACT 2601, Australia}

\author{David Tanner}
\affiliation{Department of Physics, University of Florida, Gainesville, FL 32611, USA}


%
\author{Hiro Yamamoto}
\affiliation{LIGO Laboratory, California Institute of Technology, Pasadena, CA 91125, USA}

\author{Michael Zucker}
\affiliation{LIGO Laboratory, California Institute of Technology, Pasadena, CA 91125, USA}
\affiliation{LIGO Laboratory, Massachusetts Institute of Technology, Cambridge, MA 02139, USA}

\begin{abstract}
    Gravitational-wave observations by the Laser Interferometer Gravitational-Wave Observatory (LIGO) and Virgo have provided us a new tool to explore the Universe on all scales from nuclear physics to the cosmos and have the massive potential to further impact fundamental physics, astrophysics, and cosmology for decades to come. In this paper we have studied the science capabilities of a network of LIGO detectors when they reach their best possible sensitivity, called \asharp, given the infrastructure in which they exist and a new generation of observatories that are factor of 10 to 100 times more sensitive (depending on the frequency), in particular a pair of L-shaped Cosmic Explorer observatories (one 40 km and one 20 km arm length) in the US and the triangular Einstein Telescope with 10 km arms in Europe. We use a set of science metrics derived from the top priorities of several funding agencies to characterize the science capabilities of different networks.  The presence of one or two \asharp observatories in a network containing two or one next generation observatories, respectively, will provide good localization capabilities for facilitating multimessenger astronomy and precision measurement of the Hubble parameter. Two Cosmic Explorer observatories are indispensable for achieving precise localization of binary neutron star events, facilitating detection of electromagnetic counterparts and transforming multimessenger astronomy. Their combined operation is even more important in the detection and localization of high-redshift sources, such as binary neutron stars, beyond the star-formation peak, and primordial black hole mergers, which may occur roughly $100$ million years after the Big Bang. The addition of the Einstein Telescope to a network of two Cosmic Explorer observatories is critical for accomplishing all the identified science metrics including the nuclear equation of state, cosmological parameters, the growth of black holes through cosmic history, but also make new discoveries such as the presence of dark matter within or around neutron stars and black holes, continuous gravitational waves from rotating neutron stars, transient signals from supernovae, and the production of stellar-mass black holes in the early Universe. For most metrics the triple network of next generation terrestrial observatories are a factor 100 better than what can be accomplished by a network of three \asharp\ observatories.  
\end{abstract}
\maketitle
\clearpage

\tableofcontents

\section{Introduction}
Over the past eight years since their first discovery, the Laser Interferometer Gravitational-wave Observatory \cite{LIGOScientific:2014pky} (LIGO) in the U.S. and the Virgo observatory \cite{VIRGO:2014yos} in Europe have observed $\sim{\cal O}(100)$ binary black hole mergers and a handful of neutron star binary mergers \cite{LIGOScientific:2021djp} during the first three science runs O1-O3.  The fourth science run (O4, Advanced LIGO and Virgo sensitivity \cite{LIGOScientific:2014pky}) and the fifth (O5, \aplus sensitivity \cite{KAGRA:2013rdx}) over the next two to seven years will add hundreds more to the catalog of compact binary coalescences. LIGO and Virgo will eventually be joined by KAGRA \cite{KAGRA:2020agh} and LIGO-India \cite{LIGO-India} to make many more detections and discoveries. These detections will enable electromagnetic follow-up observations, multi-messenger astronomy, compact binary population inferences, ultra-dense matter phenomenology and cosmological studies. 

The LIGO and Virgo collaborations have already developed plans for further improvements in sensitivity beyond O5 that will fully exploit what is possible at existing facilities. In particular, the \asharp\ (pronounced A-sharp) concept \cite{T2200287} is expected to improve the sensitivity by a factor of two compared to \aplus\ strain sensitivity \cite{KAGRA:2013rdx}. Accomplishing sensitivity levels significantly greater than those currently envisaged will require new facilities, with longer interferometer arms, but marginal improvements in detector technology, as described in the NSF-funded Horizon Study \cite{Evans:2021gyd} for the Cosmic Explorer (CE) project\footnote{Visit the CE project website \url{https://dcc.ligo.org/LIGO-T2200287/public} for news and sensitivity curves.}. Einstein Telescope (ET) is a similar concept currently considered for funding in Europe \cite{Punturo:2010zz,Hild:2010id,et_science_team_2011_3911261,ETDesign2020}. We shall refer to CE and ET as next-generation observatories or XG for short.

The National Science Foundation (NSF) has appointed a sub-committee\footnote{\label{fo:nggw charge}Membership of the sub-committee can be found at \url{https://www.nsf.gov/mps/phy/nggw-members.jsp} and NSF's charge to the sub-committee is at: \url{https://www.nsf.gov/mps/advisory/subcommittee_charges/mpsac-nggw-charge_signed.pdf}} of the Mathematical and Physical Sciences Advisory Committee (MPSAC) to \emph{assess and recommend configurations for a U.S. GW detection network that can operate at a sensitivity approximately an order of magnitude greater than that of LIGO \aplus\ by the middle of the next decade\textsuperscript{\ref{fo:nggw charge}}.} The sub-committee has invited White Papers from the community addressing \emph{science motivation and key science objectives, technical description of the proposed concept(s) and how different aspects are associated with key science, current and new technologies needed, risks, timelines, and approximate cost assessment, any synergies or dependencies on other multi-messenger facilities (existing or future)\footnote{The call for White Papers can be found at \url{https://www.nsf.gov/mps/phy/nggw/WhitePaperCall2.pdf}.}}. The CE project conducted a trade study to assess the relative performances of plausible detector networks operating in the 2030s and summarized the findings in the White Paper \cite{Evans:2023euw} (hereby referred to as the WP) submitted in response to that call. This document provides the details of the trade study including the populations considered, the methodology used, and the results obtained. It serves as the technical basis for what is reported in the project's submission.

Gravitational-wave (GW) observations can address questions across multiple disciplines from general relativity to relativistic astrophysics,  nuclear physics to dark matter searches and cosmology to beyond the standard model of particle physics. They can do this by observing binary black hole coalescences from an epoch when the universe was still assembling its first stars, binary neutron stars far beyond redshifts when the star formation in the universe was at its peak, stochastic backgrounds produced in the primordial universe, new sources and phenomena such as supernovae, stellar quakes and rapidly rotating neutron stars and, very likely, new phenomena and sources not imagined by anyone so far. To realize the full potential of GW astronomy, it is necessary to build longer detectors with sensitivity levels that are at least an order of magnitude better than those of A+ detectors. In this Tech Report, we describe the science that can be accomplished at the limit of current facilities and how future observatories like CE can vastly transform the field of GW astronomy, while answering many of the pressing problems in high-energy physics, astronomy and cosmology. 

To this end, we consider eight different networks, described in Sec.~\ref{sec:Networks} and summarized in Table \ref{tab:networks}, each consisting of three observatories belonging to one of four classes: \textbf{0~XG:} three upgraded LIGO detectors, two in the US and one in India (HLA), \textbf{1~XG:} two upgraded LIGO detectors, in the US or India, and one CE observatory of 40 km or 20 km arm-length (HLET, 20LA, 40LA) \textbf{2~XG:} one upgraded LIGO detector in the US together with two next generation observatories (20LET, 40LET, 2040A), or \textbf{3~XG:} three next generation observatories (4020ET). The results of the trade study are summarized in Table \ref{tab:network_science}, listed under the five key science themes that are discussed at length in Sec.~\ref{sec:science questions}. 
Our study concludes that \emph{a network of three next generation observatories composed of a 40 km arm-length Cosmic Explorer, a 20 km arm-length Cosmic Explorer and a 10 km a side Einstein Telescope triangle is two orders-of-magnitude better than the planned \aplus\ network in respect of almost every metric considered in this study.} More precisely, for most metrics the numbers in the last column of Table \ref{tab:network_science} for a network 3 XG observatories are a factor 100 better than those in the second column corresponding to a network of 3 \asharp\ observatories. A brief account of our findings is given below.

\begin{table*}
\caption{\label{tab:network_science}Comparison of eight different detector networks against five key science goals defined in the text. The comparison uses several metrics defined in column 1 for each science goal, followed by what's accomplished by each network in eight columns corresponding to the eight networks considered in this study (see, \ref{tab:networks}). Each network has 3 observatories and falls into one of four classes: 0 next generation observatories in column 2, 1 such observatory in columns 3-5, two such observatories in columns 6-8 and three such observatories in column 9. Networks that contain 2 or fewer XG observatories are populated with one or more \asharp\ observatories.}  
 \begin{tabular}{ p{.45\textwidth} | p{\colwidth} | p{\colwidth} p{\colwidth} p{\colwidth} | p{\colwidth} p{\colwidth} p{\colwidth}  |  p{\colwidth} } 
\hhline{=========}
& \multicolumn{8}{c}{\bf Network Performance}\\
\hhline{~--------}
\textbf{Science Goal Requirements} &\textbf{0 XG} & & \textbf{1 XG} & & & \textbf{2 XG} & & \textbf{3 XG} \\
\hhline{~--------}
& \textbf{HLA} & \textbf{HLET} & \textbf{20LA} & \textbf{40LA} & \textbf{20LET} & \textbf{40LET} & \textbf{4020A} & \textbf{4020ET} \\ 
\hhline{---------}
{\bfseries BHs and NSs Throughout Cosmic Time }&  & & & & & & &  \\
Measure mass function, determine formation scenarios:&  & & & & & & &  \\
$N_{\rm BNS}$/yr, $z\ge1,$ $\delta z/z \le 0.2,$ $\delta m_{1}/m_{1}\le 0.3$ & 0 & 0 & 0 & 0 & 0 & 7 & 22 & 81\\
Detect the (injected) second Gaussian feature:&  & & & & & & &  \\
$N_{\rm BNS}$/yr, $m_1\ge1.5\msun,$ $\delta m_1/m_1 \le 0.1$ & 0 & 37 & 9 & 24 & 68 & 105 & 58 & 155\\
Unveiling the elusive population of IMBH:& & & & & & & & \\
$N_{\rm IMBBH}$/yr, $z\ge 3,$ $\delta z/z \le 0.2,$ $\delta m_{1}/m_{1}\le 0.2$ & 6 & 430 & 150 & 190 & 840 & 870 & 510 & 890\\
High-$z$ BBH formation channels and mass function: &  & & & & & & &\\
$N_{\rm BBH}$/yr $z\ge 10,$ $\delta z/z \le 0.2$ $\delta m_{1}/m_{1} \le 0.2$  & 0 & 12 & 6 & 35 & 65 & 140 & 110 & 230 \\
\hhline{---------}
{\bfseries MMA and Dynamics of Dense Matter} &  & & & & & & &   \\
GW170817-like golden sample: &  & & & & & & & \\
$N_{\rm BNS}$/yr $z\le0.06,$ $\Delta \Omega\le 0.1$\,deg$^2$ & 0 & 0 &  0 &  0 & 0 & 1 & 1 & 7 \\
r-process and kilonova-triggered follow up: &  & & & & & & &  \\
$N_{\rm BNS},$ $0.06<z\le0.1,$ $\Delta\Omega\le1$\,deg$^2$ & 1 & 8 & 6 & 6  & 26 & 47 & 32  & 71 \\
Jet afterglows, large-FOVs or small-FOV mosaicking: & & & & & & & & \\
$N_{\rm BNS}$/yr, $0.1<z\le2,$ $\Delta\Omega\le 10$\,deg$^2$
& 260  & 1000 & 780 & 890 & 6000 & 9200 & 3900 & 27000 \\
Mapping GRBs to progenitors up to star-formation peak: & & & & & & & & \\
$N_{\rm BNS}$/yr, $z>2,$ $\Delta \Omega\le 100$\,deg$^2$ 
& 0 &  2 & 19 & 37 & 6200  & 25000 & 3700 & 66000\\
Pre-merger alerts 10 minutes before merger & & & & & & & & \\
$N_{\rm BNS}$/yr, $\Delta \Omega\le 100$\,deg$^2$ & 0 & 20 & 0 & 0 &  200 & 400 & 200 & 700  \\
NS EoS constraints:&  & &  & & & &   \\
$N_{\rm BNS}$/yr, SNR$\ge 100$ 
& 0 & 48 & 39 & 220 & 120 & 350 & 350 &  480\\
$N_{\rm BNS}$/yr, $\Delta R < 0.1$\,km &  0 & 160 & 20 & 72 & 280 & 450 & 320 & 740 \\
\hhline{---------}
{\bfseries New Probes of Extreme Astrophysics}&  & & & & & & &   \\
Pulsars with ellipticity $10^{-9}$ detectable in 1 year
& 1 & 3 & 3 & 5 & 5 & 11 & 9 & 21 \\
Years to detect 25 pulsars with ellipticity $10^{-9}$
& 12 & 3.5 & 3.7 & 2.3 & 2.2 & 1.7 & 1.7 & 1.3 \\
\hhline{---------}
{\bfseries Fundamental Physics and Precision Cosmology} &  & & & & & & &    \\
Constrain graviton mass:&  & & & & & &   \\
$N_{\rm BNS}$/yr, $z\ge5$  & 0 & 0 & 0 & 84 & 0 & 340 & 580 & 880\\
$N_{\rm BBH}$/yr, $z\ge5$  & 19 & 2500 &  2200 & 3900 & 3880 & 4700 & 4500 & 5100\\
Probing rare events:&  & & & & & & &   \\
$N_{\rm BBH}$/yr, SNR\,$> 100$  & 17 & 1900 & 1300 & 5000 & 3700 & 7500 &  6900 & 9500 \\
$N_{\rm BBH}$/yr, SNR\,$> 1000$ & 1 & 2 & 1 & 5 & 4 & 7 & 7 & 10 \\
Precision tests of GR  (IMR and QNM): & & & & & & & & \\
BBH root sum square total SNR & 2400  & 11,000 & 9800  & 16,000 & 15,000 & 20,000 & 19,000 & 22,000 \\
BBH root sum square post-inspiral SNR & 1900  & 5800  & 5300  & 8100  & 7800  & 9900  & 9500  & 11,000 \\
$N_{\rm BBH}$/yr, post-inspiral SNR\,$> 100$  & 6 & 319 & 290 & 1200 & 790 & 1900 & 1700 & 2500 \\
Cosmology and tests of GR:&  & &   & & & & & \\
$N_{\rm BNS}$/yr, $z \le 0.5$, $\delta d_L/d_L \le 0.1$ and $\Delta \Omega\le 10$\,deg$^2$ & 14 & 270 & 63 & 71 & 1100 & 1600 & 790 & 4800 \\
$N_{\rm BBH}$/yr, $\delta d_L/d_L \le 0.1,$ $\Delta \Omega\le 1\,\deg^2$ & 69 & 490 & 290 & 350 & 2200 & 3300 & 1300 & 6800 \\
Lensed BNS events/yr: & 1 & 25 & 11 & 65 & 47 & 91 & 87 & 110 \\
\hhline{---------}
{\bfseries Physics Beyond the Standard Model} & & & & & & & &  \\
Stochastic signal $\Omega_{\mathrm{GWBG}}$ in units of $10^{-10}$ & 2 & 0.03 & 0.5 & 0.3 & 0.03 & 0.02 & 0.06 & 0.02\\
Dark matter assisted NS implosions:$N_{\rm BBH}$/yr &  0 & 2000 & 1200 & 4600 & 4200 &  8000 & 6300 & 9700 \\
Primordial black hole mergers: $N_{\rm pBBH}$/yr, $z>25$, $\Delta\,z/z < 0.2$: &  0 & 0 & 0 & 0 &  8 & 17  & 3  & 28 \\
Pop III black hole mergers $z>10$, $\Delta\,z/z < 0.1$: $N_{\rm PopIII}$/yr: & 0 & 2 & 0 & 9 & 38 & 110 & 74 &  360 \\
Max distance (Mpc) to detectable axion clouds in BHs & 0.13 & 0.73 & 0.82 & 1.34 & 0.60 & 1.60 & 1.87 & 1.66 \\
\hhline{=========}
\end{tabular}
\end{table*}

\paragraph{Black Holes and Neutron Stars Throughout Cosmic Time} 
A network of XG observatories will build a survey of black hole mergers from epochs before the universe was assembling its first stars and observe double neutron star mergers and neutron star-black hole mergers far beyond redshifts when the star formation rate was at its peak. Four key metrics for this theme are listed in Table \ref{tab:network_science}. In particular, a network consisting of at least one XG observatory will chart hundreds of black hole mergers at $z>10,$ but a network consisting of at least two XG observatories is necessary to observe binary black holes at $z>10$ and definitively say if a sub-population exists at those redshifts. Similarly, a network of two XG observatories is key to observing neutron star mergers at $z>1$ and measuring their masses accurately enough to conclude that they are neutron stars. Moreover, XG observatories have the unique opportunity to detect intermediate-mass black hole binaries up to $z\sim5,$ accurately measure their mass- and redshift-distribution.

\paragraph{Multimessenger Astrophysics and Dynamics of Dense Matter}
Mergers of double neutron star and neutron star-black hole binaries involve dense matter in relativistic motion and observing them in the electromagnetic (EM) window will require accurate 3D localization and alerts from GW observatories. Seven rows in Table \ref{tab:network_science} illustrate the power of XG observatories in accomplishing all the science goals under this theme. Golden binary neutron star mergers such as GW170817 would be observed with a signal-to-noise ratio (SNR) $>10^3$ several times a year. Several tens of high-fidelity signals will be localized to $\Delta\Omega<1\,\deg^2,$ enabling deeper insight into the physics r-process kilonova, tens of thousands will be localized to within $10\,\deg^2$ to study jet afterglows produced by merger remnants and hundreds will be detected and localized to within $100\,\deg^2$ 10 minutes before merger, providing pre-merger alerts for EM follow-up. Tens of thousands of events will be localized to within $100\,\deg^2$ at $z>2$ providing an opportunity to correlate every short gamma-ray burst (GRBs) with binary neutron star mergers. Finally, hundreds of high-fidelity signals will enable precision measurement of the neutron star radius and the equation of state of dense matter.

\paragraph{New Probes of Extreme Astrophysics}
Next generation observatories will not be limited to observing just compact binary mergers. They will detect new classes of transient signals, e.g., from core-collapse supernovae and magnetar flares to continuous waves from rapidly rotating pulsars. We expect the network of 3 XG observatories to detect 20 pulsars with ellipticities of $10^{-9}$ or smaller and detect 25 pulsars in less than 2 years. The \asharp\ network will detect a handful of such signals in its lifetime. Continuous waves from neutron stars would provide first hints of the physics of neutron-star crust and an independent (from binary mergers) confirmation of the equation of state of cold dense matter. 

\paragraph{Fundamental Physics and Precision Cosmology}
General relativity has been a highly successful theory in explaining laboratory experiments and astronomical observations. Yet, due to some of its conceptual problems (e.g. the black hole singularity and the information paradox), it is largely expected that the theory will prove to be incompatible with high-precision observations of black holes and neutron stars in the era of  XG observatories. Signals from binary neutron stars and black holes at redshifts larger than 5 will help constrain massive graviton and other theories that require GWs to travel at sub- or super-luminal speeds. Thousands of binary black hole events with $\mbox{SNR} > 100$ and cumulative SNRs of more than 20,000 from the entire population of observed binary black hole mergers in a network of 2 XG observatories will subject general relativity to stringent tests. 

Binary coalescences are standard sirens allowing precision measurement of the luminosity distance to sources. Thousands of well-localized binary black holes and neutron stars will allow exquisite measurement of the Hubble constant and other cosmological parameters. 

\paragraph{Physics Beyond the Standard Model} 
The Standard Model of particle physics is in excellent agreement with results from the collider experiments and yet there are several conceptual problems, e.g. strong CP problem and QCD axions, which seem to suggest physics beyond the Standard Model. Gravitational observations could discover the presence of axion clouds around black holes affecting their spin distributions or the accumulation of weakly interacting massive particles in neutron stars converting them to black holes. A network of XG observatories will provide direct or indirect evidence of the existence of dark matter, probe the nature of dark energy and either detect or set stringent limits on the stochastic backgrounds from the early universe such as the electroweak phase transition or cosmic strings. Next generation observatories are sensitive to energy density in stochastic backgrounds and could detect them with $SNR>3$ at a fiducial frequency of $f_{\mathrm{ref}}=25~\mathrm{Hz}$ if $\Omega_{\mathrm{GWBG}} \ge 10^{-12}.$ 

A network of two Cosmic Explorer observatories will herald a remarkable enhancement in sensitivity, surpassing \asharp\ detectors by approximately an order of magnitude, especially at frequencies below $100,\mathrm{Hz}$. Individually, in a network with \asharp\ detectors, they can detect the majority of the cosmic population of binary black hole mergers and $\mathcal{O}(10^5)$ binary neutron star and neutron star-black hole mergers every year. The 20 km detector can be optimized to measure the post-merger signal from binary neutron stars, which will reveal useful insights into the physical processes involved in the dynamically evolving post-merger remnant \cite{Prakash:2023afe}. However, the true scientific potential is realized through the synergistic capabilities of both observatories, proving pivotal in realizing the objectives outlined in Table \ref{tab:network_science}. A network with both the 20 km and 40 km observatories will not only prove indispensable in measuring the mass spectrum of binary neutron star systems, it will also enable precise sky-localization of binary neutron star mergers, facilitating detection of multi-band electromagnetic counterparts, which is the cornerstone of multimessenger astronomy. Such a network will also have the capability to impose stringent constraints on the equation of state of dense matter, by achieving sub-$100$ m precision in measuring the neutron star radius for dozens of events annually. Additionally, the ability to detect signals from high-redshift GW sources, e.g. primordial black holes that are detectable at redshifts surpassing 25, underscores the vital role played by the two Cosmic Explorer observatories in advancing our understanding of the cosmos across diverse cosmic epochs.

The rest of the paper is organized as follows. In Sec.\,\ref{sec:Networks} we introduce the list of observatories considered in this study and their strain sensitivity and networks composed of those observatories. Sec.\,\ref{sec:pop} provides the assumptions we make about the cosmic population of binary sources. This is followed by a discussion of the Fisher information matrix approach followed in this study in Sec.\,\ref{sec:det and pe}. This section relies on a number of tables and figures to illustrate the detection and measurement capabilities of different detector networks. Finally, in Sec.\,\ref{sec:science questions} we provide a detailed account of the science questions that are of broad interest and how \asharp\ and future XG observatories can probe those questions.

\section{Gravitational-Wave Observatory Network Configurations}
\label{sec:Networks}
In this Section, we summarize the list of GW detectors that are expected to be available over the next two decades. We start with detectors with the best sensitivities that could be installed in LIGO facilities, followed by networks that include one or more XG observatories consisting of CE and/or ET.  The collection of network configurations studied is intended to be sufficiently broad without being unduly complex: broad enough to gauge all plausible network configurations but not so complex as to consider every possible scenario. Indeed, we are aided by research indicating that the critical feature of a future gravitational-wave network is the number of next-generation detectors present for most of the science metrics, while their locations are of secondary importance~\cite{Hall:2019xmm}. For the localization metrics, however, the network area has a large effect on the performance, once the  the network composition is optimized. To this end, we consider the following observatories:

\begin{table}[!tb]
    \begin{tabular}{l|l|l|l}
        \toprule
        Detector & Latitude & Longitude & Orientation \\
        \midrule
        CE-A & $46^\circ 00'00''$       & $-125^\circ 00'00''$      & $260.0^\circ $ \\ 
        CE-B & $29^\circ 00'00''$       & $-94^\circ 00'00''$       & $200.0^\circ $ \\
        ET   & $40^\circ 31'00''$       & $+9^\circ 25'00''$         & $90.0^\circ $ \\
        LLO  & $30^\circ 33'46.4196''$  & $-90^\circ 46'27.2654''$  & $197.7165^\circ $ \\
        LHO  & $46^\circ 27'18.5280''$  & $-119^\circ 24'27.5657''$ & $125.9994^\circ $ \\
        LAO  & $19^\circ 36'47.9017''$  & $+77^\circ 01'51.0997''$   & $117.6157^\circ $\\
        \toprule
    \end{tabular}
    \caption{Position and orientation of the detectors. Latitudes (Longitudes) are positive in the northern hemisphere (East of the Greenwich meridian). The orientation is the angle north of east of the $x$-arm (Note: here we follow the same convention used in Bilby \cite{Ashton:2018jfp, Ashton:2021anp, Romero-Shaw:2020owr}, which is different from what is used in Refs. \cite{LIGO-India-coord,glob-loc-coord}, where the orientations of the detectors are clockwise rotations from the local north). For L-shaped detectors, the $x$-arm is defined as the one that completes a right-handed coordinate system together with the other arm and the local, outward, vertical direction. For ET, the x-arm is defined such that the two other arms lay westward of it.}
    \label{tab:locations}
\end{table}

\begin{table}[!bt]
  \centering
  \begin{tabular}{l l l}
    \toprule
    \textsf{Number of XG } & \textsf{Network} & \textsf{Detectors in} \\ 
    \textsf{Observatories} & \textsf{Name}    & \textsf{the network} \\ \midrule
    None & HLA    & LHO, LLO, LAO     \\ \midrule
    \multirow{3}{*}{1 XG} 
    & HLET & LHO, LLO, ET \\ 
    & 20LA & CE-A 20 km, LLO, LAO \\ 
    & 40LA & CE-A 40 km, LLO, LAO \\ \midrule
    \multirow{3}{*}{2 XG} 
    & 4020A & CE-A 40 km, CE-B 20 km, LAO \\
    & 40LET & CE-A 40 km, LLO, ET         \\
    & 20LET & CE-A 20 km, LLO, ET         \\ \midrule
    3 XG & 4020ET & CE-A 40 km, CE-B 20 km, ET \\
    \bottomrule
\end{tabular}
\caption{We consider four classes of networks containing, zero to three next-generation (XG) observatories.
Each network is given a name to facilitate comparisons.
The HLA network sets the stage, representing the baseline from which CE return on investment can be assessed.  20LA and 40LA represent a single CE operating in the context of an upgraded 2G network while HLET is a single ET operating together with LLO and LHO. 4020A is the CE reference configuration, operating with an upgraded LIGO Aundha, while 20LET and 40LET represent a single CE operating with LLO and ET. 4020ET is the reference CE configuration operating with ET.\label{tab:networks}}
\end{table}

\begin{figure*}
  \centering
  \includegraphics[width=0.8\textwidth]{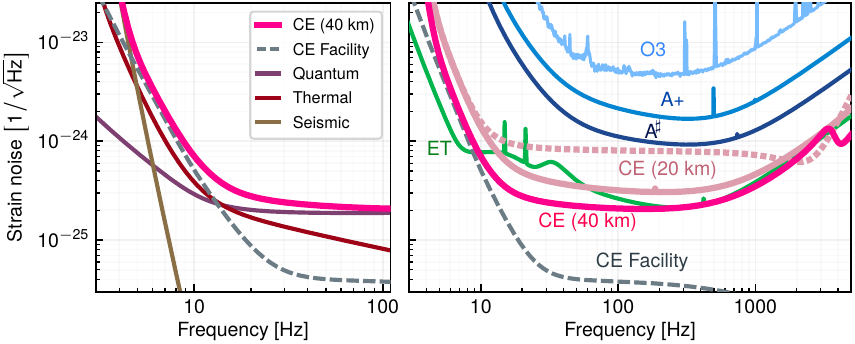}
  \caption{\emph{Left:} Estimated spectral sensitivity (solid black) of Cosmic Explorer (CE)
    and the known fundamental sources of noise that contribute to this total (other curves). \emph{Right:} Comparison of spectral sensitivities of LIGO A+, LIGO A$^\sharp$, Einstein Telescope (a triangular arrangement of six interferometric detectors), and 20 km and 40 km versions of Cosmic Explorer. The facility limit for Cosmic Explorer is also indicated. (Reproduced from the WP)}
  	\label{fig:sensitivity}
\end{figure*}

\paragraph{Cosmic Explorer Observatories (CE-A, CE-B)}
Since the locations of the CE observatories have yet to be determined, we selected two fiducial locations for CE; CE-A off the coast of Washington state, and CE-B off the coast of Texas.  These locations are intentionally unphysical to avoid impacting our ability to find a potential home for CE, but close enough to a wide range of potential sites to be representative from the point of view of gravitational-wave science (see Table \ref{tab:locations}).  The CE-A location is considered in both the $40\,$km and the $20\,$km lengths, while the CE-B location hosts only a $20\,$km observatory. The strain sensitivity of the two choices is shown in Fig.~\ref{fig:sensitivity}

\paragraph{Existing LIGO Sites (LHO, LLO, LAO)}
In order to focus on the science enabled by CE beyond what is possible in the current facilities, we model the LIGO detectors in an upgraded form (known as ``A$^\sharp$'', \cite{T2200287}, and essentially equivalent in sensitivity to ``Voyager'' \cite{Adhikari:2019zpy}) that approximately represents the limit to what is achievable in the LIGO facilities.  Furthermore, in addition to the LIGO Hanford (LHO) and LIGO Livingston (LLO) detectors, we also consider LIGO Aundha (LAO) at \asharp\ sensitivity, as it is expected to be operational starting in the early 2030s. The currently projected strain sensitivity of A$^\sharp$ is shown in Fig.~\ref{fig:sensitivity}.

\paragraph{Einstein Telescope (ET)}
The Einstein Telescope is a planned next-generation GW observatory in Europe~\cite{Punturo:2010zz}. It is currently envisioned as an underground triangular facility with 10~km arm length, housing six interferometers (however, other configurations and their scientific potentials have also been studied in Ref. \cite{Branchesi:2023mws}) . The targeted timeline calls for first observations by the mid-2030s. The underground location, which is strongly preferred in Europe, suppresses the expected seismic disturbances, thereby reducing the Newtonian noise that limits terrestrial gravitational-wave facilities a low frequencies (c.f.\ the difference between CE and ET below {8}{\,Hz} is depicted in Fig.~\ref{fig:sensitivity}). 

ET's adoption into the European Strategy Forum on Research Infrastructure (ESFRI) road map has affirmed the observatory's role in the future of gravitational-wave physics astronomy. Nevertheless, we present network configurations that do not include ET to inform the value of US investment in the absence of ET. Our models for each of these network nodes are described below and shown in Table \ref{tab:networks}. 

\section{Population of compact binaries}
\label{sec:pop}
\subsection{Binary black holes}
To analyze the science capabilities of the different detector network configurations, we simulate populations of various types of compact binary mergers and evaluate the detection and measurement abilities of the networks for these populations. There are still large uncertainties in the properties that characterize these populations due to the low number of detections at present.  Therefore, the populations we described below are intended to represent plausible, but not necessarily exact, realizations of the true populations. This is sufficient for the purposes of this work as our goal is not to predict the actual detection rates but, instead, to compare the capabilities of different networks for the chosen populations.

\subsubsection{Local population}
The local population of binary black hole (BBH) mergers is consistent with what has been inferred up to GWTC-3 \cite{LIGOScientific:2021djp,KAGRA:2021duu}. One difference is that we do not consider precession for any of the populations. As precession, in general, is expected to improve the estimation of parameters~\cite{Vecchio:2003tn}, the measurability estimates presented in this work will be on the conservative side. Specifically, for the local population we use:

\begin{itemize}
    \item \textbf{Primary mass}: \texttt{POWER LAW + PEAK} \cite{KAGRA:2021duu} model with the following values for the model parameters: $\alpha=-3.4$, $m_{min} = 5\msun$, $m_{max}=87\msun$, $\lambda=0.04$, $\mu_{peak}=34\msun$, $\sigma_{peak}=3.6$, $\delta_m=4.8\msun.$
    \item \textbf{Mass ratio}: $p(q) \propto q^\beta$ with $\beta=1.1$, and enforcing $m_{min} = 5\msun.$
    \item \textbf{Spin magnitude}: Independently and identically distributed (IID) spins following a beta distribution (see, e.g., \cite{Wysocki:2018mpo}) with $\alpha_\chi=2$, $\beta_\chi=5$ (see Eq. (10) in Ref. \cite{Wysocki:2018mpo}), but restricted to aligned spins.
    \item \textbf{Redshift}:  Merger rate following the Madau-Dickinson star formation rate~\cite{Madau:2014bja,2019PhRvD.100d3030T},
    \begin{equation}
        \psi(z|\gamma, \kappa, z_p) = \frac{(1 + z)^\gamma}{1 + (\frac{1 + z}{1 + z_p})^\kappa},
    \end{equation}
    with $\gamma=2.7$, $z_{p}=1.9$, and $\kappa=5.6$. We choose a local merger rate density of $24$ $\mathrm{Gpc}^{-3}$ $\mathrm{yr}^{-1}$ \cite{LIGOScientific:2020kqk}.
\end{itemize}

\subsubsection{Population III black holes}

We also consider a population of high-redshift BBHs, representing BBHs formed from Pop-III stars. As no uncontroversial detection of these objects exists, the uncertainty on their parameters is substantial. We use:

\begin{itemize}
    \item \textbf{Primary mass}: A fixed value of $20\msun$.
    \item \textbf{Mass ratio}: A fixed value of $0.9$.
    \item \textbf{Spin magnitude}: Same as the local BBH population.
    \item \textbf{Redshift}: The merger rate follows the distribution introduced in Ref.~\cite{Ng:2020qpk} (Eq. C15)
    with $a_{III} = 0.66$, $b_{III} = 0.3$ and $z_{III} = 11.6$.
\end{itemize}

\subsubsection{Primordial black holes}

In addition, we consider a population of even higher redshift sources that could be representative of primordial black holes. For these too, our knowledge of the masses and spins of the companion black holes is very limited. Nonetheless, we use:

\begin{itemize}
    \item \textbf{Primary and secondary mass}: The lognormal distribution of Ref.~\cite{Ng:2022agi} (Eq. 1) centered at $M_c=30\msun$ and with $\sigma=0.3\msun$.
    \item \textbf{Spin magnitude}: Zero spins.
    \item \textbf{Redshift}: Merger rate distribution that increases as the age of the universe decreases (Ref.~\cite{Ng:2022agi}, Eq.~5).
\end{itemize}

While the prescriptions above fix the characteristics for each formation channel, for Pop III and primordial black holes we need two more parameters to fix the relative importance of these channels. We follow Refs.~\cite{Ng:2020qpk} and ~\cite{Ng:2022agi} and work with $N_{\mathrm{III}}=2400$ and $N_{\mathrm{pbh}}=600$ mergers per year in the two channels.

\subsubsection{Intermediate mass binary black holes}
We would also like to know how well the next generation of GW observatories can characterize a population of intermediate-mass binary black hole (IMBBHs) binaries, especially with the improved sensitivity at low frequencies. We use:
\begin{itemize}
    \item \textbf{Masses}: A power-law distribution for the two masses with $\alpha = -2.5$. Further, we choose the smallest and largest masses in the distribution to be $m_{\mathrm{min}} = 100\msun$ and $m_{\mathrm{max}} = 1000\msun,$ respectively. Masses larger than about $10^3\msun$ are not likely to be observed by next-generation observatories as they could merge well before reaching the sensitivity of CE or ET.
    \item \textbf{Spins}: The spins for both the BHs are chosen to follow a uniform distribution between $[-0.9,0.9]$. 
    \item \textbf{Redshift}: Same as the local BBH population, but with a local merger rate density of $1$ $\mathrm{Gpc}^{-3}$ $\mathrm{yr}^{-1}$.
\end{itemize}

\subsection{Binary neutron stars}
\label{sec:3B}
We simulate a single population of binary neutron stars (BNSs), whose merger rate peaks at cosmic noon, and is consistent with the local merger rate as measured by the LIGO-Virgo-KAGRA (LVK). We choose the following parameters:

\begin{itemize}
    \item \textbf{Primary and secondary mass}: A double Gaussian distribution, $p(m)= w \mathcal{N}(\mu_L,\sigma_L) +(1-w) \mathcal{N}(\mu_R,\sigma_R)$. We use parameters equal to the median values of Ref.~\cite{2020RNAAS...4...65F}: $\mu_L=1.35\msun$, $\sigma_L=0.08\msun$, $\mu_R=1.8\msun$, $\sigma_R=0.3\msun$, and $w = 0.64$. Each normal distribution is independently truncated and normalized in the range $[1,2.2]~\msun$.
    \item \textbf{Spin magnitude}: Uniform in the range $[0,0.1]$.
    \item \textbf{Redshift}: Same as local BBHs, but with a local merger rate density of $320$ $\mathrm{Gpc}^{-3}$ $\mathrm{yr}^{-1}$ \cite{LIGOScientific:2020kqk}.
    \item \textbf{Equation of state}: We use \texttt{APR4} \cite{Akmal:1998cf} as the equation of state of the neutron star. Note that the maximum mass of the NS listed above corresponds to the maximum mass allowed by the \texttt{APR4}.
\end{itemize}

While there is some evidence that the population probed by LVK via GWs differs from the galactic neutron-star population from which this bimodal mass distribution is derived, simulating a structured mass distribution allows us to verify if and how precisely XG observatories can characterize the population (see \ref{tab:network_science}).

\subsection{Neutron star-black hole mergers}
Due to low number of detections, the properties of neutron star-black hole (NSBH) mergers are not well known. Because of this uncertainty, we will adopt a semi-agnostic approach to define the population for NSBH mergers. The specifications are as follows:

\begin{itemize}
    \item \textbf{Black Hole Mass}: The \texttt{POWER LAW + PEAK} distribution, same as the primary mass of the local BBH population.
    \item \textbf{Neutron Star Mass}: Uniform between [1,2.2] $\msun$.
    \item \textbf{Spins}: For the BH, the spin is assumed to be aligned with the orbital angular momentum and follows a Gaussian distribution with $\mu = 0$ and $\sigma = 0.2$. The NS is assumed to be slowly spinning, following a uniform distribution between $[-0.1,0.1]$. 
    \item \textbf{Redshift}: Same as the local BBH population, but with a local merger rate density of $45$ $\mathrm{Gpc}^{-3}$ $\mathrm{yr}^{-1}$ \cite{KAGRA:2021duu,LIGOScientific:2021qlt}.
\end{itemize}

For all other GW parameters for all the cases (i.e., sky location, orbital orientation, polarization angle, coalescence time and phase) we use uninformative distributions. We assume all sources are quasi-circular, i.e., we ignore orbital eccentricity.

\section{Detection and parameter estimation}
\label{sec:det and pe}
Having introduced different network configurations and population models, we next wish to address the detectability of these source classes and how precisely the parameters of these sources can be extracted with different detector configurations. Detectability is quantified in terms of the SNR $\rho$ defined as 
\begin{equation}
    \rho^2=4\int_{f_{\rm low}}^{f_{\rm upper}}\frac{|{\tilde h}_A|^2}{S_n^{\rm A}}\,df,
\end{equation}
where ${\tilde h}_A$ is the waveform of the signal at detector A, $S_n^{\rm A}$ is the one-sided noise power spectral density (PSD) of detector A and $f_{\rm low}$ and $f_{\rm upper}$ denote the lower and upper cut of frequencies of the integration.

Similarly, we use the Fisher information matrix to compute the statistical uncertainties associated with measuring binary parameters. The Fisher matrix $\Gamma_{ab}$ is related to the derivatives of the waveform with respect to the set of source parameters ${\bm {\lambda}}$ as
\begin{equation}
 \Gamma_{ab}=   2\int_{f_{\rm low}}^{f_{\rm upper}}\frac{{\tilde h_{{\rm A},a}}{\tilde h^{\star}_{{\rm A},b}}+{\tilde h^{\star}_{{\rm A},a}}{\tilde h_{{\rm A},b}}}{S_n^{\rm A}}\,df,
\end{equation}
where $\star$ denotes the operation of complex conjugation and the comma denotes differentiation with respect to various elements of the parameter space $\bm{\lambda}$. The inverse of the Fisher matrix is called covariance matrix $\Sigma_{ab}$ and the square root of the diagonal elements provides the $1\sigma$ (68\% CL) uncertainty range for the measurement of different parameters for a given detector A
\begin{equation}
    \sigma_a=\sqrt{\Sigma_{aa}}.
\end{equation}
All measurement uncertainties mentioned here are at 68\% credibility except the angular resolution $\Delta\Omega,$ which is reported at 90\% credibility.

For the computation of errors, we use \texttt{GWBENCH}~\cite{Borhanian:2020ypi}, a publicly available Python-based software that computes the Fisher matrix for various waveform families available in the LIGO Algorithms Library (LAL) \cite{lalsuite}. Fisher matrix is then numerically inverted to deduce the statistical errors associated with the measurement of various parameters in the waveform. {\tt GWBENCH} can perform parameter estimation for different combinations of detector networks, thereby facilitating a detailed assessment of their performance. 

Here, we use the IMRPhenomXHM \cite{Garcia-Quiros:2020qpx} waveform to deduce errors in the case of BBHs and NSBHs whereas we use IMRPhenomPV2\_NRtidalv2 \cite{Dietrich:2019kaq} for BNSs. The former is a non-precessing waveform that covers the inspiral-merger-ringdown phases of a compact binary merger and has higher-order spherical harmonic modes. On the other hand, IMRPhenmPV2\_NRtidalv2 has only the leading $\ell=2,m=2$ mode of the gravitational waveform but accounts for tidal effects up to 6PN order. 

The parameter space spanned by BBH and NSBH signals is given by
\begin{eqnarray}
{\bm{\lambda}}=\{{\cal M},\eta,\chi_{1z},\chi_{2z}\,D_L,\iota, \alpha,\delta, \psi,t_c,\phi_c\},
\end{eqnarray}
which denote, chirp mass, symmetric mass ratio, projections of the spins of the primary and secondary along the orbital angular momentum axis, luminosity distance to the source, inclination angle, right ascension and declination, polarization angle, time and phase at coalescence, respectively (see Sec. IIB of Ref.~\cite{Gupta:2023evt} for details). For BNS systems, in addition to these parameters, effective tidal deformability $\lambda_{\rm eff}$ is added as an additional parameter to be estimated.  The lower cut-off frequency is taken to be $5\,$Hz for all the network configurations. The upper cut-off frequency for all cases is the frequency above which the signal has no power.

Using \texttt{GWBENCH} we next study the detection efficiency and detection rates of various classes of sources introduced earlier for the different detector configurations. We then discuss the parameter uncertainties and their implications for astrophysics and fundamental physics.

\subsection{Detection efficiency and detection rate}
The detection efficiency $\epsilon(z)$ of an observatory is a function of the luminosity distance or, equivalently, redshift. It is the fraction of all events at a redshift $z$ that are observed with an expected SNR greater than a preset threshold SNR. Instead of an SNR threshold, one could define the efficiency at a fixed false alarm rate but in the Fisher matrix approach the two are equivalent. We quote the efficiency at two different SNR thresholds: a network SNR $\rho=10$ and $\rho=100,$ where the SNR of a network is root-mean-square SNR obtained for all the detectors in the network. In addition, we also require each detector in the network to record a minimum SNR of 5 to say that it is detected. An SNR of $\rho=10$ serves as the smallest SNR at which a confident detection can be made while an SNR of $\rho=100$ is an SNR at which exquisite measurement can be made. 

\begin{figure*}[htbp] 
\centering
\includegraphics[width=0.95\textwidth]{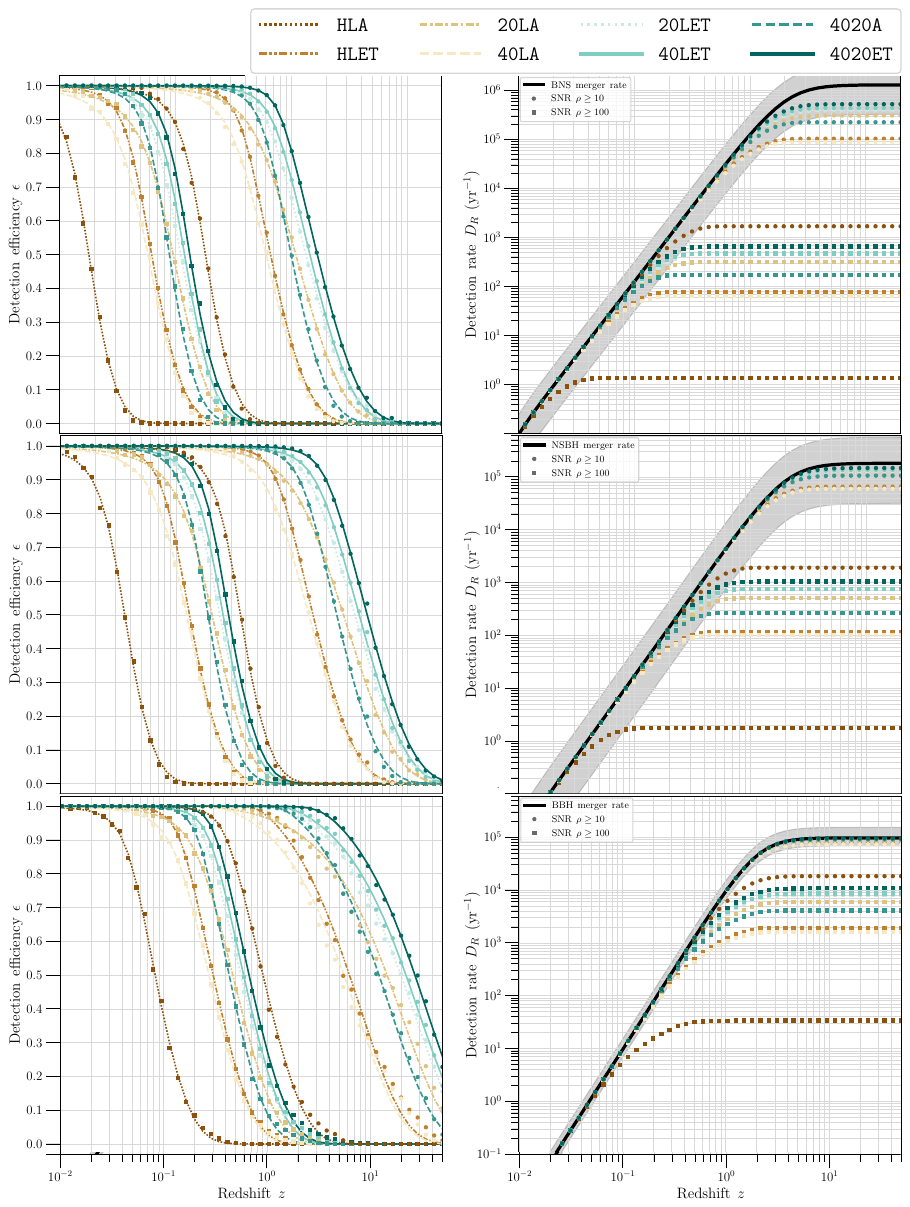}
\vspace{-0.4cm}
\caption{\label{fig:local_eff_rate}The network efficiency (left panels) and detection rate (right panels) for the eight GW detector networks.  For each network, we plot the efficiency and rate at two threshold SNRs, one at 10 (circles) and the other at 100 (squares).  Solid lines in the left panels are the \textit{best-fit} sigmoid functions of the efficiency. Black solid lines on the right are the total merger rate and the gray shaded area is the variation in the rate due to the uncertainty in the local merger rate density determined from current observations \cite{LIGOScientific:2020kqk,KAGRA:2021duu}. From top to bottom, the panels correspond to BNS, NSBH and BBH systems.}
\end{figure*}

The detection efficiency of different detector networks is reported in Fig.~\ref{fig:local_eff_rate} and in Table \ref{tab:reach_and_pop_snr}. The left panels of Fig.~\ref{fig:local_eff_rate} show the detection efficiency as a function of redshift while the right panels show the detection rate per year for BNS, NSBH and BBH populations, respectively. The grey region shows the error bars due to the uncertainties in their current merger rates. Table~\ref{tab:reach_and_pop_snr} quantifies the capabilities even better in terms of redshift reach as well as the number of detections that these configurations can make based on our current understanding of their rates.

With an SNR of 10, 50\% of the BNS merger may be detected at a redshift of 1.7 with the 4020ET 3 XG network, whereas an HLA network can observe these sources only up to a redshift of 0.18, almost a factor of 10 smaller. There is unlikely to be any BNS detection with an SNR above 100 with HLA whereas it is seen that $\mathcal{O}(100)$ such detections could be made with the 4020ET network.

Similarly, for NSBH mergers, compared to a redshift reach of 0.18(0.04) at an SNR of 10(100) with HLA, the 4020ET network can see up to a redshift of 4.5(0.3), which highlights the benefits XG detectors bring in. Also, given our current rates, it is unlikely that an HLA network would detect any NSBH merger with an SNR above 100 whereas a 4020ET network is likely to detect $\mathcal{O}(100)$ of these. Performance of 1 XG and 2 XG detectors lie in between these two extremes which may see ${\cal O}(1)$ and ${\cal O}(10)$ such sources per year, respectively.

It is impressive to note that the 4020ET network, which has a redshift reach of 27 with an SNR threshold of 10, would detect almost all BBHs within the horizon of $z=5$. On the other hand, the detectability of BBHs is complete only up to a redshift of 0.01 for the HLA network. The 4020ET network would see around 6000 BBHs per year which has an SNR of 100 or above. This will provide us with an unprecedented opportunity to probe the diverse classes of BBHs and infer their astrophysical properties and formation mechanism, among other things. Networks with 1 XG and 2 XG networks can also detect hundreds to a couple of thousands of these golden BBHs.

\begin{table*}
  \centering
  \caption{\label{tab:reach_and_pop_snr}The redshift reach $z$ at which the detection efficiency of a network is 50\% corresponding to binary merger events for the eight detector networks (column 1) for threshold SNRs of $\rho_{*} = 10$ and $\rho_{*} = 100$ (columns 2 and 3) are listed together with the number events every year with SNRs greater than $10$ (column 4), $30$ (column 5), and $100$ (column 6). The lower and upper bounds in these columns are calculated using the uncertainty in the local merger rate density as determined by current observations.}
  \renewcommand{\arraystretch}{1.5} 
    \begin{tabular}{l | P{2cm} P{2cm} |  P{3cm}P{3cm}P{3cm}}
    \hhline{======}
    Network & $z\,(\rho_{*} = 10)$ & $z\,(\rho_{*} = 100)$ &  $N\,(\rho > 10)$ & $N\,(\rho > 30)$ & $N\,(\rho > 100)$ \\
    \hhline{------}
    \multicolumn{6}{c}{\bf BNS: Cosmic merger rate is $\mathbf{1.2^{+2.0}_{-0.9} \times 10^6}$ yr$^{-1}$}\\
    \hhline{------}
    HLA & 0.18 & 0.018 
    & $1.3^{+1.9}_{-1.0} \times 10^{3}$ & $2.7^{+6.6}_{-2.3} \times 10^{1}$ & $0$ \\
    HLET & 0.66 & 0.062 
    & $8.5^{+13.0}_{-6.4} \times 10^{4}$ & $2.5^{+3.9}_{-1.9} \times 10^{3}$ & $4.8^{+7.4}_{-3.7} \times 10^{1}$ \\
    20LA & 0.61 & 0.058
    & $7.1^{+11.0}_{-5.4} \times 10^{4}$ & $2.1^{+3.1}_{-1.6} \times 10^{3}$ & $3.9^{+6.7}_{-3.3} \times 10^{1}$ \\
    40LA & 1.1 & 0.096 
    & $2.7^{+4.1}_{-2.0} \times 10^{5}$ & $1.1^{+1.7}_{-0.8} \times 10^{4}$ & $2.2^{+3.3}_{-1.8} \times 10^{2}$ \\
    20LET & 1 & 0.089 
    & $1.9^{+2.9}_{-1.4} \times 10^{5}$ & $5.9^{+9.0}_{-4.4} \times 10^{3}$ & $1.2^{+1.9}_{-1.0} \times 10^{2}$ \\
    40LET & 1.4 & 0.12 
    & $3.9^{+5.9}_{-2.9} \times 10^{5}$ & $1.7^{+2.6}_{-1.2} \times 10^{4}$ & $3.5^{+5.5}_{-2.9} \times 10^{2}$ \\
    4020A & 1.3 & 0.11 
    & $3.6^{+5.5}_{-2.7} \times 10^{5}$ & $1.7^{+2.6}_{-1.3} \times 10^{4}$ & $3.5^{+5.6}_{-2.9} \times 10^{2}$ \\
    4020ET & 1.7 & 0.13 
    & $4.7^{+7.2}_{-3.5} \times 10^{5}$ & $2.3^{+3.6}_{-1.8} \times 10^{4}$ & $4.8^{+7.7}_{-3.9} \times 10^{2}$ \\
    \hhline{======}
    \multicolumn{6}{c}{\bf NSBH: Cosmic merger rate is $\mathbf{1.8^{+3.8}_{-1.5} \times 10^5}$ yr$^{-1}$} \\
    \hhline{------}
    HLA & 0.36 & 0.036 
    & $1.5^{+3.1}_{-1.2} \times 10^{3}$ & $3.6^{+9.2}_{-3.3} \times 10^{1}$ & $0.0^{+1.0}_{-0.0} \times 10^{0}$ \\
    HLET & 1.5 & 0.13 
    & $5.9^{+12.4}_{-4.8} \times 10^{4}$ & $3.7^{+8.2}_{-3.1} \times 10^{3}$ & $8.4^{+17.2}_{-7.2} \times 10^{1}$ \\
    20LA & 1.4 & 0.12 
    & $5.3^{+11.3}_{-4.4} \times 10^{4}$ & $3.2^{+7.1}_{-2.7} \times 10^{3}$ & $7.4^{+14.6}_{-6.7} \times 10^{1}$ \\
    40LA & 2.8 & 0.21 
    & $1.0^{+2.2}_{-0.9} \times 10^{5}$ & $1.5^{+3.3}_{-1.3} \times 10^{4}$ & $3.9^{+8.0}_{-3.2} \times 10^{2}$ \\
    20LET & 2.5 & 0.19 
    & $9.8^{+20.7}_{-8.1} \times 10^{4}$ & $8.8^{+19.1}_{-7.3} \times 10^{3}$ & $2.2^{+4.3}_{-1.8} \times 10^{2}$ \\
    40LET & 3.8 & 0.26 
    & $1.3^{+2.7}_{-1.1} \times 10^{5}$ & $2.2^{+4.7}_{-1.8} \times 10^{4}$ & $6.1^{+12.0}_{-5.1} \times 10^{2}$ \\
    4020A & 3.5 & 0.24 
    & $1.2^{+2.6}_{-1.0} \times 10^{5}$ & $2.2^{+4.6}_{-1.8} \times 10^{4}$ & $6.1^{+12.3}_{-5.1} \times 10^{2}$ \\
    4020ET & 4.5 & 0.28 
    & $1.4^{+3.0}_{-1.2} \times 10^{5}$ & $2.9^{+6.2}_{-2.4} \times 10^{4}$ & $8.4^{+17.1}_{-7.1} \times 10^{2}$ \\
    \hhline{======}
    \multicolumn{6}{c}{\bf BBH: Cosmic merger rate is $\mathbf{9.6^{+5.7}_{-2.8} \times 10^4}$ yr$^{-1}$}\\
    \hhline{------}
    HLA & 0.92 & 0.083 & $1.6^{+9.3}_{-0.5} \times 10^{4}$ & $1.1^{+6.3}_{-0.3} \times 10^{3}$ & $1.7^{+1.2}_{-0.5} \times 10^{1}$ \\
    HLET & 6.3 & 0.3   & $7.7^{+4.5}_{-2.2} \times 10^{4}$ & $2.3^{+1.3}_{-0.7} \times 10^{4}$ & $1.6^{+9.0}_{-0.5} \times 10^{3}$ \\
    20LA & 5.6 & 0.28  & $7.1^{+4.1}_{-2.1} \times 10^{4}$ & $2.1^{+1.2}_{-0.6} \times 10^{4}$ & $1.3^{+7.3}_{-0.4} \times 10^{3}$ \\
    40LA & 15 & 0.47   & $8.5^{+4.9}_{-2.5} \times 10^{4}$ & $4.3^{+2.5}_{-1.2} \times 10^{4}$ & $5.0^{+3.0}_{-1.5} \times 10^{3}$ \\
    20LET & 12 & 0.43  & $8.9^{+5.2}_{-2.6} \times 10^{4}$ & $3.8^{+2.3}_{-1.1} \times 10^{4}$ & $3.3^{+2.0}_{-1.0} \times 10^{3}$ \\
    40LET & 22 & 0.60  & $9.2^{+5.4}_{-2.7} \times 10^{4}$ & $5.5^{+3.2}_{-1.6} \times 10^{4}$ & $7.3^{+4.3}_{-2.2} \times 10^{3}$ \\
    4020A & 20 & 0.56  & $9.1^{+5.3}_{-2.7} \times 10^{4}$ & $5.1^{+3.0}_{-1.5} \times 10^{4}$ & $6.9^{+4.0}_{-2.0} \times 10^{3}$ \\
    4020ET & 27 & 0.67 & $9.5^{+5.5}_{-2.8} \times 10^{4}$ & $6.1^{+3.6}_{-1.8} \times 10^{4}$ & $9.2^{+5.4}_{-2.7} \times 10^{3}$ \\
    \hhline{======}
    \end{tabular}
\end{table*}

\subsection{Measurement uncertainty of source parameters}
We are now in a position to study the statistical uncertainties with which different compact binary parameters will be estimated by various detector networks. Figures \ref{fig:bbh_cdf_measure}, \ref{fig:bns_cdf_measure} and \ref{fig:nsbh_cdf_measure} summarize our results for BBH, BNS and NSBH populations, respectively. 
\subsubsection{Binary Black Holes}
\begin{figure*}
\centering
\includegraphics[scale=0.33]{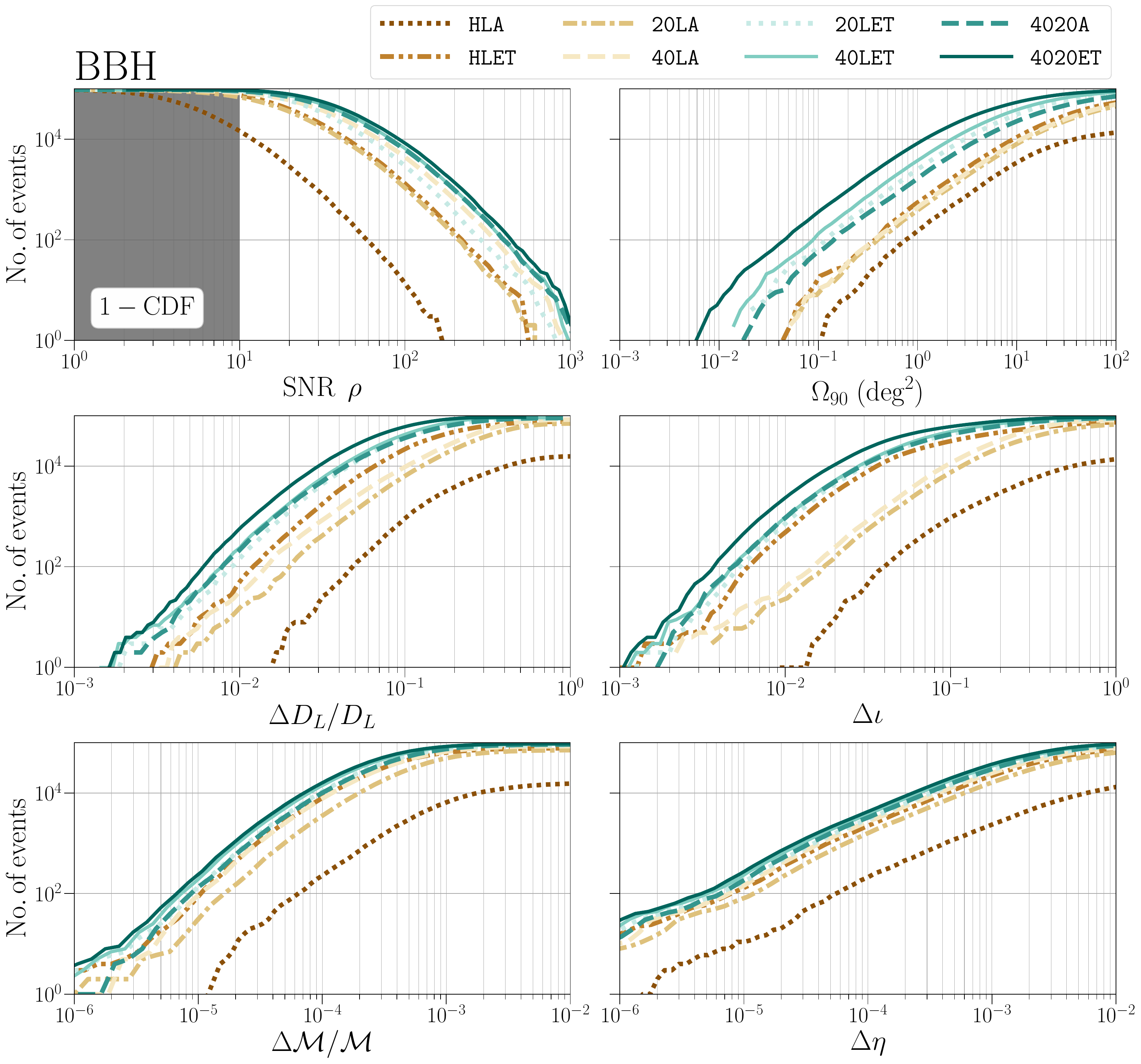}
\caption{\label{fig:bbh_cdf_measure}The scaled cumulative density function plots showing the trends in SNR $\rho$ and sky-localization $\Delta \Omega$ of the detected BBH events. It also shows the plots for fractional errors in chirp mass and luminosity distance, i.e., $\Delta \mathcal{M}/\mathcal{M}$ and $\Delta D_L/D_L$, and absolute errors in inclination angle, and symmetric mass ratio, i.e., $\Delta \iota$ and $\Delta \eta$, respectively.}
\end{figure*}

Figure ~\ref{fig:bbh_cdf_measure} shows the number of detections as a function of SNR, errors in source localization, luminosity distance, inclination angle, chirp mass and symmetric mass ratio for the population we synthesized and analysed for different network configurations. Overall, as expected, the performance of the 3 XG network is the best followed by 2 XG, 1 XG and 0 XG clearly suggesting the crucial role 3G detectors will play in the precision measurement of source parameters. 

It is seen that a 4020ET network would detect ${\cal O}(10)$ events with an SNR of 1000. These high-fidelity sources are going to play an extremely crucial role in astrophysics, cosmology and fundamental physics as they facilitate very precise inference of source parameters. The number of high-fidelity sources in a 1 XG or 0 XG network is considerably less whereas 2 XG numbers would fare comparably with 3 XG, though the numbers are slightly less.  In parameter inference, the 3 XG network performs significantly better for the angular resolution and moderately better for luminosity distance and inclination angle measurements compared to the second-best configuration 40LET thereby underscoring the importance of a 3 XG network.

Regarding individual measurements, it is interesting to note that a golden subpopulation of ${\cal O}(100)$ BBHs would allow the measurement of chirp mass and symmetric mass ratio of ${\cal O}(10^{-5})$ by a 3 XG network. Similarly, the 4020ET network would allow localization of a subpopulation of BBHs to $\leq 0.01 {\rm deg^2}$ and luminosity distance error $\leq 1\%$. These sources could allow precise measurement of Hubble constant~\cite{Borhanian:2020vyr}. Likewise, the precisely localized subpopulation would also be useful in searching for a potential EM counterpart~\cite{Graham:2020gwr}. 

\subsubsection{Binary Neutron Stars}
\begin{figure*}
\centering
\includegraphics[scale=0.33]{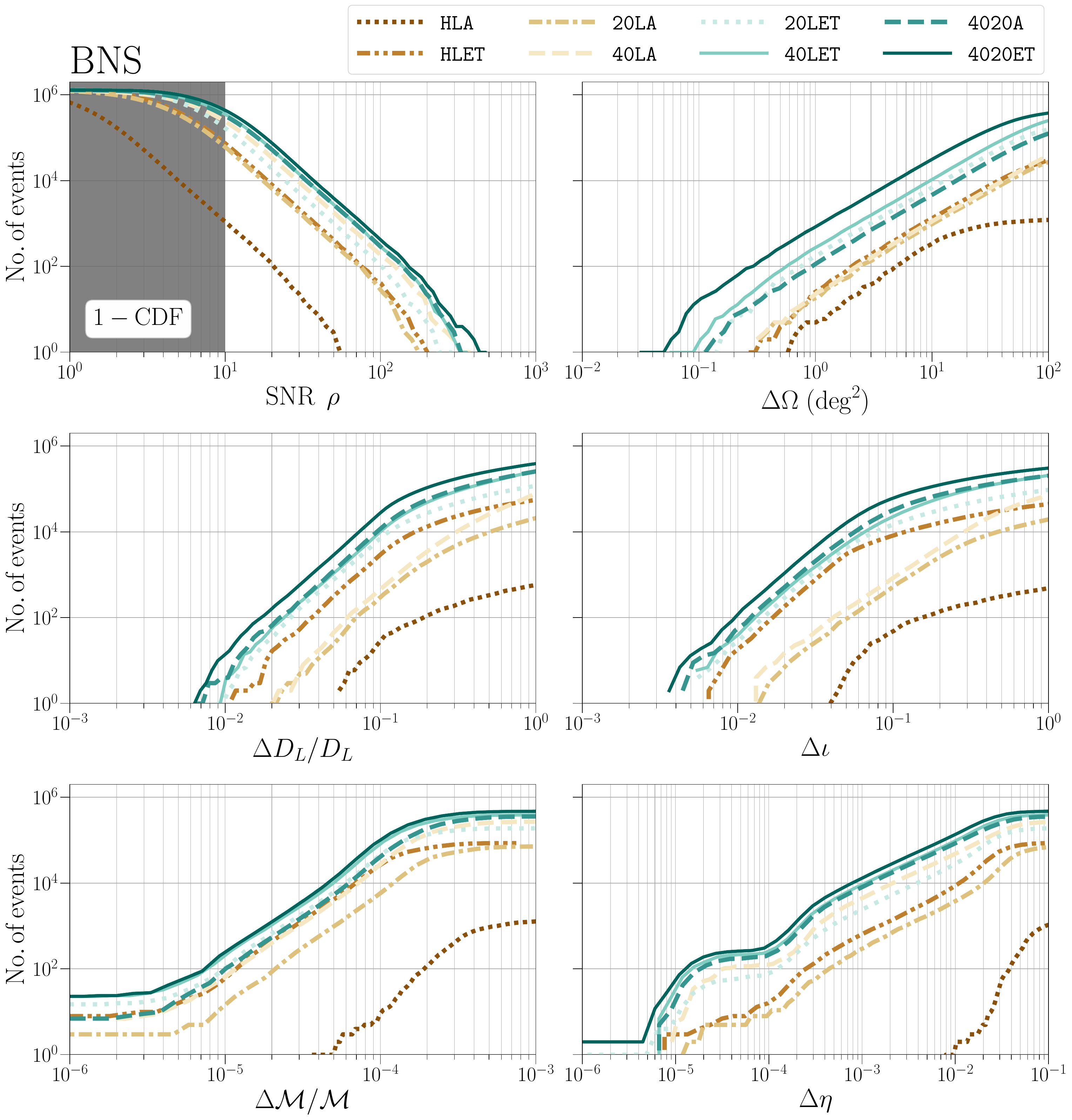}
\caption{\label{fig:bns_cdf_measure}The scaled cumulative density function plots showing the trends in SNR $\rho$ and sky-localization $\Delta \Omega$ of the detected BNS events. It also shows the plots for fractional errors in chirp mass and luminosity distance, i.e., $\Delta \mathcal{M}/\mathcal{M}$ and $\Delta D_L/D_L$, and absolute errors in inclination angle, and symmetric mass ratio, i.e., $\Delta \iota$ and $\Delta \eta$, respectively.}
\end{figure*}

Figure \ref{fig:bns_cdf_measure} displays the parameter inference in the context of BNS mergers using various network configurations.
The trends in terms of various detector networks remain as in the case of BBHs where 3 XG network performs the best followed by 2 XG, 1 XG and 0 XG networks. In terms of sources with SNR higher than 100, 3 XG and 2 XG configurations fare comparably as they detect a few hundred sources. A 3 XG network may be able to localize around 100 BNS mergers to about $0.2\,{\rm deg^2}$ whereas the best-localized 100 sources by a 2 XG network may have an angular resolution which is a factor of 3 worse at $0.6\,{\rm deg^2}$. Likewise, 3 XG detectors will measure luminosity distance of about 100 sources to less than 2\% and the performance of a 2 XG network is comparable though slightly worse, as expected. 

The inference of inclination angle is very important in this case as it may help in better understanding the structure of the relativistic jets potentially associated with these mergers and may be of immense help in the multimessenger modeling of BNS mergers. It is impressive to note that both 3 XG and 2 XG detectors will be able to measure $\Delta \iota \leq 0.01\,{\rm deg}$ for about 100 sources. The mass measurement uncertainties are even more exquisite as a few tens of sources with 3 XG detectors will permit measurement of chirp mass to $10^{-4}\%$ and close to 100 sources will be able to measure $\Delta\eta\leq 10^{-5}$. These mass measurements would potentially be the most precise measurements of NS masses. These measurements also carry a lot of importance for understanding the equation of the state of the neutron stars as well as in the measurement of their radii.

\subsubsection{Neutron star--Black holes}
\begin{figure*}
\centering
\includegraphics[scale=0.33]{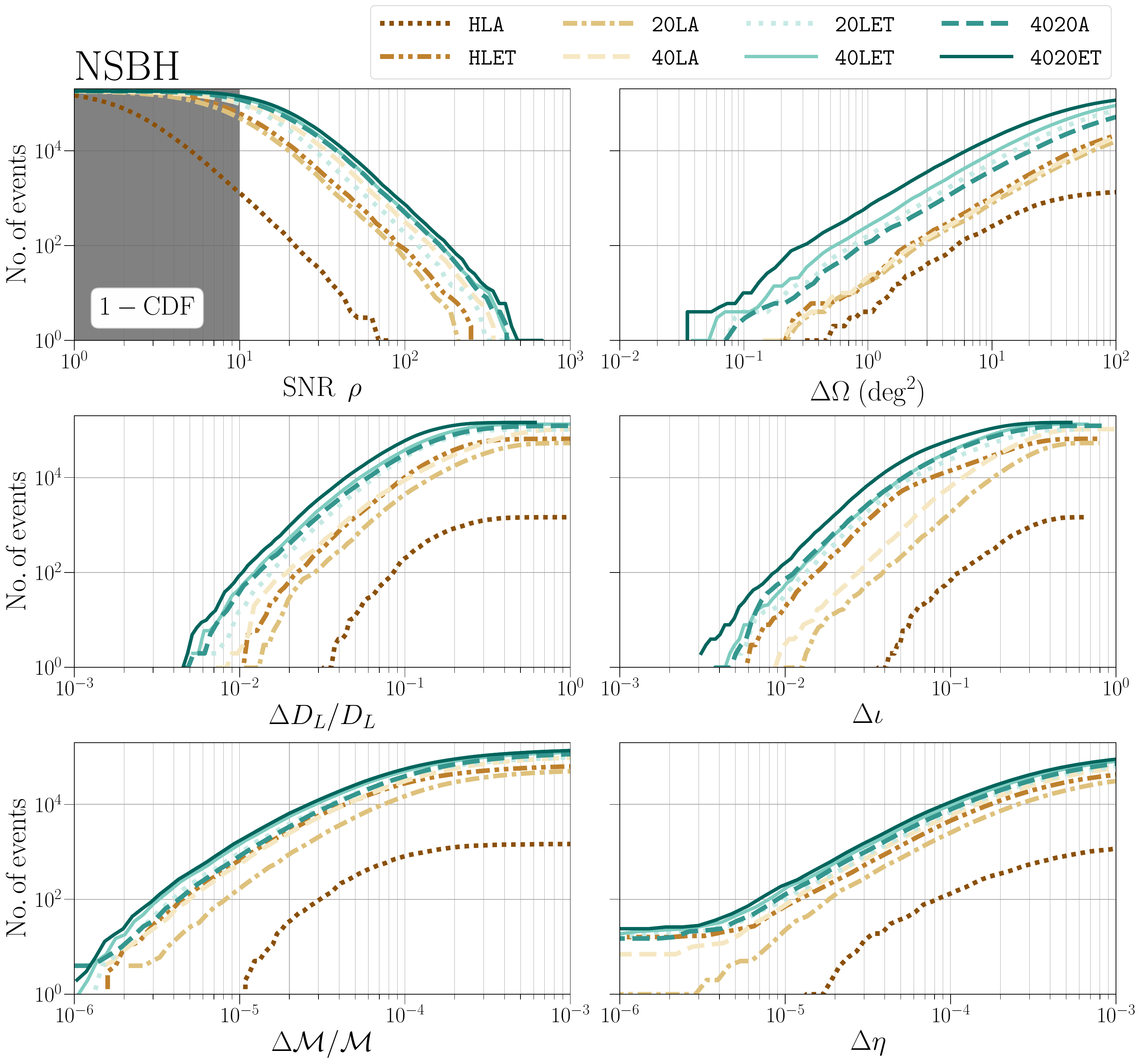}
\caption{\label{fig:nsbh_cdf_measure}The scaled cumulative density function plots showing the trends in SNR $\rho$ and sky-localization $\Delta \Omega$ of the detected NSBH events. It also shows the plots for fractional errors in chirp mass and luminosity distance, i.e., $\Delta \mathcal{M}/\mathcal{M}$ and $\Delta D_L/D_L$, and absolute errors in inclination angle, symmetric mass ratio, and spins of the BH and the NS i.e., $\Delta \iota$, and $\Delta \eta$, respectively.}
\end{figure*}

NSBH mergers, like BNS mergers, are of importance because a subset of them would be multimessenger sources. In terms of performance hierarchy, the trends seen in Fig.~\ref{fig:nsbh_cdf_measure} for NSBHs are similar to that of BBH and BNSs. The number of events with an SNR of 100 or above is close to 1000 with 3 XG and 2 XG networks while it is around 100 with 1 XG networks.
High-fidelity NSBHs with an SNR above 400 are of the order of 10 even with 3 XG and 2 XG networks.

The number of sources that are localized to better than $0.02\,{\rm deg^2}$ is close to 100 for 3 XG while it is of the order of a few tens for 2 XG and even lesser for 1 XG. Similar trends are seen for luminosity distance, where ${\cal O}(100)$ sources would allow better than 1\% measurement of luminosity distance with 3 XG and comparable, but slightly less, number of sources with 2 XG. The inclination angle may be estimated to be better than $0.01\,{\rm deg}$ for 100 sources with 3 XG and 2 XG networks. Regarding mass measurements, again, around 100 sources will permit measurement of chirp mass and symmetric mass ratio to better than $10^{-5}$. These measurements will be crucial for inferring the NS equation of state as well as understanding the ``low mass gap'' black holes with masses less than $5M_{\odot}$ which are not observed in galactic X-ray binaries.

To summarize, for all the three classes of populations any combination of networks with at least one 3G detector performs significantly better than three detectors at $A^{\sharp}$ sensitivity with 4020ET, a network with two CE and one ET, being the winner in all the metrics considered.

\subsection{3D localization of sources and early-warning} 
\label{ss:3d_loc}
In addition to the source parameters, one can also infer the sky location and the luminosity distance associated with the source from the GW data. Precise localization of the source is critical for multiple science objectives. Assuming that the cosmology is known, accurate distance estimation enables the calculation of source-frame masses of the binary objects, which are important for unraveling the mass spectrum and distinguishing between formation channels (see section \ref{ss:bh_ns_cosmos}). Localization of the source plays a crucial role in enabling multimessenger astronomy (MMA) (see section \ref{ss:mma}) and inference of cosmological parameters (see section \ref{ss:tgr_cosmology}). The localization of the source from GW observations is communicated to electromagnetic (EM) telescopes, which allows them to capture EM transients that may follow the binary merger. While the field of view (FOV) of EM telescopes is, in general, smaller than $10$ $\mathrm{deg}^2$ (see Table \ref{tab:telescope_FOVs}), they can cover multiple patches in the sky to observe large sky areas. Thus, precise localization and timely communication are necessary to facilitate MMA. 

\begin{table}
    \caption{\label{tab:telescope_FOVs}The field of view (FOV) of some of the existing and planned (in italics) electromagnetic (EM) telescopes. The space telescopes are in bold and will operate for a limited lifetime of the Cosmic Explorer facility.}
  \renewcommand{\arraystretch}{1.5} 
    \begin{tabular}{l c}
    \hhline{==}
    Telescope & FOV ($\mbox{deg}^2$)\\
    \hhline{--}
        \textit{Rubin} \cite{web:Rubin,LSST:2008ijt} & 9.6 \\
    \textit{\bf EUCLID} \cite{Euclid:2021icp} & 0.54 \\
    \textit{\bf Athena} \cite{2013arXiv1308.6785R} & 0.35 \\
    \textit{\bf Roman} \cite{Hounsell:2017ejq,Chase:2021ood} & 0.28 \\
    ngVLA \cite{Murphy2018} (2.4\,GHz; FWHM) & 0.17 \\
    \textit{\bf Chandra X-ray} \cite{web:Chandra} & 0.15 \\
    \textit{\bf Lynx} \cite{2019JATIS...5b1001G} & 0.13 \\
    \textit{Swift–XRT} \cite{2000SPIE.4140...64B} & 0.12 \\
    Keck \cite{bundy2019fobos} & 0.11 \\
    GMT \cite{GMT} & 0.11 \\
    ELT\footnote{For a brief description of the Extremely Large Telescope see \url{https://www.eso.org/sci/publications/messenger/archive/no.127-mar07/messenger-no127-11-19.pdf}} & 0.03 \\
    Jansky VLA \cite{Perley2011} (3\,GHz; FWHM) & 0.0625 \\
    \hhline{==}
    \end{tabular}
\end{table}

\begin{figure*}
  \includegraphics[scale=0.5]{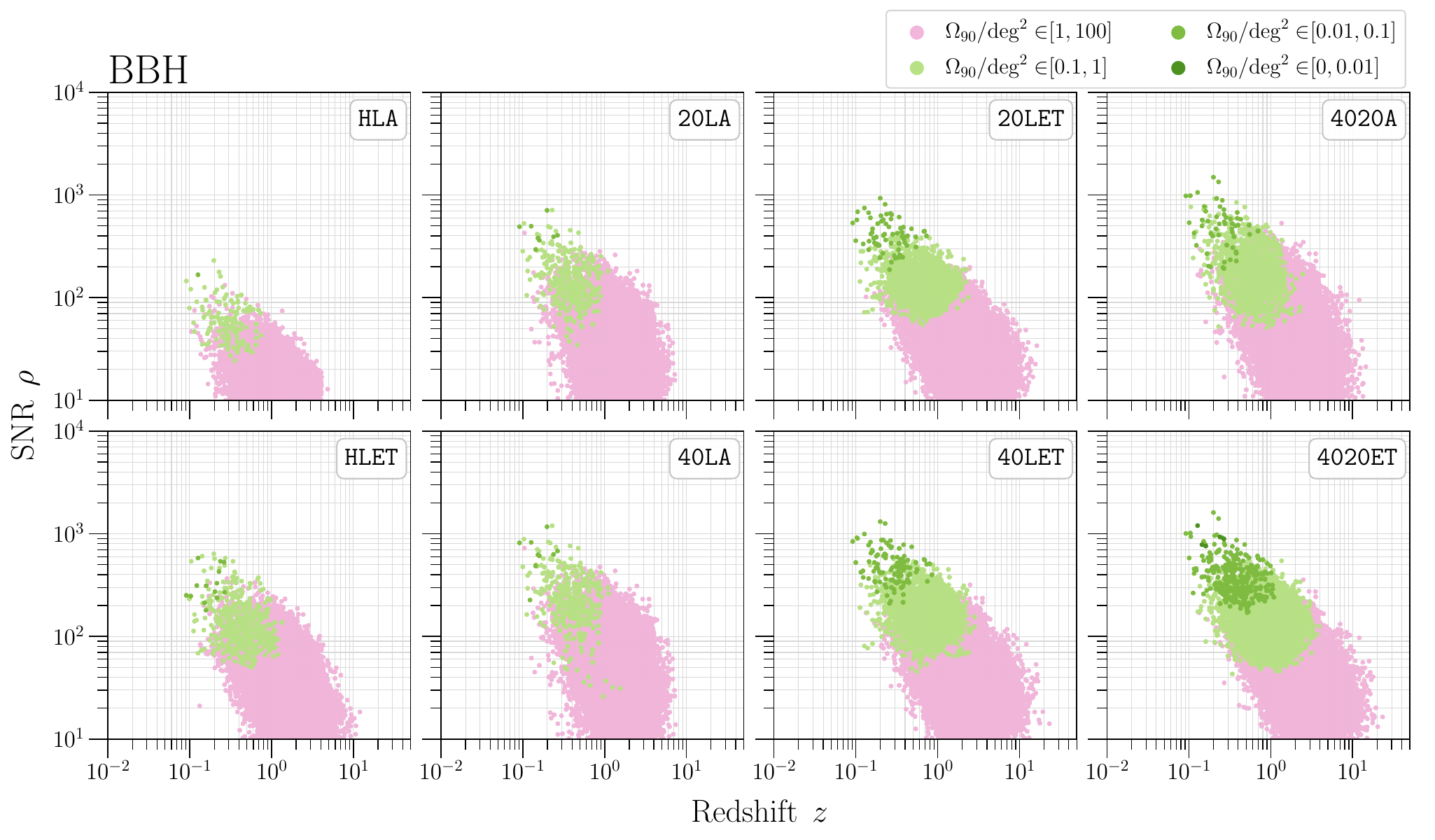}
  \caption{\label{fig:bbh_pink_scatter_no_mma}Plot showing the relationship between SNR $\rho$, sky localization $\Delta \Omega$ and the redshift $z$ for events belonging to the Pop-1 population, corresponding to the eight GW detector networks. Each marker is an event detected by the corresponding detector network in an observation time of $1$ year. The color of the marker conveys how well that event can be localized in the sky using GW observation.}
\end{figure*}

Table \ref{tab:bbh_pop_sa90_logDL} shows the number of BBH detections every year for varying precision of sky-localization and luminosity distance measurement. Without any XG detectors, a network with three \asharp detectors is only able to localize $\sim 1\%$ of all BBH mergers to a smaller area than $100$ $\mathrm{deg}^2$ in the sky. Having just one XG detector enhances this fraction to $\sim 50\%$, whereas a network with three XG detectors is able to localize $\sim 95\%$ of all BBH mergers to $\Delta\Omega\leq100\,\mathrm{deg}^2$. Further, networks with at least two XG detectors localize $\mathcal{O}(1000)$ BBH events every year to better than $1\,\mathrm{deg}^2$, which is an order of magnitude more events compared to a network containing only one XG detector. In addition, Fig. \ref{fig:bbh_pink_scatter_no_mma} also shows that only networks with three XG detectors are able to localize events to $\Delta\Omega\leq0.1\,\mathrm{deg}^2$. This metric is of particular relevance to host-galaxy identification, as the number of galaxies lying within an observation volume scales linearly with sky area. 

The luminosity distance measurement is also aided by the improved sensitivity of the XG detectors. For a network with three \asharp detectors, we can expect about 100 BBH mergers every year for which the error in luminosity distance is within $10\%$. However, luminosity distance cannot be measured to $1\%$ precision for any of the events. For networks with two or more XG detectors, not only will they detect thousands of BBH mergers every year for which $\Delta D_L/D_L \leq 0.1$, but they will also detect tens of events for which luminosity distance is measured to sub-percent precision.

For compact binary mergers involving one or more NS, localization is important to facilitate EM follow-up. GRBs, if they occur, can be detected up to large distances. In Tab. \ref{tab:bns_specific}, we give the number of BNS mergers, and the median and maximum redshifts corresponding to a particular $\Delta \Omega$ threshold. With networks that have just one XG detector, BNS events can be localized to $100$ $\mathrm{deg}^2$ in the sky up to a redshift of $z=2.2$. At a similar redshift, 4020ET can localize the event to $1$ $\mathrm{deg}^2$, and events as far as $z=8.8$ can be localized to $100$ $\mathrm{deg}^2$. Thus, XG detector networks will be capable of informing EM telescopes with precise sky localizations even for far-away events allowing them to follow up, and potentially observe, GRBs that accompany BNS mergers \cite{Ronchini2022}. 

\begin{figure*}
  \includegraphics[width=\textwidth]{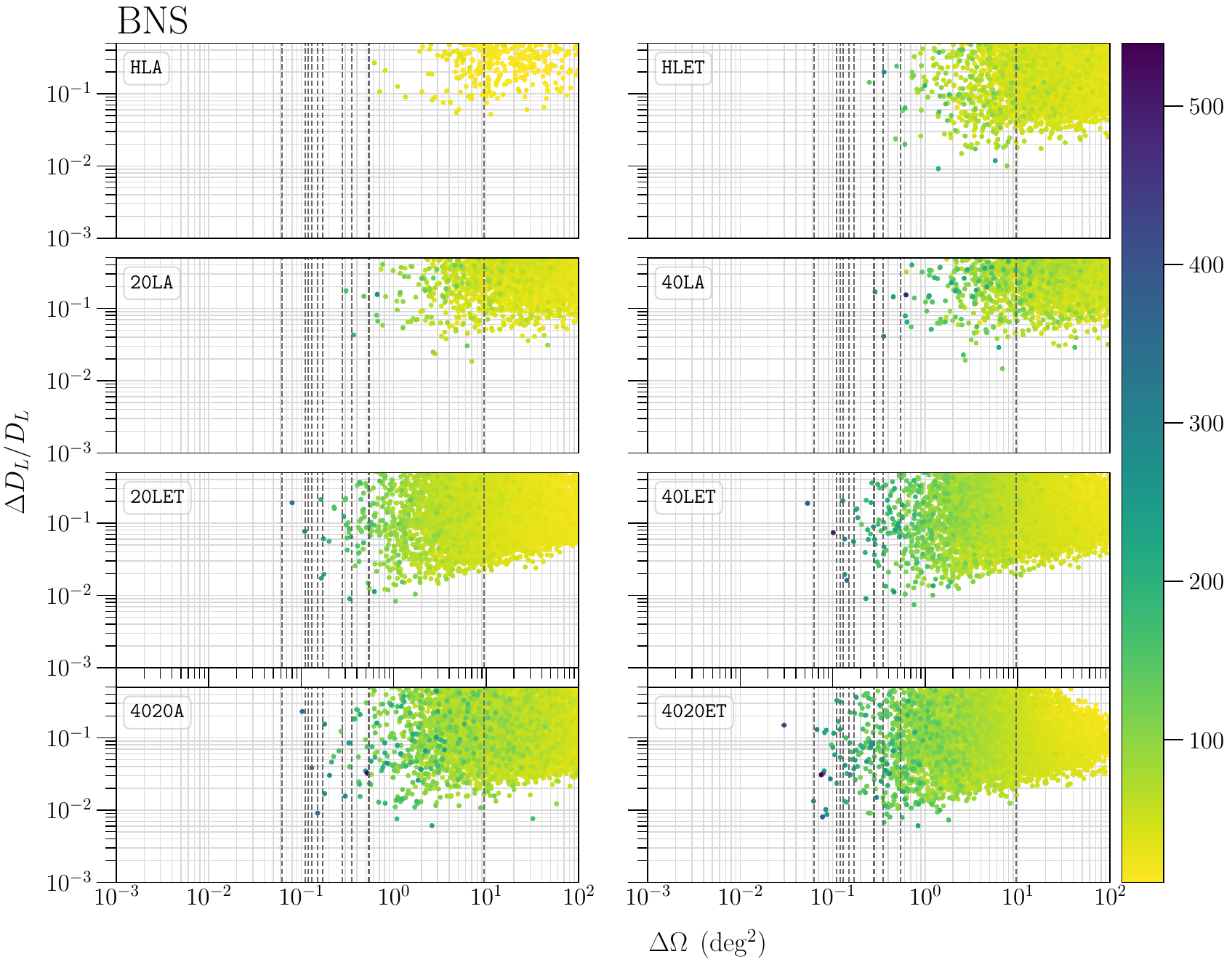}
  \caption{\label{fig:bns_yellow_scatter}The figure shows the relationship between the fractional error in the luminosity distance $\Delta D_L/D_L$, $90\%$-credible sky area $\Delta \Omega$ and the SNR (denoted by the color bar) for BNS events up to $z=0.5$. Each of these events, detected in an observation span of $1$ year, appears as a spot placed according to the associated measurement errors in luminosity distance and sky position. The color of the dots represents the SNR with which that particular event was detected in a GW detector network. The vertical dotted lines correspond to the FOV of the EM telescopes listed in Table \ref{tab:telescope_FOVs}.}
\end{figure*}

\begin{figure*}[htbp]
  \includegraphics[scale=0.6]{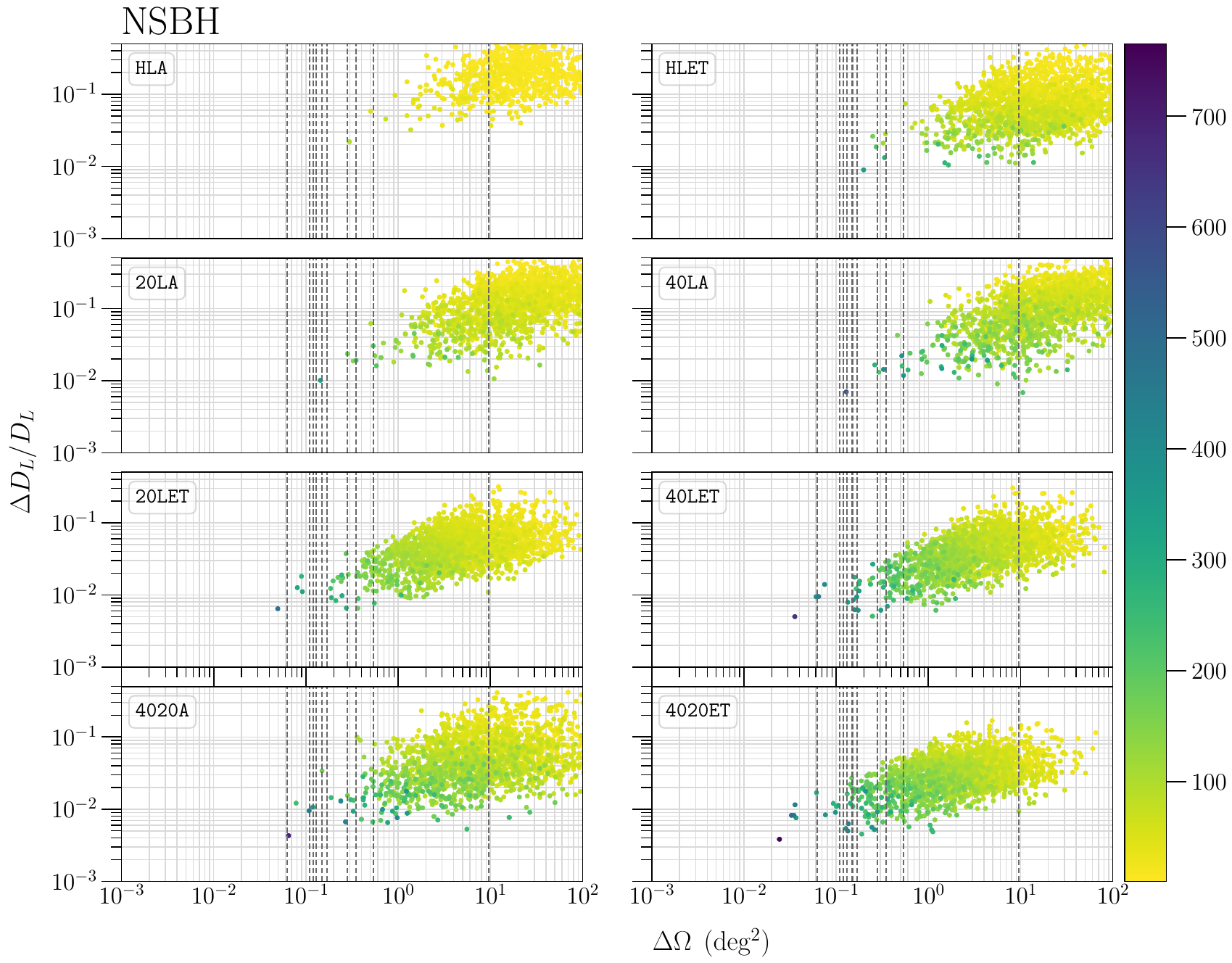}
  \caption{\label{fig:nsbh_yellow_scatter} The figure shows the relationship between the fractional error in luminosity distance $\Delta D_L/D_L$, $90\%$-credible sky area $\Delta \Omega$ and the SNR (denoted by the color bar) of NSBH events for which $z<0.5$. Each of these events, detected in an observation span of $1$ year, appears as a spot placed according to the associated measurement errors in luminosity distance and sky position. The color of the dots represents the SNR with which that particular event was detected in a GW detector network. The vertical dotted lines correspond to the FOV of multiple EM telescopes listed in Table \ref{tab:telescope_FOVs}}
\end{figure*}

Considering the specifications of the current and planned EM telescopes, kilonovae are not expected to be detected beyond $z=0.5$ \cite{Branchesi:2023mws, Gupta:2023evt}. Thus, for BNS and NSBH mergers, we select the sub-population of events that lie within $z=0.5$ and look at the ability of the different detector networks in terms of 3D localization as well as early-warning alerts. In particular, Figs. \ref{fig:bns_yellow_scatter} and \ref{fig:nsbh_yellow_scatter} show the localization of BNS and NSBH events belonging to the sub-population, respectively. The corresponding numbers are presented in Tables \ref{tab:bns_mma_sa90_logDL} and \ref{tab:nsbh_mma_sa90_logDL}. While a network with three \asharp observatories will detect $~1000$ BNS and NSBH mergers every year that are localized to within $100$ $\mathrm{deg}^2$, a network with at least two XG observatories will detect almost all BNS and NSBH mergers within $z=0.5$ with this precision. We do not see a considerable improvement in the number of events detected with $\Delta \Omega \leq 100\,\mathrm{deg}^2$ when going from networks with one XG observatory to a network with three XG observatories. A drastic improvement is only seen when we consider smaller localizations, e.g., networks with one XG observatory detect $\mathcal{O}(10)$ events with $\Delta \Omega \leq 1\,\mathrm{deg}^2$, networks with two or more XG observatories detect $\mathcal{O}(100)$ such events. A similar trend is seen for luminosity distance errors, where networks with only one XG observatory detect $\mathcal{O}(1)$ events for which luminosity distance is measured to sub-percent precision, whereas this number increases to $\mathcal{O}(10)$ events for a three-XG network. ET is particularly good at sky localization due to its better low-frequency sensitivity compared to others as signals last longer in its sensitivity band, which in turn causes amplitude modulation due to the changing antenna pattern in the direction of the source.

\begin{figure*}
    \centering
    \includegraphics[width=\textwidth]{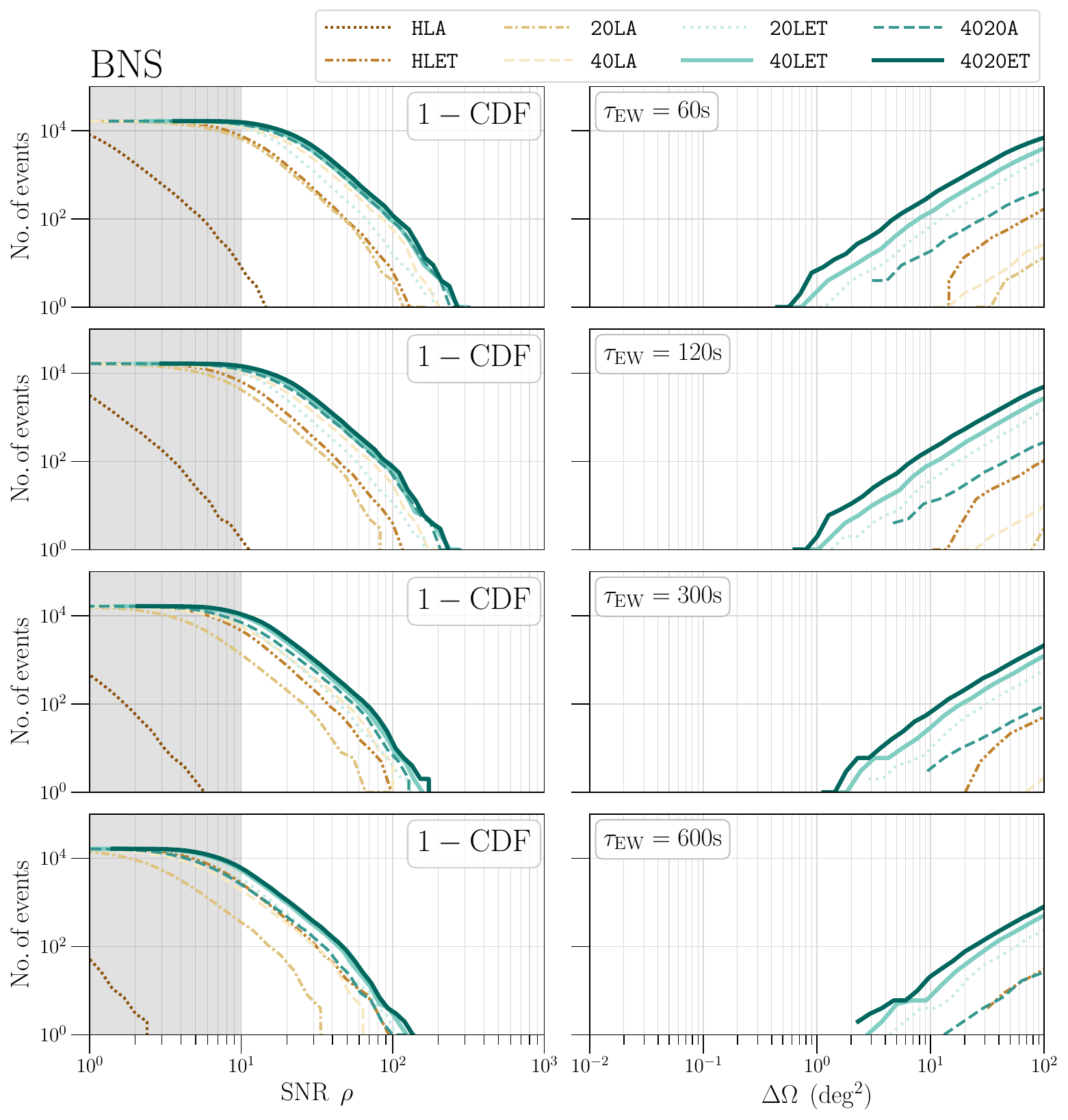}
    \caption{The scaled CDF plots for BNS events belonging to the multimessenger sub-population for which early-warning alerts can be sent $1$ minute, $2$ minutes, $5$ minutes, and $10$ minutes before their respective mergers.}
    \label{fig:bns_ew}
\end{figure*}

\begin{figure*}[htbp]
  \includegraphics[width=\textwidth]{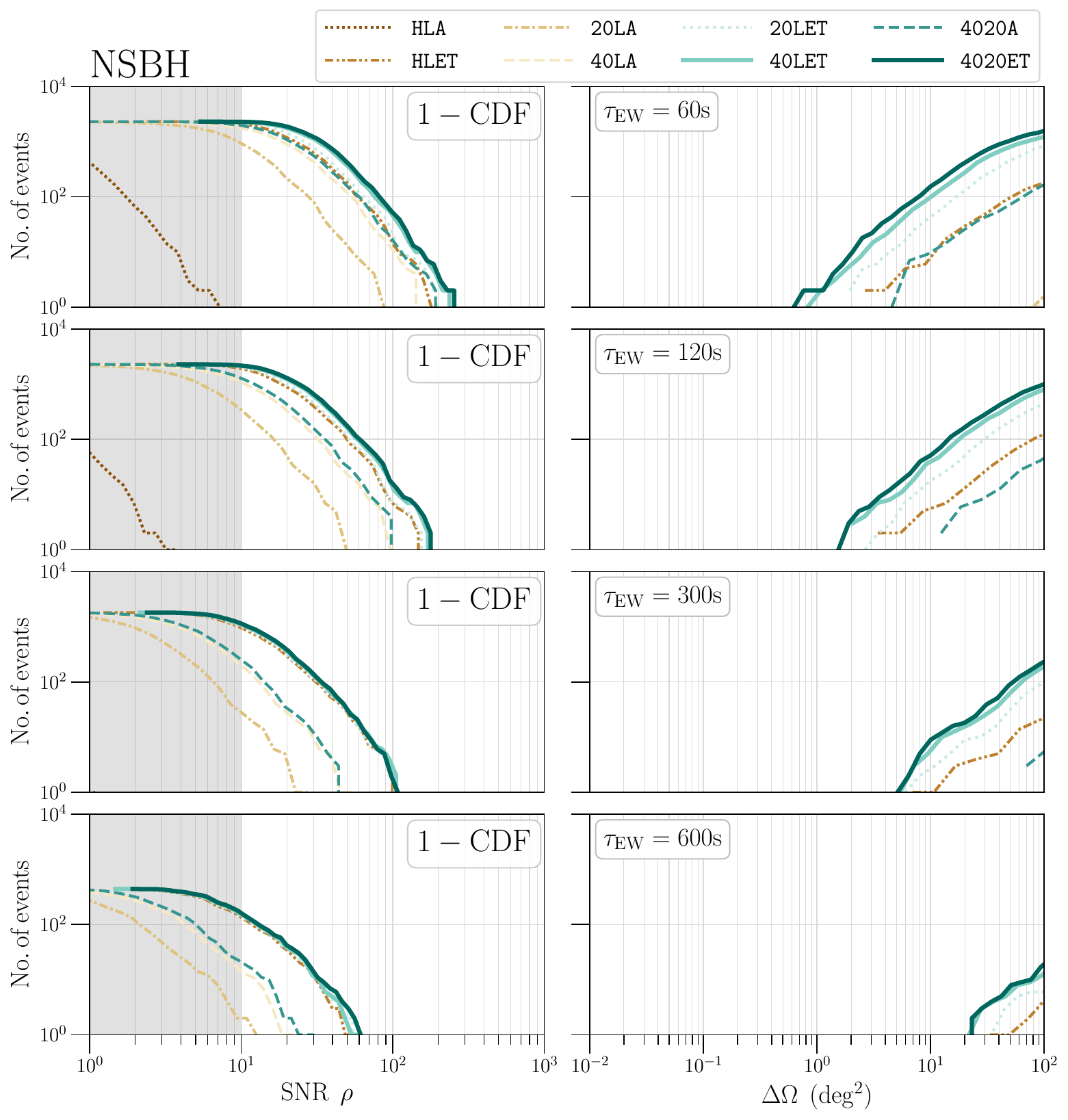}
  \caption{\label{fig:nsbh_ew_CDFs}The scaled CDF plots for NSBH events belonging to the multimessenger sub-population for which early-warning alerts can be sent $1$ minute, $2$ minutes, $5$ minutes, and $10$ minutes before their respective mergers.}
\end{figure*}

Another important step towards increasing the efficiency of the EM follow-up is the timely communication of the localization of the merger event to EM telescopes. BNS and NSBH mergers can remain in the sensitive band of the GW networks long enough such that the telescopes can be alerted even before the detection. However, the earlier the alert is sent, the lesser the amount of information that was extracted from the GW signal, leading to worse sky localization compared to if the alert was sent at the time of merger. This results in the trade-off between how early an alert is sent, and how well the event can be localized at that time. In Figs. \ref{fig:bns_ew} and \ref{fig:nsbh_ew_CDFs}, we show the number of events BNS and NSBH events for which the alert can be sent $60\,$s, $120\,$s, $300\,$s, and $600\,$s before the merger, and the corresponding SNR and $\Delta \Omega$. The corresponding numbers are listed in Tables \ref{tab:bns_ew_sa90} and \ref{tab:nsbh_ew_sa90}. We see that the prospects of sending early-warning alerts with the \asharp network are not promising, with only 10 BNS events and no NSBH events for which the alert can be sent $1$ minute before the merger. This improves drastically for networks with two or more XG detectors, where the alerts can be sent $60\,$s prior to the merger for $\mathcal{O}(1000)$ BNS and NSBH detections, for each of which $\Delta \Omega \leq 100\,\mathrm{deg}^2$. In fact, for the same sky localization, alerts can be sent for $\mathcal{O}(100)$ BNS mergers and $\mathcal{O}(10)$ NSBH mergers $10$ minutes before the merger. 

\begin{table*}
  \centering
  \caption{\label{tab:bbh_pop_sa90_logDL}The number of BBH detections per year for the six detector networks with $90\%$-credible sky area less than $10$, $1$, $0.1$ and $0.01$ $\mbox{deg}^2$ and fractional error in luminosity distance less than $0.1$ and $0.01$.}
  \renewcommand{\arraystretch}{1.5} 
    \begin{tabular}{l | P{2.1cm} P{2.1cm} P{2.1cm} P{2.3cm} P{2cm} | P{2.1cm} P{2.3cm}}
    \hhline{========}
    Metric & \multicolumn{5}{c|}{$\Delta \Omega\mbox{ (deg)}^2$} & \multicolumn{2}{c}{$\Delta D_L / D_L$} \\
    \hhline{--------}
    Quality & $\leq 100$ & $\leq 10$ & $\leq 1$ & $\leq 0.1$ & $\leq 0.01$ & $\leq 0.1$ & $\leq 0.01$ \\
    \hhline{--------}
    HLA & $1.3^{+7.8}_{-0.4} \times 10^{4}$ & $3.1^{+2.0}_{-0.9} \times 10^{3}$ & $1.3^{+7.2}_{-0.4} \times 10^{2}$ & $1.0^{+4.0}_{-0.0} \times 10^{0}$ & $0$ & $8.4^{+4.5}_{-2.4} \times 10^{2}$ & $0$ \\
    HLET & $5.2^{+3.0}_{-1.5} \times 10^{4}$ & $9.9^{+5.8}_{-2.9} \times 10^{3}$ & $4.8^{+2.7}_{-1.4} \times 10^{2}$ & $1.4^{+1.2}_{-0.8} \times 10^{1}$ & $0$ & $2.1^{+1.2}_{-0.6} \times 10^{4}$ & $4.2^{+2.2}_{-1.4} \times 10^{1}$ \\
    20LA & $4.4^{+2.5}_{-1.3} \times 10^{4}$ & $7.3^{+4.3}_{-2.1} \times 10^{3}$ & $3.4^{+2.0}_{-1.0} \times 10^{2}$ & $8.0^{+5.0}_{-6.0} \times 10^{0}$ & $0$ & $5.9^{+3.3}_{-1.7} \times 10^{3}$ & $1.3^{+5.0}_{-0.5} \times 10^{1}$ \\
    40LA & $4.8^{+2.8}_{-1.4} \times 10^{4}$ & $8.1^{+4.8}_{-2.4} \times 10^{3}$ & $3.8^{+2.3}_{-1.1} \times 10^{2}$ & $9.0^{+6.0}_{-7.0} \times 10^{0}$ & $0$ & $9.0^{+5.2}_{-2.6} \times 10^{3}$ & $2.8^{+1.2}_{-0.7} \times 10^{1}$ \\
    20LET & $8.1^{+4.7}_{-2.3} \times 10^{4}$ & $2.9^{+1.7}_{-0.8} \times 10^{4}$ & $2.3^{+1.4}_{-0.7} \times 10^{3}$ & $6.6^{+4.3}_{-2.3} \times 10^{1}$ & $0$ & $3.6^{+2.1}_{-1.1} \times 10^{4}$ & $1.2^{+7.0}_{-0.4} \times 10^{2}$ \\
    40LET & $8.5^{+5.0}_{-2.5} \times 10^{4}$ & $3.7^{+2.1}_{-1.1} \times 10^{4}$ & $3.4^{+2.1}_{-1.0} \times 10^{3}$ & $1.1^{+6.8}_{-0.4} \times 10^{2}$ & $0.0^{+1.0}_{-0.0} \times 10^{0}$ & $4.1^{+2.4}_{-1.2} \times 10^{4}$ & $2.1^{+1.2}_{-0.7} \times 10^{2}$ \\
    4020A & $7.0^{+4.1}_{-2.0} \times 10^{4}$ & $2.1^{+1.2}_{-0.6} \times 10^{4}$ & $1.4^{+8.9}_{-0.4} \times 10^{3}$ & $5.2^{+3.3}_{-2.2} \times 10^{1}$ & $0$ & $3.4^{+2.0}_{-1.0} \times 10^{4}$ & $1.9^{+1.2}_{-0.5} \times 10^{2}$ \\
    4020ET & $9.1^{+5.3}_{-2.7} \times 10^{4}$ & $5.2^{+3.0}_{-1.5} \times 10^{4}$ & $7.6^{+4.5}_{-2.2} \times 10^{3}$ & $3.1^{+1.8}_{-0.9} \times 10^{2}$ & $5.0^{+5.0}_{-3.0} \times 10^{0}$ & $5.9^{+3.4}_{-1.7} \times 10^{4}$ & $5.1^{+2.8}_{-1.5} \times 10^{2}$ \\
    \hhline{========}
    \end{tabular}
\end{table*}

\begin{table*}[htbp] 
  \centering
  \caption{\label{tab:bns_specific}\# of BNS mergers every year, and the median and maximum redshift up to which the events are detected associated with the particular $\Delta \Omega$ criteria. These numbers were calculated using the median local merger rates for BNS (320 $\mbox{Gpc}^3\mbox{yr}^{-1}$).}
  \renewcommand{\arraystretch}{1.5} 
    \begin{tabular}{P{3cm}| P{1.5cm} | P{1.5cm} P{1.5cm} P{1.5cm} | P{1.5cm} P{1.5cm} P{1.5cm} | P{1.5cm}}
    \hhline{=========}
    \multirow{2}{*}{Quantity} & \textbf{0 XG} & \multicolumn{3}{c|}{\textbf{1 XG}} & \multicolumn{3}{c|}{\textbf{2 XG}} & \textbf{3 XG} \\
    \hhline{~--------}
    & HLA & HLET & 20LA & 40LA & 20LET & 40LET & 4020A & 4020ET \\
    \hhline{=========}
    \multicolumn{9}{c}{$\Delta \Omega \leq 1$ $\mathrm{deg}^2$} \\
    \hhline{---------}
    Number & $5$ & $24$ & $17$ & $18$ & $160$ & $250$ & $100$ & $754$ \\
    Median $z$ & $0.06$ & $0.09$ & $0.07$ & $0.08$ & $0.10$ & $0.13$ & $0.11$ & $0.19$ \\
    Maximum $z$ & $0.10$ & $0.16$ & $0.13$ & $0.13$ & $0.23$ & $0.287$ & $0.243$ & $0.503$ \\
    \hhline{---------}
    \multicolumn{9}{c}{$\Delta \Omega \leq 10$ $\mathrm{deg}^2$} \\
    \hhline{---------}
    Number & $320$ & $1200$ & $870$ & $980$ & $6200$ & $9400$ & $4000$ & $28\,000$ \\
    Median $z$ & $0.15$ & $0.21$ & $0.22$ & $0.20$ & $0.36$ & $0.41$ & $0.33$ & $0.60$ \\
    Maximum $z$ & $0.36$ & $0.53$ & $0.52$ & $0.53$ & $1.1$ & $1.30$ & $1.06$ & $2.12$ \\
    \hhline{---------}
    \multicolumn{9}{c}{$\Delta \Omega \leq 100$ $\mathrm{deg}^2$} \\
    \hhline{---------}
    Number & $1200$ & $28\,000$ & $25\,000$ & $34\,000$ & $150\,000$ & $230\,000$ & $110\,000$ & $360\,000$ \\
    Median $z$ & $0.21$ & $0.60$ & $0.63$ & $0.68$ & $1.03$ & $1.19$ & $0.98$ & $1.35$ \\
    Maximum $z$ & $0.52$ & $2.12$ & $2.23$ & $2.23$ & $3.68$ & $5.89$ & $3.65$ & $8.80$ \\
    \hhline{=========}
    \end{tabular}
\end{table*}

\begin{table*}
  \caption{\label{tab:bns_mma_sa90_logDL}For the BNS sub-population with events for which $z<0.5$, the table lists the number of detections per year for the six detector networks with $90\%$-credible sky area $\Delta \Omega < 10$, $1$, $0.1$ and $0.01$ $\mbox{deg}^2$ and fractional error in luminosity distance $\Delta D_L / D_L < 0.1$ and $0.01$.}
  \renewcommand{\arraystretch}{1.5} 
    \begin{tabular}{l | P{2.1cm} P{2.1cm} P{2.1cm} P{2.3cm} P{2cm} | P{2.1cm} P{2.3cm}}
    \hhline{========}
    Metric & \multicolumn{5}{c|}{$\Delta \Omega\mbox{ (deg)}^2$} & \multicolumn{2}{c}{$\Delta D_L / D_L$} \\
    \hhline{--------}
    Quality & $\leq 100$ & $\leq 10$ & $\leq 1$ & $\leq 0.1$ & $\leq 0.01$ & $\leq 0.1$ & $\leq 0.01$ \\
    \hhline{--------}
    HLA & $1.2^{+1.8}_{-0.9} \times 10^{3}$ & $3.2^{+4.7}_{-2.5} \times 10^{2}$ & $5.0^{+11.0}_{-5.0} \times 10^{0}$ & $0$ & $0$ & $2.6^{+4.2}_{-2.3} \times 10^{1}$ & $0$ \\
    HLET & $1.0^{+1.5}_{-0.8} \times 10^{4}$ & $1.2^{+1.8}_{-0.9} \times 10^{3}$ & $2.4^{+4.7}_{-2.1} \times 10^{1}$ & $0.0^{+3.0}_{-0.0} \times 10^{0}$ & $0$ & $2.3^{+3.4}_{-1.7} \times 10^{3}$ & $1.0^{+2.0}_{-1.0} \times 10^{0}$ \\
    20LA & $8.6^{+13.3}_{-6.4} \times 10^{3}$ & $8.6^{+12.9}_{-6.8} \times 10^{2}$ & $1.7^{+3.3}_{-1.5} \times 10^{1}$ & $0$ & $0$ & $2.4^{+4.2}_{-1.9} \times 10^{2}$ & $0$ \\
    40LA & $9.8^{+15.1}_{-7.3} \times 10^{3}$ & $9.7^{+14.6}_{-7.6} \times 10^{2}$ & $1.8^{+3.8}_{-1.6} \times 10^{1}$ & $0$ & $0$ & $3.1^{+5.4}_{-2.4} \times 10^{2}$ & $0.0^{+2.0}_{-0.0} \times 10^{0}$ \\
    20LET & $1.5^{+2.3}_{-1.1} \times 10^{4}$ & $4.9^{+7.4}_{-3.7} \times 10^{3}$ & $1.6^{+2.4}_{-1.3} \times 10^{2}$ & $1.0^{+6.0}_{-1.0} \times 10^{0}$ & $0$ & $4.4^{+6.9}_{-3.3} \times 10^{3}$ & $2.0^{+6.0}_{-2.0} \times 10^{0}$ \\
    40LET & $1.6^{+2.4}_{-1.2} \times 10^{4}$ & $6.3^{+9.7}_{-4.8} \times 10^{3}$ & $2.5^{+3.8}_{-2.0} \times 10^{2}$ & $1.0^{+9.0}_{-1.0} \times 10^{0}$ & $0$ & $4.9^{+7.7}_{-3.7} \times 10^{3}$ & $2.0^{+9.0}_{-2.0} \times 10^{0}$ \\
    4020A & $1.4^{+2.1}_{-1.0} \times 10^{4}$ & $3.4^{+5.3}_{-2.6} \times 10^{3}$ & $9.7^{+15.7}_{-7.7} \times 10^{1}$ & $0.0^{+4.0}_{-0.0} \times 10^{0}$ & $0$ & $4.5^{+6.9}_{-3.4} \times 10^{3}$ & $4.0^{+11.0}_{-4.0} \times 10^{0}$ \\
    4020ET & $1.6^{+2.5}_{-1.2} \times 10^{4}$ & $1.0^{+1.5}_{-0.8} \times 10^{4}$ & $7.5^{+11.4}_{-5.8} \times 10^{2}$ & $1.3^{+2.9}_{-1.2} \times 10^{1}$ & $0.0^{+2.0}_{-0.0} \times 10^{0}$ & $8.5^{+13.1}_{-6.4} \times 10^{3}$ & $1.2^{+2.2}_{-1.2} \times 10^{1}$ \\
    \hhline{========}
    \end{tabular}
\end{table*}

\begin{table*}
  \centering
  \caption{\label{tab:nsbh_mma_sa90_logDL}For the NSBH sub-population with events for which $z<0.5$, the table lists the number of detections per year for the six detector networks with $90\%$-credible sky area $\Delta \Omega < 10$, $1$, $0.1$ and $0.01$ $\mbox{deg}^2$ and fractional error in luminosity distance $\Delta D_L / D_L < 0.1$ and $0.01$.}
  \renewcommand{\arraystretch}{1.5} 
    \begin{tabular}{l | P{2.1cm} P{2.1cm} P{2.1cm} P{2.3cm} P{2cm} | P{2.1cm} P{2.3cm}}
    \hhline{========}
    Metric & \multicolumn{5}{c|}{$\Delta \Omega\mbox{ (deg)}^2$} & \multicolumn{2}{c}{$\Delta D_L / D_L$} \\
    \hhline{--------}
    Quality & $\leq 100$ & $\leq 10$ & $\leq 1$ & $\leq 0.1$ & $\leq 0.01$ & $\leq 0.1$ & $\leq 0.01$ \\
    \hhline{--------}
    HLA & $9.4^{+19.7}_{-8.1} \times 10^{2}$ & $2.2^{+5.3}_{-1.9} \times 10^{2}$ & $4.0^{+13.0}_{-4.0} \times 10^{0}$ & $0$ & $0$ & $1.4^{+3.6}_{-1.3} \times 10^{2}$ & $0$ \\
    HLET & $2.0^{+4.3}_{-1.7} \times 10^{3}$ & $6.9^{+15.0}_{-6.0} \times 10^{2}$ & $1.6^{+6.8}_{-1.6} \times 10^{1}$ & $0.0^{+1.0}_{-0.0} \times 10^{0}$ & $0$ & $1.5^{+3.2}_{-1.3} \times 10^{3}$ & $1.0^{+17.0}_{-1.0} \times 10^{0}$ \\
    20LA & $1.9^{+4.1}_{-1.6} \times 10^{3}$ & $5.3^{+11.0}_{-4.5} \times 10^{2}$ & $1.5^{+4.4}_{-1.5} \times 10^{1}$ & $0$ & $0$ & $7.4^{+16.0}_{-6.3} \times 10^{2}$ & $0.0^{+6.0}_{-0.0} \times 10^{0}$ \\
    40LA & $2.0^{+4.2}_{-1.6} \times 10^{3}$ & $5.8^{+12.3}_{-5.0} \times 10^{2}$ & $1.8^{+5.5}_{-1.8} \times 10^{1}$ & $0$ & $0$ & $9.7^{+20.7}_{-8.2} \times 10^{2}$ & $3.0^{+31.0}_{-3.0} \times 10^{0}$ \\
    20LET & $2.3^{+4.9}_{-1.9} \times 10^{3}$ & $1.6^{+3.4}_{-1.3} \times 10^{3}$ & $1.4^{+3.4}_{-1.2} \times 10^{2}$ & $4.0^{+7.0}_{-4.0} \times 10^{0}$ & $0$ & $2.1^{+4.5}_{-1.7} \times 10^{3}$ & $1.4^{+6.1}_{-1.3} \times 10^{1}$ \\
    40LET & $2.3^{+4.9}_{-1.9} \times 10^{3}$ & $1.8^{+3.8}_{-1.5} \times 10^{3}$ & $2.3^{+5.4}_{-2.0} \times 10^{2}$ & $4.0^{+10.0}_{-4.0} \times 10^{0}$ & $0$ & $2.1^{+4.6}_{-1.8} \times 10^{3}$ & $3.5^{+11.4}_{-2.8} \times 10^{1}$ \\
    4020A & $2.2^{+4.8}_{-1.8} \times 10^{3}$ & $1.3^{+2.6}_{-1.1} \times 10^{3}$ & $1.0^{+2.2}_{-0.9} \times 10^{2}$ & $2.0^{+4.0}_{-2.0} \times 10^{0}$ & $0$ & $1.8^{+4.0}_{-1.6} \times 10^{3}$ & $2.3^{+9.1}_{-2.0} \times 10^{1}$ \\
    4020ET & $2.3^{+4.9}_{-1.9} \times 10^{3}$ & $2.1^{+4.5}_{-1.7} \times 10^{3}$ & $5.0^{+10.9}_{-4.2} \times 10^{2}$ & $9.0^{+46.0}_{-9.0} \times 10^{0}$ & $0$ & $2.3^{+4.9}_{-1.9} \times 10^{3}$ & $6.2^{+19.8}_{-5.4} \times 10^{1}$ \\
    \hhline{========}
    \end{tabular}
\end{table*}

\begin{table*}[htbp] 
  \centering
  \caption{\label{tab:bns_ew_sa90}The number of BNS detections per year for the GW detector networks for which an EW alert can be sent $60$ s, $120$ s, $300$ s and $600$ s before the merger, with $90\%$-credible sky area measured to be better than $100$, $10$, $1$ $\mbox{deg}^2$ at the time when the alert is sent.}
  \renewcommand{\arraystretch}{1.5} 
    \begin{tabular}{l P{2.1cm}P{2.2cm}P{2.2cm} P{2.1cm}P{2.2cm}P{2.2cm}}
    \hhline{=======}
    EW Time & \multicolumn{3}{c}{$\tau_{\mathrm{EW}} = 60\mbox{ s}$} & \multicolumn{3}{c}{$\tau_{\mathrm{EW}} = 120\mbox{ s}$} \\
    \hhline{-------}
    $\Delta \Omega\,(\mbox{deg}^2)$ & $\leq 100$ & $\leq 10$ & $\leq 1$ & $\leq 100$ & $\leq 10$ & $\leq 1$ \\
    \hhline{-------}
    HLA & $0.0^{+1.0}_{-0.0} \times 10^{0}$ & $0$ & $0$ & $0$ & $0$ & $0$ \\
    HLET & $1.3^{+2.4}_{-1.1} \times 10^{2}$ & $1.0^{+10.0}_{-1.0} \times 10^{0}$ & $0$ & $8.3^{+15.7}_{-6.9} \times 10^{1}$ & $1.0^{+5.0}_{-1.0} \times 10^{0}$ & $0$ \\
    20LA & $5.0^{+10.0}_{-4.0} \times 10^{0}$ & $0$ & $0$ & $2.0^{+1.0}_{-1.0} \times 10^{0}$ & $0$ & $0$ \\
    40LA & $7.0^{+19.0}_{-6.0} \times 10^{0}$ & $0$ & $0$ & $3.0^{+6.0}_{-2.0} \times 10^{0}$ & $0$ & $0$ \\
    20LET & $2.0^{+3.2}_{-1.6} \times 10^{3}$ & $4.9^{+9.7}_{-4.0} \times 10^{1}$ & $1.0^{+3.0}_{-1.0} \times 10^{0}$ & $1.2^{+1.8}_{-0.9} \times 10^{3}$ & $3.0^{+5.5}_{-2.4} \times 10^{1}$ & $0.0^{+2.0}_{-0.0} \times 10^{0}$ \\
    40LET & $3.4^{+5.2}_{-2.6} \times 10^{3}$ & $1.2^{+1.9}_{-0.9} \times 10^{2}$ & $2.0^{+4.0}_{-2.0} \times 10^{0}$ & $2.3^{+3.5}_{-1.7} \times 10^{3}$ & $7.4^{+12.0}_{-6.3} \times 10^{1}$ & $1.0^{+2.0}_{-1.0} \times 10^{0}$ \\
    4020A & $3.7^{+6.2}_{-2.8} \times 10^{2}$ & $1.5^{+2.3}_{-1.2} \times 10^{1}$ & $0$ & $2.2^{+3.1}_{-1.7} \times 10^{2}$ & $1.1^{+1.0}_{-0.9} \times 10^{1}$ & $0$ \\
    4020ET & $6.3^{+9.4}_{-4.7} \times 10^{3}$ & $2.7^{+4.5}_{-2.1} \times 10^{2}$ & $5.0^{+12.0}_{-4.0} \times 10^{0}$ & $4.4^{+6.6}_{-3.3} \times 10^{3}$ & $1.5^{+2.6}_{-1.2} \times 10^{2}$ & $1.0^{+4.0}_{-1.0} \times 10^{0}$ \\
    \hhline{=======}
    EW Time & \multicolumn{3}{c}{$\tau_{\mathrm{EW}} = 300\mbox{ s}$} & \multicolumn{3}{c}{$\tau_{\mathrm{EW}} = 600\mbox{ s}$} \\
    \hhline{-------}
    $\Delta \Omega\,(\mbox{deg}^2)$ & $\leq 100$ & $\leq 10$ & $\leq 1$ & $\leq 100$ & $\leq 10$ & $\leq 1$ \\
    \hhline{-------}
    HLA & $0$ & $0$ & $0$ & $0$ & $0$ & $0$ \\ 
    HLET & $4.2^{+7.9}_{-3.5} \times 10^{1}$ & $0.0^{+2.0}_{-0.0} \times 10^{0}$ & $0$ & $2.4^{+4.3}_{-1.9} \times 10^{1}$ & $0.0^{+1.0}_{-0.0} \times 10^{0}$ & $0$ \\ 
    20LA & $0$ & $0$ & $0$ & $0$ & $0$ & $0$ \\ 
    40LA & $0$ & $0$ & $0$ & $0$ & $0$ & $0$ \\ 
    CE20LET & $4.7^{+7.6}_{-3.6} \times 10^{2}$ & $7.0^{+26.0}_{-6.0} \times 10^{0}$ & $0$ & $2.0^{+3.2}_{-1.6} \times 10^{2}$ & $4.0^{+11.0}_{-4.0} \times 10^{0}$ & $0$ \\ 
    40LET & $1.0^{+15.9}_{-0.8} \times 10^{3}$ & $2.2^{+53.0}_{-1.7} \times 10^{1}$ & $0.0^{+1.0}_{-0.0} \times 10^{0}$ & $4.1^{+6.7}_{-3.2} \times 10^{2}$ & $6.0^{+22.0}_{-5.0} \times 10^{0}$ & $0$ \\ 
    4020A & $6.2^{+8.5}_{-5.2} \times 10^{1}$ & $2.0^{+0.0}_{-2.0} \times 10^{0}$ & $0$ & $1.9^{+2.0}_{-1.6} \times 10^{1}$ & $0$ & $0$ \\ 
    4020ET & $1.8^{+28.6}_{-1.4} \times 10^{3}$ & $5.2^{+9.3}_{-4.3} \times 10^{1}$ & $0.0^{+2.0}_{-0.0} \times 10^{0}$ & $6.8^{+11.2}_{-5.3} \times 10^{2}$ & $1.5^{+3.7}_{-1.2} \times 10^{1}$ & $0.0^{+1.0}_{-0.0} \times 10^{0}$ \\    
    \hhline{=======}
    \end{tabular}
\end{table*}

\begin{table*}[htbp] 
  \centering
  \caption{\label{tab:nsbh_ew_sa90}The number of NSBH detections per year for the GW detector networks for which an EW alert can be sent $60$ s, $120$ s, $300$ s and $600$ s before the merger, with $90\%$-credible sky area measured to be better than $100$, $10$, $1$ $\mbox{deg}^2$ at the time when the alert is sent.}
  \renewcommand{\arraystretch}{1.5} 
    \begin{tabular}{l P{2.1cm}P{2.2cm}P{2.2cm} P{2.1cm}P{2.2cm}P{2.2cm}}
    \hhline{=======}
    EW Time & \multicolumn{3}{c}{$\tau_{\mathrm{EW}} = 60\mbox{ s}$} & \multicolumn{3}{c}{$\tau_{\mathrm{EW}} = 120\mbox{ s}$} \\
    \hhline{-------}
    $\Delta \Omega\,(\mbox{deg}^2)$ & $\leq 100$ & $\leq 10$ & $\leq 1$ & $\leq 100$ & $\leq 10$ & $\leq 1$ \\
    \hhline{-------}
    HLA & $0$ & $0$ & $0$ & $0$ & $0$ & $0$ \\
    HLET & $1.6^{+3.2}_{-1.4} \times 10^{2}$ & $6.0^{+22.0}_{-5.0} \times 10^{0}$ & $0$ & $9.2^{+20.5}_{-8.2} \times 10^{1}$ & $3.0^{+10.0}_{-3.0} \times 10^{0}$ & $0$\\
    20LA & $0.0^{+1.0}_{-0.0} \times 10^{0}$ & $0$ & $0$ & $0$ & $0$ & $0$ \\
    40LA & $0.0^{+1.0}_{-0.0} \times 10^{0}$ & $0$ & $0$ & $0$ & $0$ & $0$ \\
    20LET & $7.7^{+16.5}_{-6.6} \times 10^{2}$ & $3.5^{+10.8}_{-3.4} \times 10^{1}$ & $0.0^{+1.0}_{-0.0} \times 10^{0}$ & $3.9^{+8.7}_{-3.4} \times 10^{2}$ & $1.3^{+4.3}_{-1.2} \times 10^{1}$ & $0.0^{+1.0}_{-0.0} \times 10^{0}$ \\
    40LET & $1.2^{+2.5}_{-1.0} \times 10^{3}$ & $8.1^{+20.9}_{-7.2} \times 10^{1}$ & $2.0^{+4.0}_{-2.0} \times 10^{0}$ & $7.2^{+15.4}_{-6.2} \times 10^{2}$ & $3.3^{+10.6}_{-3.0} \times 10^{1}$ & $0.0^{+1.0}_{-0.0} \times 10^{0}$ \\
    4020A & $1.3^{+3.1}_{-1.2} \times 10^{2}$ & $9.0^{+14.0}_{-7.0} \times 10^{0}$ & $0$ & $3.9^{+10.6}_{-3.2} \times 10^{1}$ & $0.0^{+3.0}_{-0.0} \times 10^{0}$ & $0$ \\
    4020ET & $1.5^{+3.3}_{-1.2} \times 10^{3}$ & $1.3^{+3.1}_{-1.1} \times 10^{2}$ & $2.0^{+7.0}_{-2.0} \times 10^{0}$ & $9.3^{+19.8}_{-7.9} \times 10^{2}$ & $4.8^{+14.1}_{-4.5} \times 10^{1}$ & $0.0^{+1.0}_{-0.0} \times 10^{0}$ \\
    \hhline{=======}
    EW Time & \multicolumn{3}{c}{$\tau_{\mathrm{EW}} = 300\mbox{ s}$} & \multicolumn{3}{c}{$\tau_{\mathrm{EW}} = 600\mbox{ s}$} \\
    \hhline{-------}
    $\Delta \Omega\,(\mbox{deg}^2)$ & $\leq 100$ & $\leq 10$ & $\leq 1$ & $\leq 100$ & $\leq 10$ & $\leq 1$ \\
    \hhline{-------}
    HLA & $0$ & $0$ & $0$ & $0$ & $0$ & $0$ \\ 
    HLET & $1.9^{+68.0}_{-1.7} \times 10^{1}$ & $1.0^{+2.0}_{-1.0} \times 10^{0}$ & $0$ & $2.0^{+10.0}_{-2.0} \times 10^{0}$ & $0.0^{+2.0}_{-0.0} \times 10^{0}$ & $0$ \\ 
    20LA & $0$ & $0$ & $0$ & $0$ & $0$ & $0$ \\
    40LA & $0$ & $0$ & $0$ & $0$ & $0$ & $0$ \\
    20LET & $7.6^{+19.6}_{-6.8} \times 10^{1}$ & $2.0^{+7.0}_{-2.0} \times 10^{0}$ & $0$ & $7.0^{+29.0}_{-7.0} \times 10^{0}$ & $0.0^{+2.0}_{-0.0} \times 10^{0}$ & $0$ \\ 
    40LET & $1.6^{+34.7}_{-1.4} \times 10^{2}$ & $4.0^{+11.0}_{-4.0} \times 10^{0}$ & $0$ & $1.1^{+5.2}_{-1.1} \times 10^{1}$ & $0.0^{+2.0}_{-0.0} \times 10^{0}$ & $0$ \\
    4020A & $2.0^{+8.0}_{-2.0} \times 10^{0}$ & $0$ & $0$ & $0.0^{+1.0}_{-0.0} \times 10^{0}$ & $0$ & $0$ \\ 
    4020ET & $2.0^{+44.5}_{-1.8} \times 10^{2}$ & $5.0^{+22.0}_{-5.0} \times 10^{0}$ & $0$ & $1.7^{+6.1}_{-1.7} \times 10^{1}$ & $0.0^{+3.0}_{-0.0} \times 10^{0}$ & $0$ \\    
    \hhline{=======}
    \end{tabular}
\end{table*}

\section{Open Science Questions Uniquely Addressed by Gravitational-Wave Observations}
\label{sec:science questions}
A number of White Papers and design study reports have documented the scientific potential of current and future GW observatories. For recent reviews see the following references \cite{Maggiore:2019uih, Kalogera:2021bya, Bailes:2021tot}. In this Section, we summarize the science questions of interest to a diverse community of physicists and astronomers and could be addressed by GW observations. In later sections, we will match these questions to specific networks that can answer them effectively.

\subsection{Black holes and neutron stars throughout the cosmos} \label{ss:bh_ns_cosmos}

The improved sensitivities of the next-generation networks will not only allow the detection of compact binaries up to larger distances but will also result in the improved estimation of binary parameters (see Sec. \ref{sec:det and pe}). The precise measurement of binary parameters is essential to infer the source properties which will inform us about the formation of such binaries. There is already some evidence that the observed variety of source properties is likely the result of multiple astrophysical formation channels~\cite{Zevin:2020gbd,Wong:2021,StevensonClarke2022,Godfrey:2023}. The peak of the mass function, and its variation with redshift, contains crucial clues about binary evolution and the final stages of the life of massive stars~\cite{Fryer:2012, Fryer:2022, Olejak:2022, vanSon:2022, Schneider:2023, DCC_P2300002, 2019ApJ...873..100C}.

\subsubsection{BBH, BNS and NSBH mergers}

From Fig. \ref{fig:local_eff_rate} and Tab. \ref{tab:reach_and_pop_snr}, we see that a network with no XG detectors will only observe $\sim 0.1\%$, $1\%$ and $15\%$ of the cosmic BNS, NSBH and BBH population, respectively. This is drastically improved for a network with 2 CE detectors, detecting $\sim 30\%$, $66\%$ and $95\%$ of all BNS, NSBH and BBH mergers, respectively. Further, a network with at least 2 XG detectors will be able to detect BNS systems up to the peak of the star formation rate $(z\sim2)$, and NSBH mergers well beyond it. In fact, BBH mergers with such a network can be observed beyond $z = 20$, i.e., even prior to $~200$ Myr after the Big Bang. It is important to note that CE40 really shines compared to CE20 and ET when it comes to detecting events at large distances. This is evident in Tab. \ref{tab:reach_and_pop_snr}, where the reach and the detection rates for 40LA are better than not just 20LA and HLET, but even the 20LET network.  The improvement in the depth of observation will provide information about the delay time distribution between the formation and the merger of the compact binary~\cite{Safarzadeh:2019pis}, and thereby allow the inference of the history of chemical evolution in the universe beyond the reach of multi-messenger astronomy~\cite{chruslinska2022_review}.

Apart from the detection itself, the source-frame masses of compact objects can be measured with unprecedented precision with the help of XG detectors. From Tab. \ref{tab:network_science}, we note that only a network with at least 2 XG detectors can detect BNS mergers beyond $z\geq1$ such that the uncertainty in the redshift and the source-frame mass measurements is within $20\%$ and $30\%$, respectively. Such measurements are important in order to measure the mass function associated with this class of binaries close to the star formation peak. Similarly, astrophysical channels for the formation and merger of BBH systems at high redshifts can be studied using GW detections with precise mass measurements for BBH systems beyond $z=10$. Table \ref{tab:network_science} shows that the number of such mergers increases from 0 to $\mathcal{O}(10)$ to $\mathcal{O}(100)$ every year with a network with no XG, 1 XG and 3 XG detectors, respectively.  In addition, recall that we chose the mass spectrum of the BNS population to follow a double Gaussian (see Sec. \ref{sec:pop}) in order to see if the second Gaussian feature $(\mu = 1.8\msun,\sigma=0.3\msun)$ could be inferred with GW detections. From Tab. \ref{tab:network_science}, we conclude that for BNS systems with $m_1 \geq 1.5 \msun$, only networks with at least 1 XG detector can detect systems such the source-frame mass of the primary component is measured to better than $10\%$ precision. 

\subsubsection{IMBBH, Pop-III BBH and PBH mergers}
\begin{figure*}
  \includegraphics[scale=0.5]{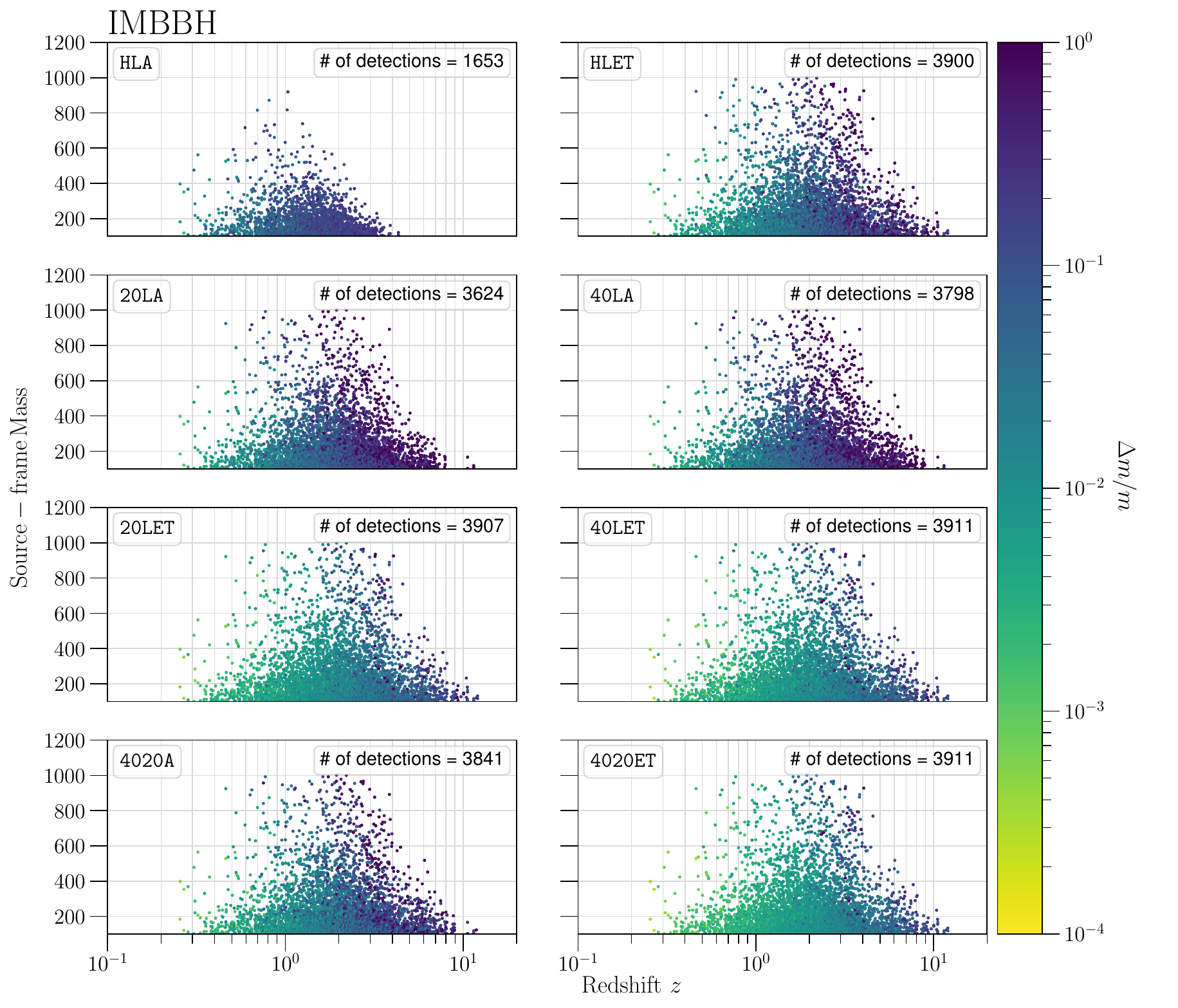}
  \caption{\label{fig:imbh}The plot shows the fractional error in the measurement of source-frame component masses for IMBBH mergers as a function of the source-frame mass and the redshift, along with the number of such systems detected by each detector network in the span of 1 year.}
\end{figure*}

\begin{figure*}
  \includegraphics[scale=0.5]{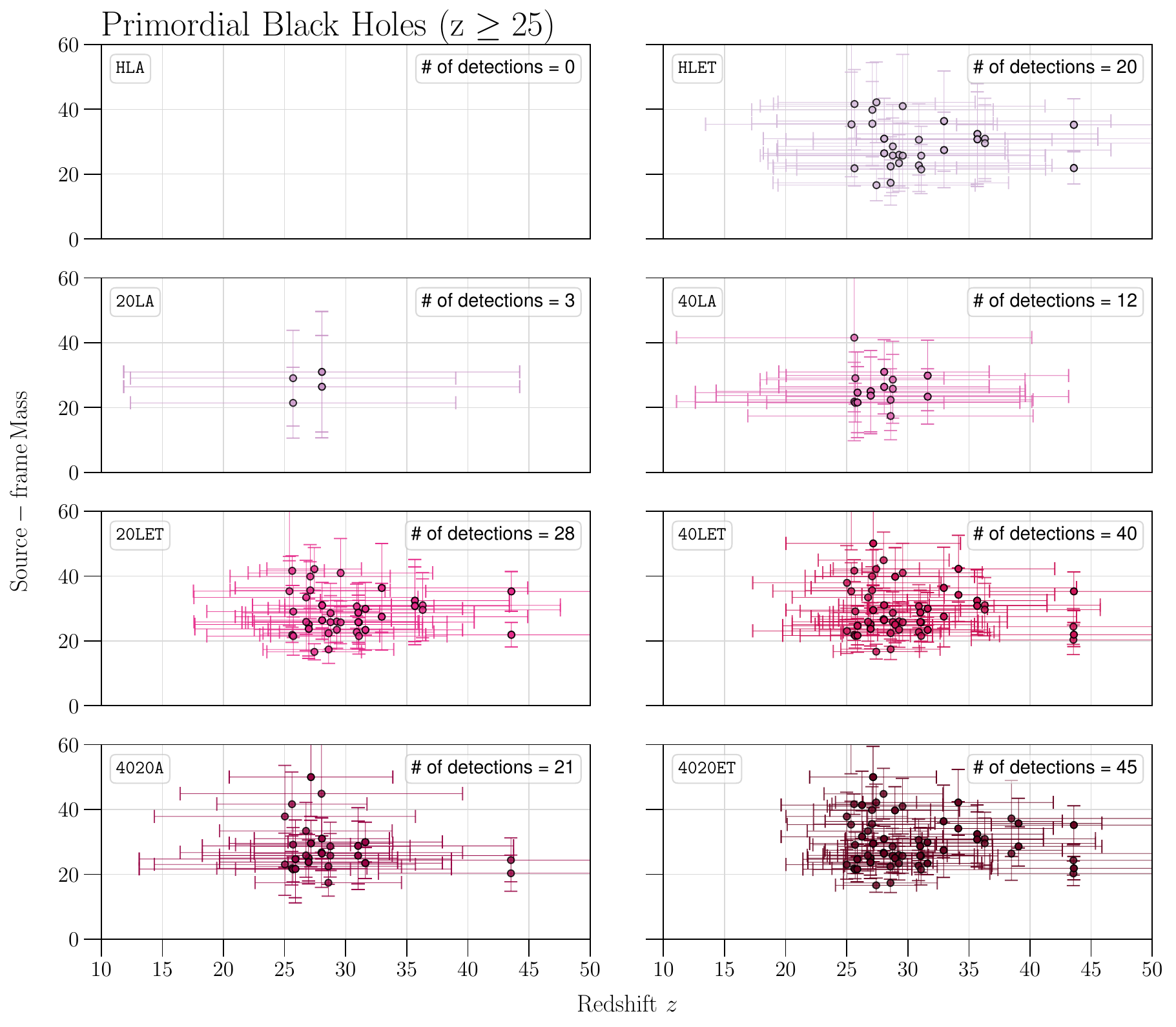}
  \caption{\label{fig:pbh} The plot shows the number of primordial black holes lying further than $z=25$ that are detected by each detector network, along with the measurement errors in the inference of the source-frame masses and the redshift for each event. The numbers correspond to an observation span of $1$ year.}
\end{figure*}

\begin{figure*}
  \includegraphics[scale=0.5]{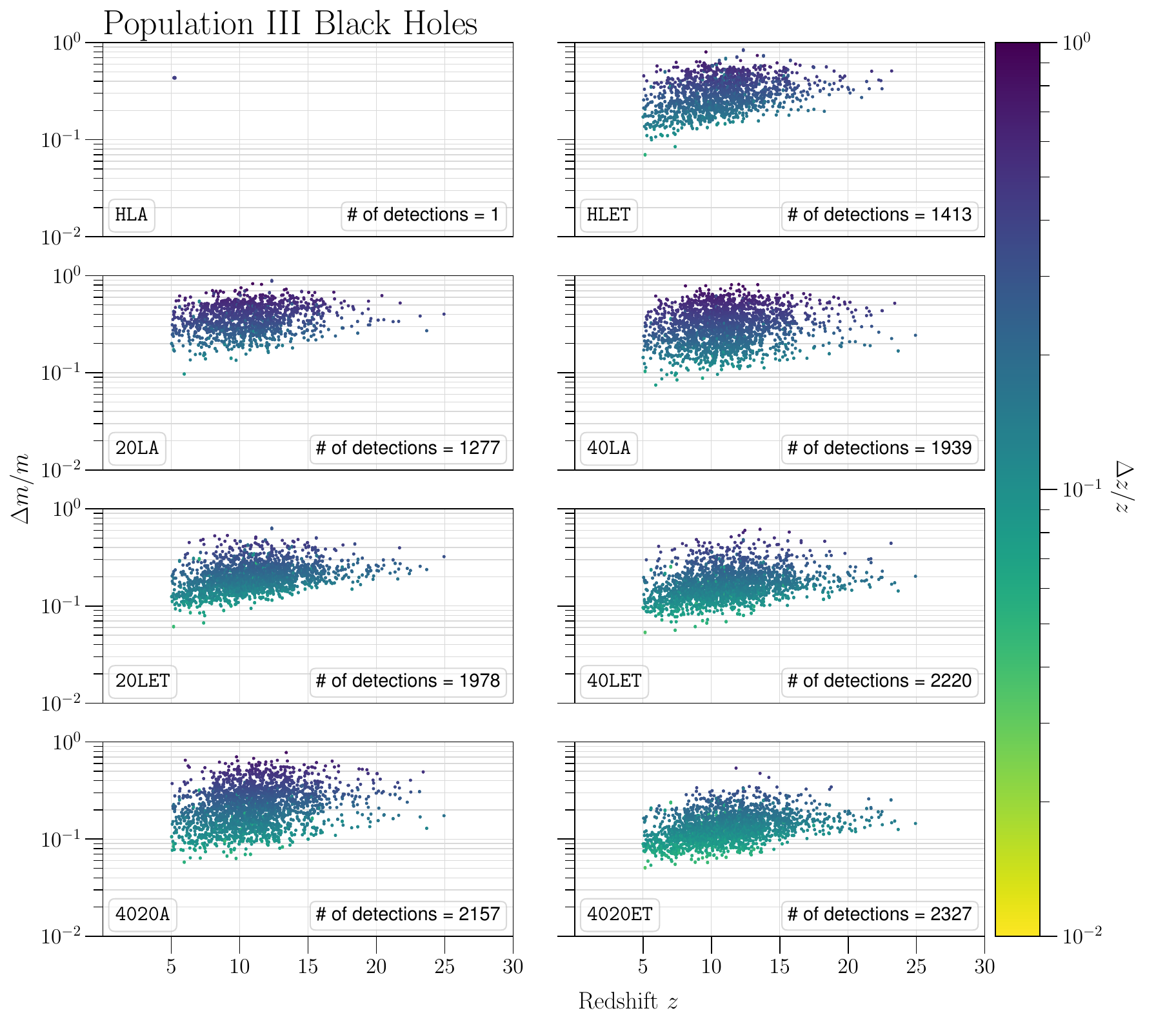}
  \caption{\label{fig:popIII}The plot shows the number of Pop-III binary black holes that are detected by each detector network, along with the measurement errors in the inference of redshift as a function of the fractional error in source-frame masses and the redshift corresponding to each event. The numbers correspond to an observation span of $1$ year.}
\end{figure*}

Constraining the BH mass function above 50~$\msun$ will allow for a better understanding of the pair instability supernova mass gap (and of the nuclear physics processes that lead to it)~\cite{Woosley:2016hmi, DCC_2300016}; the rate of hierarchical mergers~\cite{Kimball:2020qyd, Gerosa:2021mno}; and IMBH~\cite{Greene:2019vlv, DCC_P2300011}. Massive BBH systems can also result from high-redshift formation channels, e.g., PBHs created during the inflationary epoch of the universe~\cite{Carr:1974nx, Bird:2016,  Raidal:2018bbj, Franciolini:2023opt}, and Pop III stars \cite{Santoliquido:2023wzn}. The mass functions of both PBH and remnants of Pop III stars are uncertain; but they might be the seeds that formed the supermassive BHs found at the centers of most galaxies~\cite{Madau:2001sc, Volonteri:2012by, Natarajan:2014rim, Greene:2019vlv, Valiante:2020zhj}. Thus, detecting these mergers will allow us to explore one of the most pressing open questions in galaxy and structure formation. 

The major hurdle in detecting IMBBH systems or binaries at large redshifts is the large detector-frame chirp mass, which leads to low merger frequencies. These mergers lie predominantly in the range where the detectors are less sensitive. The low-frequency $(\sim 10$ Hz) sensitivity increases by more than an order of magnitude when going from \asharp to XG detectors (see Fig. \ref{fig:sensitivity}), thus, improving the chances of detecting these events. However, the detection itself is only the first step. For IMBBH mergers, we are also interested in seeing if GW networks can precisely measure the source-frame masses so as to unravel the mass spectrum of these events. On the other hand, for PBH and Pop III mergers, along with the source-frame mass, one would also need a precise measurement of the redshift in order to differentiate these high-redshift mergers from IMBBH mergers at $z \leq 10$. 

Figure \ref{fig:imbh} shows the error in the source-frame mass of the components of IMBBH mergers as a function of the injected source-frame mass and redshift. Note that while the fractional error on the mass measurement can be as low as $\sim 0.01\%$, such precision is likely to be achievable only with networks containing at least 2 XG detectors. The \asharp network is able to detect $\sim 40\%$ of the mergers, whereas networks with at least 1 XG detector can detect more than $90\%$ of these events, with HLET itself being able to detect $\sim 98\%$. The stellar performance of ET is attributed to better sensitivities in the $f < 10$ Hz region compared to CE40 and CE20. While the CE detectors do not seem to help ET significantly in detecting IMBBH sources, they aid in the precise estimation of component masses. 

Figures \ref{fig:pbh} and \ref{fig:popIII} show the errors in source-frame components masses and the redshift for PBH  systems beyond $z=25$ and Pop III BBH systems, respectively. It is improbable that a network with no XG detectors will be able to detect any such mergers. In the case of PBH, we see that $40$ $(45)$ mergers will be detected every year with 40LET (4020ET) such that the redshift measurement excludes $z=15$ $(z=20)$ at the $1\sigma$ level. Even one such detection will be the smoking gun evidence in favor of the existence of PBHs. For the Pop III case, we note that CE40 outperforms CE20 and ET. The 40LA network is itself able to detect the furthest simulated merger detected by 4020ET, which occurs around $z=25$. However, the merit of having more than 1 XG detector in the network becomes apparent when we consider the measurement accuracy of source-frame masses and redshift. Beyond $z=10$, 40LA detects just $1$ event with $\Delta m/m \leq 10\%$, whereas 4020A and 4020ET detect $\mathcal{O}(10)$ and $\mathcal{O}(100)$ such events, respectively. In fact, beyond this redshift, 40LA detects $\mathcal{O}(10)$ mergers with $\Delta z/z \leq 10\%$, whereas 4020A and 4020ET detect $8$ and $40$ times as many mergers, respectively.

The era of the XG GW observatories will observe the cosmic population of compact binary mergers like BBH, BNS and NSBH, only a glimpse of which is already seen by current networks. In addition, only networks with XG observatories will be capable of unraveling elusive compact binary populations, like those corresponding to Pop III and primordial BHs, which will greatly improve our understanding of various astrophysical processes involved in star, galaxy and structure formation, and the universe as a whole.

\subsection{Multimessenger astrophysics and dynamics of dense matter}\label{ss:mma}
NSs are among the most exotic objects in the stellar graveyard. They are characterized by a unique relationship between the associated pressure and the energy density, called the equation of state (EoS). With the EoS, one can link the mass with the radius of the NS by solving the Tolman-Oppenheimer-Volkoff equation. NSs in binary configurations with a companion NS or BH can get tidally disrupted by the gravity of their companion close to the merger. The effect of the disruption on the phase of the GW waveform near merger can be described, to leading order, using the tidal deformability $(\Lambda)$ of the NS. $\Lambda$ can be uniquely determined with the knowledge of the EoS and the mass of the NS. Inversely, the measurement of $\Lambda$ and the mass of the NS from GW observations can be used to obtain constraints on the EoS that governs NS.  

The disruption of merging NSs in binaries can result in the production of non-relativistic to mildly-relativistic neutron-rich debris, and relativistic jets. These ejecta can power a variety of EM counterparts, including UV-optical-IR kilonovae, late-time radio flares from fast kilonova tails, gamma-ray bursts (GRBs), and their radio-to-X-ray afterglows (e.g., Refs.~\cite{Li:1998bw,Goodman:1986az,Eichler:1989ve,Kobayashi2003,Belczynski2006,Nakar2011,Piran2013,Metzger:2019zeh,Kyutoku:2021icp}). As demonstrated spectacularly by the case of GW170817~\cite{Abbott2017a,Abbott2017b,Abbott2017e}, multi-messenger observations can paint a very detailed picture of BNS progenitors and ejecta \cite{Abbott2017a,Abbott2017b,Abbott2017e,Arcavi2017,Coulter2017,Drout2017,Goldstein2017,Fong2017,Hallinan2017,Haggard2017,Kasliwal2017,Margutti2017,Smartt2017,Pian2017,Troja2017,Villar2017,Lazzati2018,Mooley2018}.

In the WP, we have shown how XG GW detectors will enable major breakthroughs in multi-messenger astrophysics and in our understanding of the dynamics of dense matter. In the remainder of this Section we summarize our previous findings, highlighting the specific results of the trade study presented here that support the conclusions drawn in the WP.

\subsubsection{Multimessenger observations and early warnings}

Gamma-ray observations of GW170817 have confirmed that at least some short GRBs are associated with BNS mergers~\cite{Abbott2017e,Goldstein2017,Fong2017}. These observations have also enabled measurement of the time delay between the merger (as determined by the GW signal) and the onset of the GRB emission (as determined by the gamma-ray light curve) ~\cite{Abbott2017e,Goldstein2017}. As discussed in Ref.~\cite{Lazzati2020} (and references therein), this delay encodes key physics of the GRB central engine and the ejecta, besides enabling fundamental physics tests \cite{Abbott2017e}. 

As evident from the results reported in Table \ref{tab:bns_specific}, XG detectors will probe GWs from BNS mergers up to the peak of star formation ($z\approx 2$), hence mapping GRBs (and their electromagnetic afterglows) to their progenitors, and measuring the GW-GRB time delays in a systematic fashion~\cite{Abbott2017e,Goldstein2017,Fong2017}. We stress that systematically mapping GRBs to their progenitors up to the star formation peak is beyond the reach of 4-km-long GW detectors, and will offer new insight into what physical properties allow for the launch of successful relativistic jets in short GRBs.

XG detectors will also provide exquisite sky localizations ($\lesssim 1\,$deg$^2$) for BNS mergers in the local universe (Table \ref{tab:bns_specific}), hence building a golden sample of GW170817-like events. This golden sample will be critical to explore the diversity of merger outcomes, and map the properties of progenitors and merger remnants to those of their EM counterparts. 
While 4\,km GW detectors can build a sample of GW-kilonova associations in the local universe taking advantage of wide field-of-view optical telescopes such as Rubin and the Zwicky Transient Facility \cite{Chase:2021ood,Andreoni2022,Petrov2022}, on theoretical grounds we expect a zoo of EM counterparts to exist, ranging from optically bright and blue kilonovae associated with magnetar remnants and (perhaps) choked jets, to red and dim kilonovae associated with successful jet afterglows (e.g., \cite{Margalit2019,Kasliwal2017,Sarin2021}). The last, when viewed off-axis, could be more easily unveiled at radio wavelengths \cite{Hotokezaka2016,Hallinan2017}.  The The operation of XG observatories will sync with the era of EM telescopes such as the Extremely Large Telescope (ELT) and Giant Magellan Telescope (GMT) that can probe cosmic distances.  Hence, having a few BNS mergers per year with GW sky localizations accessible to the smaller fields of view of the most sensitive EM telescopes \cite{Perley2011,Murphy2018} (Table \ref{tab:telescope_FOVs}) will probe the diversity of BNS mergers in an optically unbiased way. 

Finally, XG detectors will also enable early warnings, thanks to their ability to localize BNSs even before the merger (Table \ref{tab:bns_ew_sa90}). This is key to potentially opening new discovery space in multi-messenger astrophysics. In fact, several GRBs are preceded by so-called gamma-ray precursors whose origin remains unclear \cite{Troja2010}. Moreover, several theoretical models predict prompt EM emission associated with various mechanisms including NS magnetic field interactions during the inspiral or the collapse of a short-lived NS remnant into a BH (e.g., Refs.~\cite{Rowlinson2019, Curtin2022, Moroianu2023} and references therein). These scenarios motivate precise sky localization and early-warning alerts to telescopes before the merger \cite{Sachdev:2020lfd,Banerjee:2022gkv}.

\subsubsection{Measuring the radius of the neutron star} \label{ss:ns_radius}
As shown in the WP, XG detectors will revolutionize our knowledge of high-density matter by detecting hundreds of BNS mergers per year with
SNR$> 100$. For these, measurements of NS tides will constrain their radii to better than 100\,m, i.e., about one part in a hundred. This will enable population-wide constraints (for the common NS EoS) at the 10\,m-level.

Constraining NS radii is of significant importance because it provides crucial insights into the properties of the NS and the nature of matter inside it (for e.g., see Refs.~\cite{Sabatucci:2022qyi,Rose:2023uui}). Universal relations are empirical relationships between various physical properties of NSs, which are instrumental in obtaining the NS radii from GW data \citep{LIGOScientific:2017ync}. GWs contain information on tidal parameters like $\Tilde{\Lambda}$, which can be used along with two universal relations to constrain the radius. We describe the procedure for this below.

We use the {\tt GWBENCH} formalism to calculate the covariance matrix of parameters corresponding to BNS events. We use the covariance matrices to generate multi-dimensional Gaussian distribution of the parameters mentioned in Sec. \ref{sec:3B}. The universal relation between symmetric and asymmetric combinations of component tidal deformability, described in Refs. \cite{Chatziioannou:2018vzf, Yagi:2016bkt}, is then used to calculate individual tidal deformabilities $\Lambda_1$ and $\Lambda_2$ from the samples of the combined tidal deformability $\Tilde{\Lambda}$ and the mass ratio $q$. The universal relation between compactness and tidal deformability defined in Ref. \cite{Yagi:2016bkt} is used to infer the NS radii from the component tidal deformabilities. EoS-specific corrections have been applied to the tidal deformability and radii distributions which are described in Ref. \cite{Kashyap:2022wzr}. 

\begin{figure}[htbp]
  \includegraphics[width=1\columnwidth]{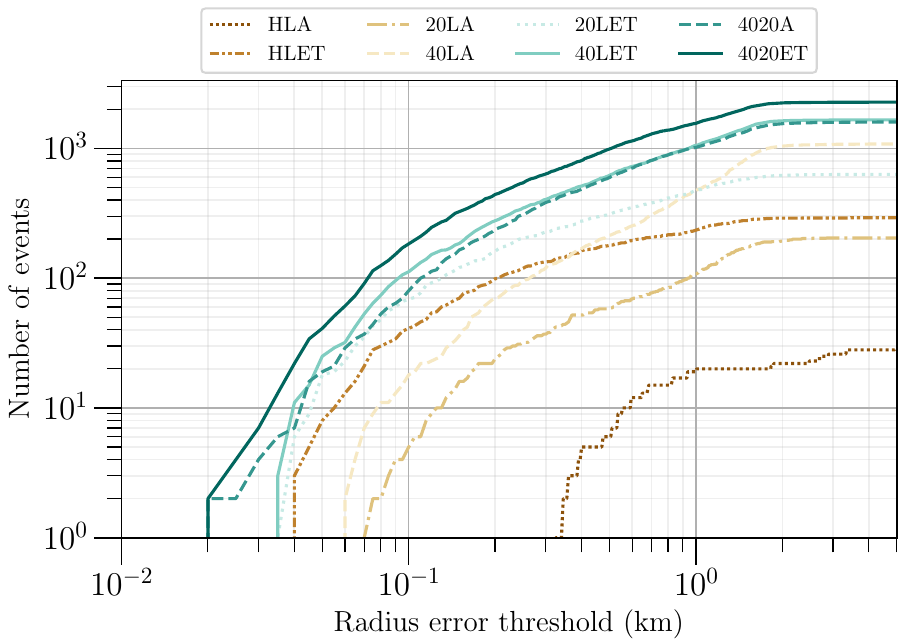}
  \caption{\label{fig:cum_raderr}The cumulative number of events as a function of combined radius error of the BNS population for all detector networks. We consider events with SNR greater than 50 for all detectors except for HLA for which we choose the events with SNR above 25.}
\end{figure}

We combine the events with similar masses to get an effective radius error as a function of mass. For this, we make 20 mass bins from $1 \msun$ to the maximum mass allowed by the EoS used in our study. We use the $\sqrt{N}$ relationship for combining errors in radii in each mass bin separately \cite{Huxford:2023qne} and report the cumulative number of events with respect to the radius errors in Fig.~\ref{fig:cum_raderr}.


The present analysis provides preliminary evidence that NS radii can be measured below 1\% by using at least one XG detector in combination with the planned upgrades of LIGO-Virgo detectors. CE design with either a 40\,km or 20\,km arm will observe up to 10\% of the BNS events where a radius error is better than 100\,m and a few events with errors of even a few tens of meters. With thousands of BNS observations, we are optimistic that a combined error measurement will improve by at least an order of magnitude and aid our understanding of nuclear physics.

\subsubsection{Post-merger NS Physics} \label{sec:post-merger}
XG detectors will unveil post-merger GWs (with network SNR $\,>\,5$) and, via those detections, provide accurate measurements of the post-merger GW frequencies. These measurements will impact our understanding of the composition and behavior of matter at its most extreme.

To examine the potential of different network configurations to detect the post-merger signal of BNS mergers, we choose a representative signal from the CoRe database; the signal corresponds to a 1.35-1.35$\msun$ merger with an SLy EoS~\cite{Dietrich:2018phi}. For each network configuration in this trade study, we simulate a population of neutron star mergers and calculate the sky and orientation averaged detection range at a threshold of SNR$\,=\,5$. The sky-location and frequency-dependent sensitivity of long-arm detectors~\cite{Essick:2017wyl} are taken into account. For this calculation, we also use the kilohertz-focused configuration~\cite{Evans:2021gyd} of the CE20 observatory which optimizes the post-merger sensitivity. The results are shown in Figs. 2 and 4 of the WP.

\subsection{New sources, new probes and extreme astrophysics} \label{ss:new_sources}

NSs and BHs can emit GWs through a wide variety of mechanisms other than binary mergers and post-mergers~\cite{Glampedakis2018, Kalogera2019}.
Although not yet detected, these other signals (with durations from a fraction of a second to longer than a human lifetime) have great discovery potential.
When detected, especially in combination with signals carried by other messengers, these GW signals will reveal different populations of compact objects and probe extreme astrophysics in a regime largely different from that probed by compact binary mergers and terrestrial colliders.
Here we summarize scenarios for detection of and extraction of information from several predicted types of signals.
We also note that the history of opening new windows of astronomy indicates that unexpected signals are to be expected.

\subsubsection{Continuous waves}

Spinning neutron stars produce \textit{continuous GWs,} signals with low amplitude compared to binary mergers but lasting many years~\cite{Riles2023, Wette2023}.
This allows for greatly enhanced detectability with matched filtering and similar techniques.
Continuous GW emission likely is dominated by either a mass quadrupole (sustained by elastic or magnetic stresses) or a mass current quadrupole (produced by an unstable or weakly stabilized $r$-mode, a rotational mode with a frequency comparable to the star's spin frequency).
Free precession can also produce GWs via a changing mass quadrupole, but based on electromagnetic pulsar observations it is likely to be rare.
For a given quadrupole, GW emission is stronger for rapidly rotating NSs~\cite{Shklovskii1969}, and the $r$-mode instability to GW emission~\cite{Andersson1998} is more likely to overcome various dissipation mechanisms at higher frequencies~\cite{Lindblom1998}.
Continuous GW searches are more sensitive when using the sky location, spin frequency, and other timing information of the source (if known).
Sensitivities can be expressed in terms of a \textit{sensitivity depth}~\cite{Dreissigacker2018, Wette2023}, which factors out the noise amplitude from everything else (methods, amount of data, etc.) and is convenient for extrapolating current searches to new detectors as we do here.

Accreting NSs are of particular interest as continuous wave sources since accretion tends to spin them up and to generate asymmetries through electron capture layers and lateral temperature gradients~\cite{Bildsten1998, Hutchins2023}, magnetic bottling of accreted material~\cite{Melatos2005}, or the GW-driven $r$-mode instability~\cite{Andersson1999}.
In fact, one popular theory posits that the spins of accreting neutron stars are limited to relatively low values (compared to the maximum allowed for most equations of state) by the spin-down torque due to GW emission balancing the spin-up torque due to accretion~\cite{Papaloizou1978}.
In this case the gravitational-wave strain of an accreting neutron star is expected to be proportional to the square-root of the observed x-ray flux~\cite{Wagoner1984}, meaning that the brightest GW emitters are Sco X-1 and other low mass x-ray binaries with no observed pulsations and thus no confirmed spin frequency~\cite{Watts2008}.
These sources exhibit stochastic x-ray variability, meaning that the accretion torque and spin frequency also fluctuate.
Despite these obstacles, a recent GW search~\cite{LVKO3ScoX1} achieved a sensitivity comparable to the strain implied by torque balance, even under pessimistic assumptions, albeit only in a narrow frequency band.
Extrapolations from this sensitivity are made in Ref.~\cite{Owen2023}, which we summarize here.
Using the sensitivity depth of Ref.~\cite{LVKO3ScoX1} (a conservative 39~Hz$^{-1/2}$) with the network noise curves from Table~\ref{tab:networks} and average bolometric fluxes estimated in Ref.~\cite{Watts2008}, Ref.~\cite{Owen2023} finds that the HLA network can detect GWs at the torque balance limit of Sco X-1 at GW frequencies up to about 800\,Hz.
This corresponds to spin frequencies up to about 400\,Hz for mass quadrupole emission or about 550\,Hz for $r$-modes.
Since accreting neutron stars are known to spin above 700\,Hz in some cases, HLA is not guaranteed detection even of Sco X-1.
With the 40LA configuration network, CE is sensitive enough to detect at the torque balance limit up to 1400\,Hz, high enough to cover almost all known neutron stars.
The 4020ET configuration is sensitive up to almost 2\,kHz, well beyond the GW frequency of any known neutron star.
4020ET is also sensitive to GX 5\textminus1 and several other neutron stars up to GW frequencies of almost 1\,kHz.
At this point even non-detection is very interesting since it strongly confronts the torque balance theory.

After accretion ends, the neutron star is believed to become a \textit{millisecond pulsar} with high spin frequency and slow spin-down~\cite{Manchester2017}.
The latter indicates a small external magnetic dipole and small internal mass quadrupole by ruling out large torques due to EM radiation and GWs respectively, and is usually believed to be dominated by magnetic dipole radiation.
However, in recent years it has become apparent that millisecond pulsar spin-downs exhibit a cutoff whose frequency dependence is quadrupolar rather than dipolar~\cite{Woan2018}.
The implied minimum quadrupole is about $10^{-9}$ times the moment of inertia, consistent with buried magnetic fields of order $10^{11}$\,G, consistent with the fields of young pulsars and with theoretical predictions~\cite{Melatos2005,Palomba:1999su}.
The buried magnetic field may survive for a long time under the accreted material~\cite{Vigelius2009}.
Millisecond pulsars which are observed regularly in radio or EM waves can be timed precisely enough to allow narrow, deep GW searches.
Based on previous examples, the sensitivity depth of such a search can be conservatively estimated as 500\,Hz$^{-1/2}$ for a year of observation~\cite{Wette2023} and scales as the square root of the observation time.
Then assuming an ellipticity of $10^{-9}$~\cite{Woan2018} and taking data from the ATNF pulsar catalog~\cite{Manchester2005}, the GW amplitude is simple to determine and compare to the search sensitivities of various networks.
In Table~\ref{tab:network_science} of this paper we quote numbers from Ref.~\cite{Owen2023} of millisecond pulsars detectable with various networks with one year of observation and for years of observation to detect 25 millisecond pulsars.
The WP plots these numbers in its Fig.~2 and Fig.~4.
These numbers only include currently known pulsars.
By the time Cosmic Explorer is operational the Square Kilometre Array, next generation Very Large Array, and other instruments are expected to detect several new pulsars for each one currently known~\cite{Smits2009, Beasley2019}.
Thus the number of detectable pulsars should improve accordingly.
Conversely, non-detection would severely constrain the theory that millisecond pulsars' original magnetic fields survive buried under accreted material.

All sky broadband continuous GW surveys for yet unknown neutron stars are another popular type of search~\cite{Riles2023, Wette2023}.
In this case, recent population simulations~\cite{Pagliaro2023} for the ET indicate that it might detect more than 100 sources on its own in an all sky survey.
With its better sensitivity, a 40\,km CE will detect even more than ET.
Any new continuous GW source detected by such surveys will be followed up with a year or more of observation, resulting in arcsecond sky localization (the diffraction limit for two astronomical units' aperture) even with one interferometer, and a frequency measurement to a precision of tens of nHz~\cite{Jaranowski1999}.
With such precise guidance, the source is likely to be detected by electromagnetic pulsar searches.

The combination of continuous GWs and EM observations will open new windows into neutron star interiors and for a population distinct from the progenitors of binary mergers~\cite{Jones2022}.
The ratio of GW frequency to spin frequency immediately yields insight into the GW emission mechanism (mass quadrupole, free procession, or $r$-mode).
In the case of a mass quadrupole it might reveal the timescale of any coupling between crust and core leading to glitches (see below), and in the case of $r$-modes it can yield a measure of the neutron star's compactness to a few percent~\cite{Idrisy2015} and thus on the EoS in a low-temperature regime inaccessible to colliders.
In some cases gravitational-wave parallax arising from a $2\times\rm 1\,AU$ baseline over a period of six months can yield a distance measurement~\cite{Sieniawska2023}, and in others, the distance can be obtained from electromagnetic astronomy.
With the distance the magnitude of the quadrupole can be measured, and long-term timing may indicate whether a mass quadrupole is sustained by elastic or magnetic forces~\cite{Lu2023}.
A large elastic quadrupole is only possible if the ``neutron'' star has an exotic composition~\cite{Owen2005}, a magnetic quadrupole measurement yields an approximation of the star's internal magnetic field, and an $r$-mode saturation amplitude is tied to viscosity and other microphysics of the stellar interior~\cite{Arras2003}.

\subsubsection{Core collapse supernovae}

Core-collapse supernovae generate short bursts of GWs from rapid motions of high density matter in their central regions~\cite{Kalogera2019}.
Unlike binary mergers, these motions cannot be predicted with sufficient precision for the use of matched filtering to detect the signals; but other techniques exist for detecting less modeled bursts.
Simulations indicate that the most common events (with little rotation) will be detectable only in the Milky Way even with CE, with uncommon events detectable in the Magellanic Clouds and very rare events perhaps detectable further away~\cite{Srivastava2019}.
Therefore the overall detection rate is expected to be of order one over the planned fifty-year lifetime of the CE facilities.
Even one detected supernova would be a tremendous opportunity for multi-messenger astronomy, as Supernova 1987A was before GW astronomy existed.
On the GW side, CE's improved sensitivity over the current generation of detectors would lead to improved waveform reconstruction~\cite{Szczepanczyk2022}.
This would provide a unique window into the explosion's central engine~\cite{Kalogera2019}, revealing for example the EoS of a newly formed protoneutron star via frequency measurements of $f$-mode and $g$-mode oscillations driven by fallback accretion~\cite{Radice2019} or diagnosing the formation of a rapidly rotating BH~\cite{Ott2011}.
It would also allow measurement of the progenitor core's angular momentum distribution in cases of rapid rotation~\cite{Abdikamalov2014}.
Neutrinos should be detected coincident with the GW emission and improve detection confidence and parameter estimation~\cite{Kalogera2019}, for example revealing the spin of the core through their modulation frequency~\cite{Takiwaki2018}.
In some extreme cases (collapsars), even disk outflow instabilities or the cocoon of material carved out by a jet of escaping material could emit a stochastic burst of GWs~\cite{Siegel2022, Gottlieb2023}.
Such events could be detectable at great distances corresponding to a rate of order ten per year~\cite{Gottlieb2023}.

\subsubsection{Starquakes}

GWs can also be emitted in bursts from less than a second to minutes in duration by the many quasinormal modes of NSs triggered by impulsive events (starquakes) observed electromagnetically as pulsar glitches or magnetar flares (possibly accompanied by fast radio bursts)~\cite{Glampedakis2018}.
Both pulsars and magnetars are relatively young, drawn from a different population than the progenitors of most binary mergers.
Quasinormal modes include the $f$-modes, which radiate almost all their excitation energy as GWs within of order one second.
Pulsar glitches may also be followed by signals of up to weeks in duration as the crust and core slowly readjust~\cite{Moragues2023}.
Even aided by the sky location and EM trigger time, current GW observatories can detect such events in much of our galaxy only in the most optimistic scenarios~\cite{Corsi2011, LVKMagnetars}.
Extrapolating from recent searches~\cite{LVKMagnetars} indicates that CE can detect short GW bursts with energy comparable to the EM energy of moderate magnetar flares or large pulsar glitches over a substantial fraction of our galaxy.
The frequencies and damping times of modes contain information on the cold neutron star EoS~\cite{Andersson1998a}.
Combined with x-ray observations, they can constrain the internal magnetic fields of NSs~\cite{Ho2020}.
Detection of a long post-glitch signal would provide information on the viscous coupling between crust and core~\cite{Melatos2015}.
For the short bursts especially, it is important to have multiple XG observatories to improve detection confidence.

\subsection{Fundamental physics and precision measurement of the Hubble constant} \label{ss:tgr_cosmology}
The improved sensitivity of the CE detectors in comparison to the current generation of GW detector networks results not only in more detections up to larger distances but also in a large number of signals with high SNRs, which are of immense importance for testing fundamental physics, general relativity and precise measurements of cosmological parameters. 

\subsubsection{Testing general relativity and fundamental physics}

\begin{figure}[htbp]
  \includegraphics[width=\columnwidth]{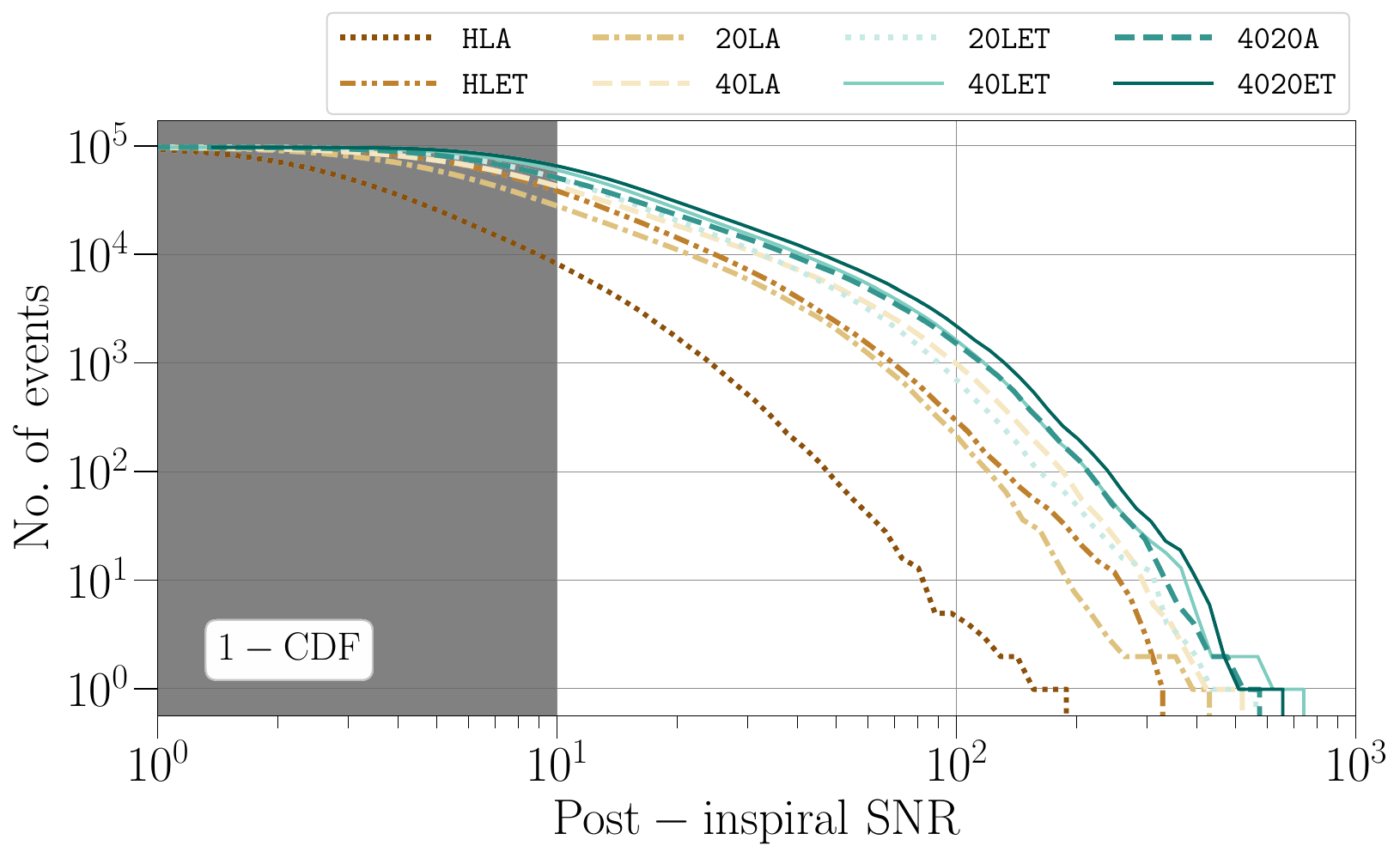}
  \caption{\label{fig:bbh_pm}The scaled CDF plots for the post-inspiral SNRs corresponding to the local BBH population.}
\end{figure}

The most general approach to testing general relativity involves the introduction of deviation parameters in the amplitude and phase of the GW binary inspiral waveform and constraining these parameters using observations \cite{Yunes:2009ke}. These deviation parameters are usually theory-agnostic but they can be mapped to specific theories if needed \cite{Yunes:2016jcc}. To a good approximation, constraints on these deviation parameters scale inversely with SNR $\rho$. When multiple GW observations are combined, the constraints on the deviation parameters also improve,
\begin{equation}
    \sigma \propto \frac{1}{\rho} \longrightarrow \sum_{i=1}^{N} \frac{1}{\sigma^2} \propto \sum_{i=1}^{N} \rho^2
\end{equation}
where $N$ is the number of GW events and $\sigma$ is the standard deviation for a fiducial deviation parameter. The bounds on the deviation parameter will be affected by both, the number of signals detected by the network as well as the SNR with which these signals are detected. In Tab. \ref{tab:network_science}, we report the effective SNR $(\,\sum \rho^2\,)$ corresponding to BBH systems for different detector networks. Just going from the \asharp network to one containing a CE detector improves the effective SNR by $\sim 4-7$ times, improving the constraints by $\sim 2-3$ times. Having at least two XG detectors in the network increases the effective SNR by two orders of magnitude compared to the \asharp network, leading to $\sim 10$ times improvement in the bounds on deviation parameters. In Tab. \ref{tab:network_science} and Fig. \ref{fig:bbh_pm}, we also report the number of BBH events with \textit{post-inspiral} SNR greater than $100$ and the effective post-inspiral SNR for each network. The post-inspiral SNR is calculated by performing the SNR calculation beginning at the ISCO (innermost stable circular orbit) frequency, instead of starting at $f_{\mathrm{low}}$. Thus, it has contributions from the merger and the ringdown phases. While the network with three \asharp detectors is only expected to detect $\mathcal{O}(10)$ events with post-inspiral SNR greater than 100, a network with CE20 will detect $\mathcal{O}(100)$ and CE40 will detect $\mathcal{O}(1000)$ such events every year. These events will allow testing general relativity in the strong-field regime close to merger and using quasinormal modes that describe the ringdown phase to test the nature of black holes \cite{Bhagwat:2023jwv}. 

These estimates can be extended to specific alternate theories of gravity (see Ref. \cite{Perkins:2020tra} for a comprehensive study). Constraints on both the dipole radiation as well as the time variation of the gravitational constant $G$ scale inversely with SNR. However, we should note that both these effects appear at low PN orders ($-1$PN and $-4$PN, respectively) and are better constrained using multiband observations with LISA \cite{Gupta:2020lxa}, instead of only using terrestrial networks. On the other hand, Lorentz  violation with non-commutative theories of gravity and parity violation with the dynamical Chern-Simon theory affect the GW phasing at $2$PN, but the constraints on these theories scale with $\rho^{-1/4}$. Moreover, theories that predict a massive graviton have a leading order effect on GW phase at $1$PN. While the constraint on the mass of the graviton scale as $\rho^{-1/2}$ with SNR, they also scale as $D_{0}^{-1/2}$, where $D_0$ is the cosmological distance. Thus, GWs from objects that are farther away can provide tighter bounds on the mass of graviton. In Tab. \ref{tab:network_science}, we list the number of BNS and BBH mergers that occur beyond $z \ge 5$ and can be detected. For the BNS case, we see that only those networks that contain a 40\,km CE can detect such far-away mergers. The number of detections increase by $7$ times when the network includes both CE40 and CE20 along with an \asharp detector, compared to only containing the CE40 with two \asharp detectors. For BBH systems, the number of detections corresponding to systems that lie beyond $z=5$ increases by two orders of magnitude when only one of the CE detectors is included, compared to a network with only \asharp detectors. Further, Fig. \ref{fig:bbh_pink_scatter_no_mma} shows that these distant events can be detected with SNR $\sim 100$ with CE detectors. Thus, GW networks with CE detectors will allow testing general relativity and fundamental physics for both theory-agnostic and theory-specific tests to unprecedented precision. 

\subsubsection{Measuring the Hubble constant with golden dark sirens and bright sirens}
Detecting GWs from compact binary mergers allows the estimation of the luminosity distance and the sky position associated with the source. As GW observations provide the distance to the source without the need for external distance calibrators, GW sources are often referred to as standard candles. Under the construct of $\Lambda$CDM cosmology,
\begin{equation} \label{eq:H0_DL}
\begin{split}
    D_L &= \frac{1+z}{H_0} \int_{1/(1+z)}^{1} \frac{dx}{x^2\sqrt{\Omega_{\Lambda}+\Omega_{m}\,x^{-3}}}\\
    &= \frac{1+z}{H_0} \int_{1/(1+z)}^{1} \frac{dx}{x^2\sqrt{1-\Omega_{m}(1-x^{-3})}} ,
\end{split}
\end{equation}
where $\Omega_{m}$ is the matter density, $\Omega_{\Lambda}$ is the dark energy density, and we have used $\Omega_{\Lambda} = 1\,-\,\Omega_m$. Thus, having obtained the distance to the source, if the redshift associated with the source can also be estimated, then these two quantities together can allow us to measure cosmological parameters, like the Hubble constant $(H_0)$. The utility of GWs in measuring $H_0$ also becomes important in light of the Hubble tension \cite{Verde:2019ivm, DiValentino:2021izs}, which is the $4\sigma-6\sigma$ discrepancy between the measurements of $H_0$ using data from the early and the late epochs of the universe \cite{Riess:2021jrx,Planck:2018vyg}, although systematics could explain, at least some of, the difference in the two measurements \cite{Sharon:2023ioz}. Using GWs to constrain $H_0$ is independent of the previously mentioned approaches and can help resolve the Hubble tension by measuring $H_0$ to better than $2\%$ precision.

Various approaches have been proposed to measure the redshift, and as a result, $H_0$, using GW observations. The NS(s) in BNS and NSBH mergers can undergo tidal disruption before the merger and lead to the generation of EM counterparts like kilonovae and short-gamma ray bursts, among others. Detecting these EM counterparts allows us to pinpoint the location of the merger and uniquely identify the host galaxy. Photometric or spectroscopic measurements of the galaxy provide the redshift associated with the source. This is referred to as the bright siren method. The BNS merger GW170817 \cite{LIGOScientific:2017vwq,LIGOScientific:2017ync,LIGOScientific:2017zic,LIGOScientific:2017pwl} was the first event that was used to measure $H_0$ with the bright siren approach, giving $H_0 = 70^{+12}_{-8}\,\,\mathrm{km}\,\mathrm{s}^{-1}\,\mathrm{Mpc}^{-1}$ \citep{LIGOScientific:2017adf}. 

In the absence of EM counterparts, as will be the case for BBH and most NSBH mergers, the sky localization of the source can be utilized to obtain redshift measurement. The first such approach was proposed in Ref. \cite{Schutz:1986gp}, also called the statistical dark siren approach. It involves combining the $H_0$ estimates from all the galaxies that lie within the localization volume associated with an event, for all the eligible detections. In doing so, the true value of $H_0$ can be isolated from the noise and inferred. Combining $8$ well-localized dark siren events, Ref. \cite{Palmese:2021mjm} obtain $H_0 = 79.8^{+19.1}_{-12.8}\,\,\mathrm{km}\,\mathrm{s}^{-1}\,\mathrm{Mpc}^{-1}$. These bounds are expected to get better with more detections. The capabilities of different detector networks w.r.t. 3D localization was discussed in section \ref{ss:3d_loc}. In comparison to the \texttt{HLA} network, the inclusion of XG detectors in the network results in drastically better localization estimates.
\begin{figure*}
    \centering
    \includegraphics[width=\textwidth]{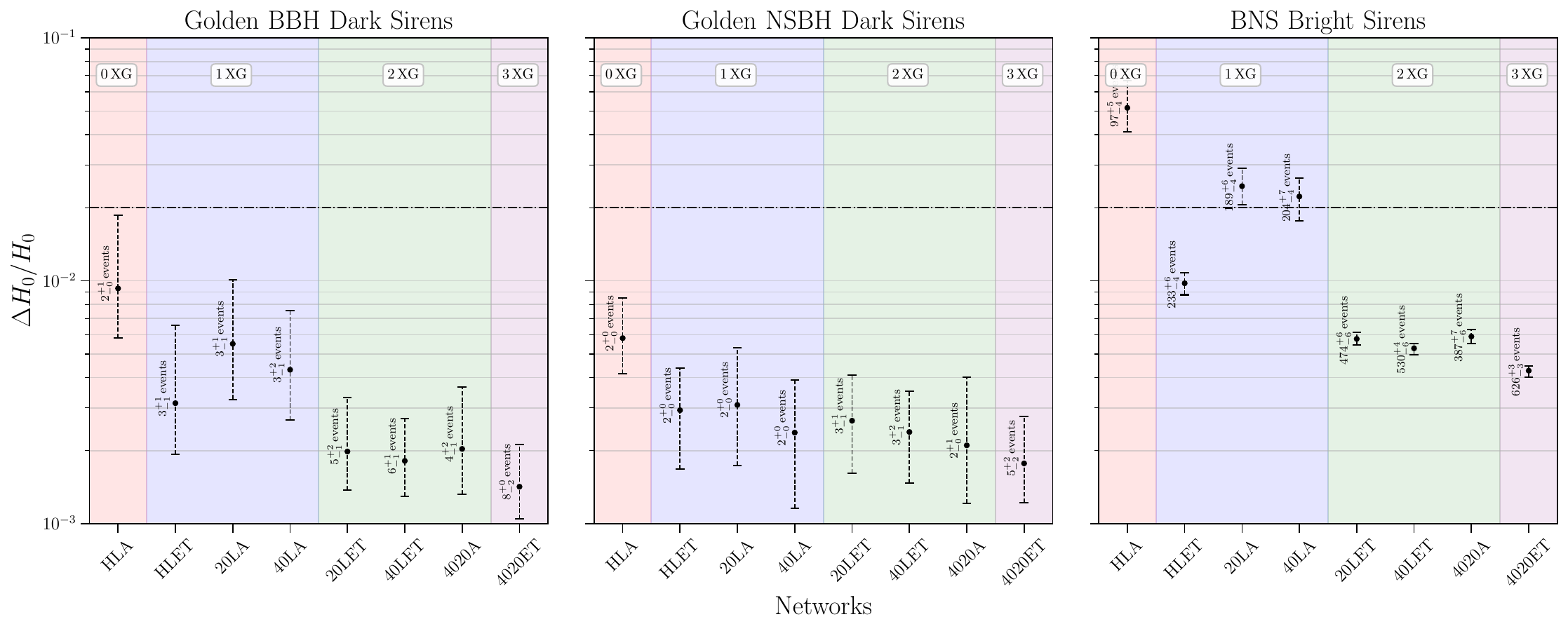}
    \caption{The accuracy in $H_0$ measurement using the bright siren approach with BNS mergers and the golden dark siren approach with BBH and NSBH mergers. The dotted black horizontal line in the three plots marks the $2\%$ precision threshold for $H_0$ measurement, which is adequate to resolve the $H_0$ tension. Alongside each error bar, we also mention the number of events that were used to obtain the corresponding bound on $H_0$.}
    \label{fig:H0}
\end{figure*}

Among these dark siren events, there will also be a fraction of events that are so well localized in the sky that only one galaxy can lie in that sky patch \cite{Nishizawa:2016ood}. This would ensure unique identification of the host galaxy and the associated redshift can be obtained. Such events are called golden dark siren events. In Fig. \ref{fig:H0}, we show the accuracy with which $H_0$ can be estimated using the golden dark siren approach and the bright siren approach for different detector networks. We follow Refs. \cite{Borhanian:2020vyr,Gupta:2022fwd} to categorize those BBH and NSBH events as golden dark sirens for which $z \leq 0.1$ and $\Delta \Omega \leq0.04\,\mathrm{deg}^2$. To calculate the fractional errors in $H_0$, we convert the luminosity distance errors to $H_0$ errors using equation \ref{eq:H0_DL}. Following Ref. \cite{Gupta:2022fwd}, the errors in the redshift measurement are neglected, but we take into account the uncertainty in the value of $\Omega_m$.  Specifically,  Planck gives $\Omega_m = 0.315 \pm 0.007$ \cite{Planck:2018vyg} and the SH0ES measurement of $q_0$ is used to give $\Omega_m = 0.327 \pm 0.016$ \cite{Riess:2021jrx}. This information is included in the Fisher analysis by applying a Gaussian prior on $\Omega_m$ standard deviation given by
\begin{equation}
    \sigma_{\Omega_m} = \sqrt{\sigma_{\mathrm{Planck}}^2 + \sigma_{\mathrm{SH0ES}}^2} = 0.017 .
\end{equation}
The Fisher matrix obtained by combining estimates from $N$ golden dark siren events is given by
\begin{equation}
    \Gamma_{ij} = \sum_{k=1}^{N} \frac{1}{\sigma_{D_L}^2} 
    \left. \left(\dfrac{\partial D_{L}}{\partial \theta_i}\right)\left(\dfrac{\partial D_{L}}{\partial \theta_j} \right)\right\rvert_k + \delta_{i2}\delta_{j2}\,\frac{1}{\sigma^2_{\Omega_m}},
\end{equation}
where $\boldsymbol{\theta} = (H_0,\Omega_m)$ and $\delta_{ij}$ refers to the Kronecker delta. From Fig. \ref{fig:local_eff_rate}, we see that the chosen redshift distribution allows for $10$ BBH and $20$ NSBH mergers within $z=0.1$ every year. To avoid making conclusions based on a specific set of events, we perform 100 realizations of the universe and calculate the combined estimates for each of these realizations. Fig. \ref{fig:H0} shows the median error in $H_0$ and the error bars portray the $68\%$ confidence interval. It is important to note that not all realizations of the universe may contain golden dark siren detections. Hence, Fig. \ref{fig:H0} only considers the realizations where at least two golden dark sirens are detected. This may lead to inflated expectations regarding the capabilities of the less sensitive 0 XG and 1 XG detector networks.

For the bright siren approach, we consider those BNS for which $z \leq 0.3$ and $\Delta \Omega \leq 10\,\mathrm{deg}^2$. The redshift range takes into account the redshift up to which a kilonova can be observed using the Rubin or the Roman telescope and the sky-area cut-off matches the field of view of the Rubin observatory. We also assume a $20\%$ duty-cycle due to the time-sensitive follow-up required for this method. While the same duty-cycle is used for all the detectors, networks with XG detectors will detect more events that can be followed up. Thus, we are implicitly assuming that the EM facilities will improve in the future and will be able to keep up with the GW detections. Following the same steps as for the golden dark siren case, we estimate the fractional errors in $H_0$ using the bright siren approach, which is also shown in Fig. \ref{fig:H0}. For both the bright siren and the golden dark siren cases, we multiply the errors by a factor of $\sqrt{2}$, to account for systematic effects that have not been included in this work.

For Fig. \ref{fig:H0}, we see that using the golden dark siren events, the tension in $H_0$ could be resolved without any XG detectors, with the \asharp network achieving sub-percent precision in the span of 1 year. With XG detectors, the bounds on $H_0$ improve by a factor of $2-8$, depending on the number of XG detectors in the network. In addition, we note that networks with at least 2 XG detectors are able to achieve $~0.2\%$ precision with a handful of events. This shows that for each event, individually, $H_0$ can be measured to the level of $1\%$ or better (assuming $\Omega_m$ is known). This also allows for the study of possible anisotropy in $H_0$ measurements, which would point to the breakdown of $\Lambda$CDM cosmology. 

Using the bright siren method with BNS detections, we see that the \asharp network will be unable to measure $H_0$ well enough to resolve the tension. However, networks with 2 or more XG detectors will be able to measure $H_0$ to sub-percent precision. Also, note that the best detector network, 4020ET, achieves a precision of $0.004$ with 626 events, i.e., with $\sim 2$ detection per day. We expect that the future EM facilities will be capable of following up on these events at the stated frequency. 

It is important to note that the actual constraints on $H_0$ will be even better if the bounds from BBH, NSBH and BNS events are combined. However, the fact that $H_0$ can be measured precisely by each of these compact binary mergers will be important as it would allow us to check the consistency of the inferred value of $H_0$ from the different types of mergers, and, consequently, different approaches. Inconsistency will help isolate, and correct for, any systematic effects that might have been ignored when inferring the value of $H_0$, or point to new physics.

\subsubsection{Measuring the LCDM with NS tides}
BNS have an intrinsic mass scale and can only exist in a narrow range of masses. This mass scale is imprinted in the tidal interaction between the component NSs. Therefore, if the nuclear EoS is known, one can determine the source-frame masses by a measurement of the tidal deformability. This, in turn, would allow the measurement of the redshift directly from a GW observation because it is the redshifted mass that is inferred from the point-particle approximation of the waveform. Such a method was first proposed in Ref.~\cite{Messenger:2011gi} and further explored in Refs.~\cite{Messenger:2013fya,Li:2013via}. The measurement of $H_0$ using a known relationship between the tidal parameter and source-frame mass was explored in Refs.~\cite{DelPozzo:2015bna,Chatterjee:2021xrm,Shiralilou:2022urk} while Ref.~\cite{Ghosh:2022muc} showed that one can simultaneously estimate both the nuclear EoS and $H_0$ using future observatories. A measurement of the dark energy EoS was explored in Refs.~\cite{Wang:2020xwn,Jin:2022qnj}. 

In this section, we explore the potential of different XG configurations to constrain the expansion history of the Universe assuming that the nuclear EoS is known. It is found in Ref.~\cite{Chatterjee:2021xrm} that up to a 15\% uncertainty in the knowledge of the EoS does not affect the measurement of the Hubble constant in a meaningful manner. We use the TaylorF2 waveform model augmented with the 5PN and 6PN tidal terms in the phase, terminating the signal at the ISCO frequency corresponding to the total mass of the binary. Additionally, we assume the APR4 EoS for the NS. We fit the logarithm (base 10) of the tidal deformability as a function of the mass of the NS using a fifth-order polynomial given by
\begin{multline}
    \log_{10}\Lambda(m) =  -5.60 m^5 + 43.2 m^4 -132 m^3 + 199 m^2 \\-151 m + 49.2,
\end{multline}
where $m$ is in units of $\msun$. We verify that the fit reproduces the slope of the curve accurately with maximum errors at a few percent around the double Gaussian from which the neutron star masses are drawn. This is crucial because it is the slope of the curve that contributes to the Fisher errors on the redshift. 

The Fisher errors from the $d_L$--$z$ space are then propagated to the space of cosmological parameters, $\Vec{\phi}$, via another Fisher matrix given by~\cite{Heavens:2014xba}
\begin{equation}
\label{eq:cosmo_fisher}
    \mathcal{G}_{ij} = \sum_{k=1}^N \frac{1}{\sigma_{d_L,k}^2} \frac{\partial d_L^k(z)}{\partial \phi^i} \frac{\partial d_L^k(z)}{\partial \phi^j} \,,
\end{equation}
where $N$ is the total number of observations in the catalog and $\sigma_{d_L,k}^2$ is the total variance in the luminosity distance for the k-th event given by
\begin{equation}
    (\sigma_{d_L})^2 = (\sigma_{d_L}^{h})^2 + (\sigma_{d_L}^{z})^2.
\end{equation}
Here, $\sigma_{d_L}^{h}$ is the contribution to the luminosity distance error due to the error in the GW amplitude while $\sigma_{d_L}^{z}$ is that due to the error in the redshift measurement, given by
\begin{equation}
\label{eq:errz_dl}
    \sigma_{D_L}^{z} = \left|\frac{\partial D_L}{\partial z}\right| \sigma_{z}.
\end{equation}
In writing Eq.~\ref{eq:cosmo_fisher}, we have neglected the correlations in the $d_L$--$z$ space for simplicity.

\begin{figure*}
    \centering
    \includegraphics[width=\columnwidth]{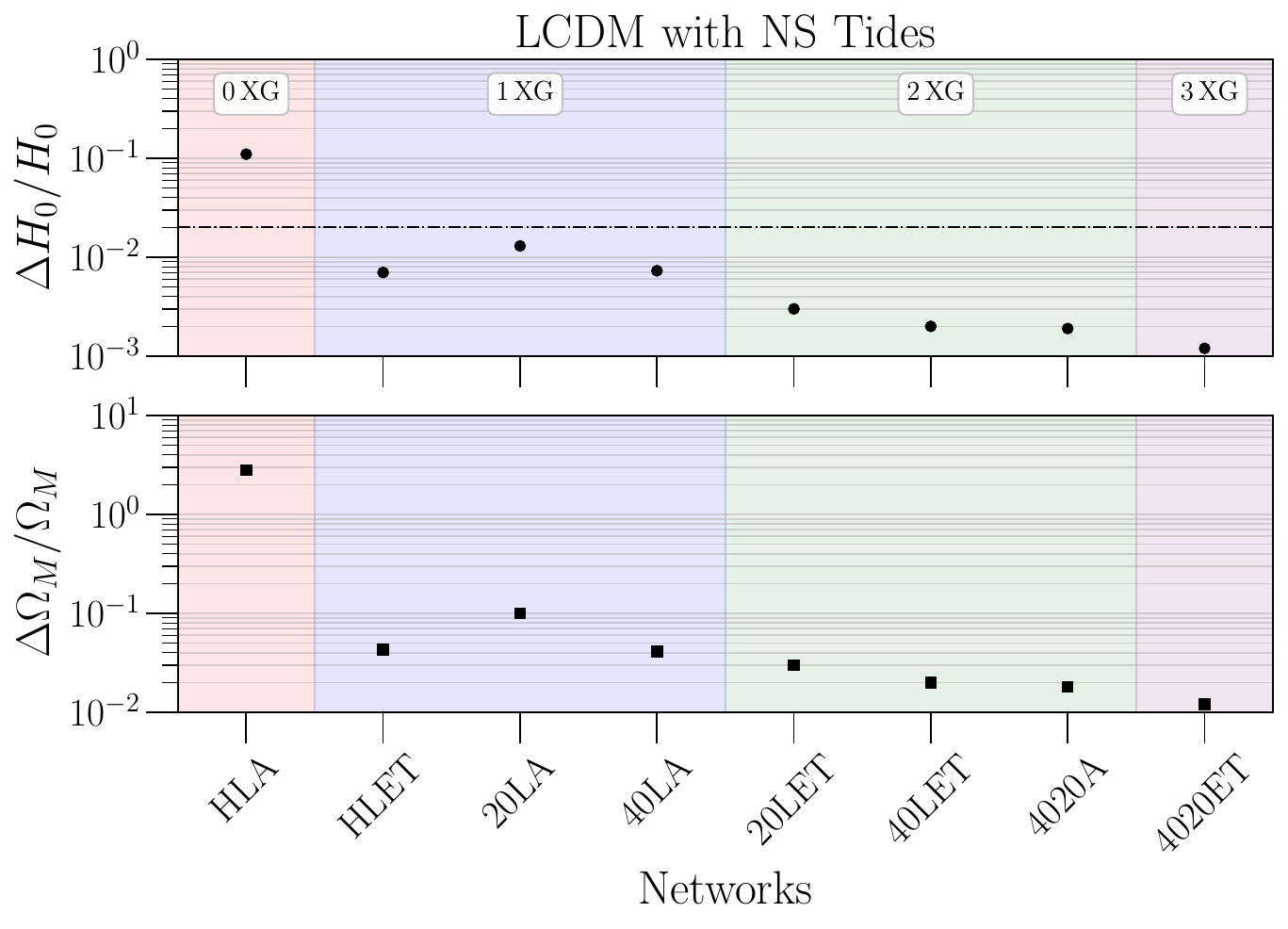}
    \includegraphics[width=\columnwidth]{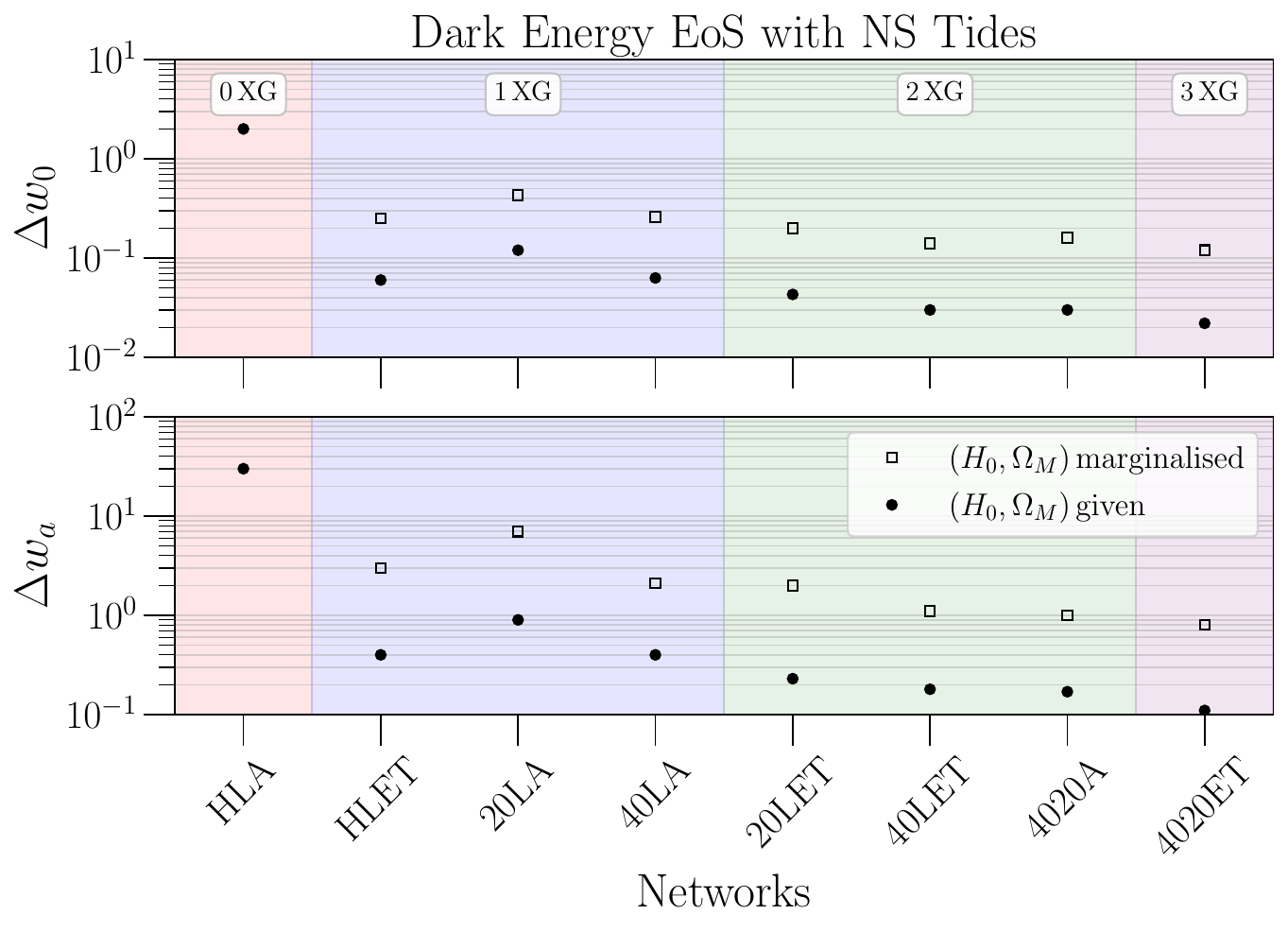}
    \caption{
    \emph{Left:} The fractional uncertainty in the Hubble constant $H_0$ and the dark matter energy density parameter $\Omega_M$ with BNS mergers for the various network configurations under consideration.
    \emph{Right:} The uncertainty in the dark energy EoS parameters $w_0$ and $w_a$ for the various network configurations under consideration. We plot the constraints obtained keeping $H_0$ and $\Omega_M$ as model parameters but marginalizing over them (empty squares) and assuming that they are known precisely from other experiments (black circles).}
    \label{fig:lcdm}
\end{figure*}

The results for $H_0$ and $\Omega_M$ are shown in the left panel of Fig.~\ref{fig:lcdm}. It is observed that $H_0$ and dark matter energy density cannot be simultaneously constrained in the absence of any XG detectors. With at least 1 XG detector, $H_0$ can be determined at the percent level while $\Omega_M$ can be measured to an accuracy of $5-10\%$. Of particular note is that an XG network consisting of CE20 is significantly worse than its 40 km counterpart and the ET observatory. With a network of 2 XG detectors, the errors decrease by a factor of $2-4$ while a full XG network consisting of 3 XG detectors further reduces the errors by another 50\%. 

\subsubsection{Measuring the dark energy with NS tides}
The results for the dark energy EoS parameters are shown in the right panel of Fig.~\ref{fig:lcdm}. In the absence of any XG detectors, dark energy EoS parameters cannot be measured. We see similar factors of improvement with the addition of each XG detector. Notably, if the $\Lambda$CDM parameters are marginalised over instead of assumed to be given from other experiments, the constraints on the dark energy EoS parameters worsen by a factor of $5-10$.

\subsubsection{Gravitational-wave lensing}

\begin{table*}[htbp] 
  \centering
  \caption{\label{tab:rate_of_lensing}{Relative rate of strong lensing detections per year for seven detector networks and variable binary compact object population models. 
  The strong lenses are generated using galaxies drawn from the SDSS galaxy catalog~\cite[see Ref. ][]{Wierda:2021upe}.} }
  \renewcommand{\arraystretch}{1.5} 
    \begin{tabular}{P{2.1cm} P{2.6cm}P{2.2cm}P{2.2cm} P{2.6cm}}
    \hhline{=====}
    Detector configuration & Local population & Population III & Primordial black holes & Binary neutron stars \\ \hline
    HLET & $6.9\times 10^{-3}$ & $2.3\times 10^{-3}$ & $2.0\times 10^{-3}$ & $2.7\times 10^{-4}$ \\
    20LA & $6.6\times 10^{-4}$ & $2.2\times 10^{-3}$ & $2.1\times 10^{-3}$ & $1.5\times 10^{-4}$ \\
    40LA & $7.3\times 10^{-4}$ & $2.4\times 10^{-3}$ & $2.1\times 10^{-3}$ & $2.4\times 10^{-4}$ \\
    40LET & $7.9\times 10^{-4}$ & $2.5\times 10^{-3}$ & $2.2\times 10^{-3}$ & $2.4\times 10^{-4}$ \\
    20LET & $7.3\times 10^{-4}$ & $2.3\times 10^{-3}$ & $2.2\times 10^{-3}$ & $2.5\times 10^{-4}$ \\
    4020A & $7.6\times 10^{-4}$ & $2.5 \times 10^{-3}$ & $2.2\times 10^{-3}$ & $2.4\times 10^{-4}$ \\
    4020ET & $8.1\times 10^{-4}$ & $2.5\times 10^{-3}$ & $2.3\times 10^{-3}$ & $2.3\times 10^{-4}$ \\
    \hhline{=====}
    \end{tabular}
\end{table*}

Gravitational lensing, a captivating phenomenon predicted by Einstein's theory of general relativity, bends light and gravitational radiation as they pass near massive intervening objects. 
The advent of XG observatories usher in a new era of gravitational lensing exploration, as it is projected that approximately one in a thousand BBHs and one in a few thousand BNSs will be strongly lensed, resulting in an annual detection rate of around $\mathcal O(50-100)$ lensed events (see Table~\ref{tab:rate_of_lensing}). 
Such lensed detections have the potential to achieve highly precise localization of BBHs with sub-arcsecond accuracy, identify new subpopulations of lensed systems, probe the fundamental properties of GWs, reconstruct gravitational lenses using GW signals, perform cosmographic measurements at submillisecond timing precision, develop comprehensive models of lens populations, and conduct multifaceted studies involving multiple messenger signals~\cite[see Refs.][and references therein]{LIGOScientific:2021izm,LIGOScientific:2023bwz}. 
Embracing this research frontier with XG observatories not only advances GW astronomy but can also pave the way for groundbreaking discoveries that enhance our knowledge of gravity, astrophysics, and the intricacies of the universe.

\subsection{Physics beyond the standard model}
\label{ss:dark_matter_early_uni}
\begin{figure*}
    \centering
    \includegraphics[width=0.99\columnwidth]{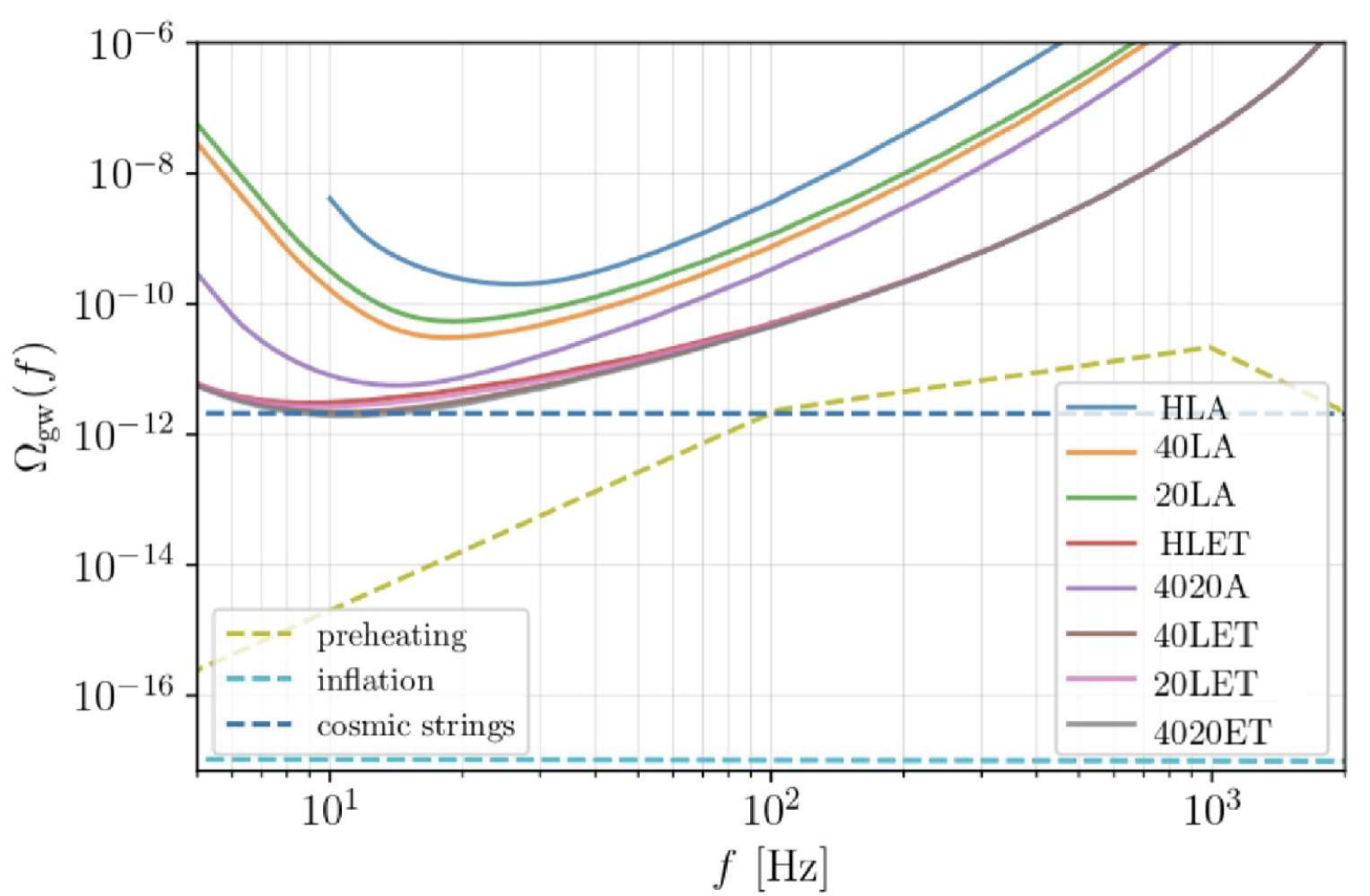}
    \includegraphics[width=0.99\columnwidth]{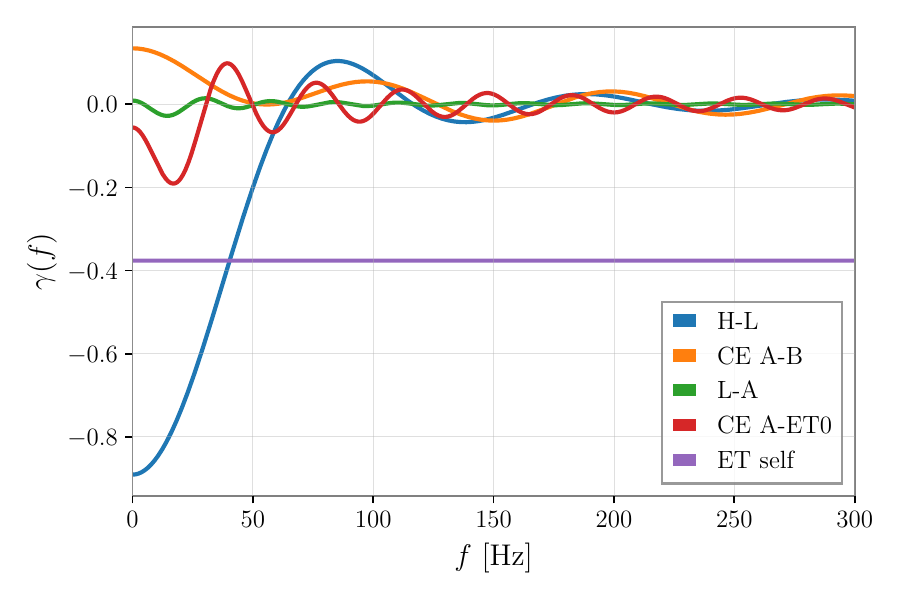}
    \caption{\textit{Left:} Power-law integrated (PI) curves showing the sensitivity of the various detector networks considered in this work to the stochastic gravitational-wave background. Any background whose spectrum crosses the PI curve would be detected with SNR=3 after one year of observing time. Dashed lines show the expected backgrounds for cosmic strings ($G\mu = 10^{-11}$ with fiducial model parameters from Ref.~\cite{Sanidas:2012ee}), preheating (for hybrid inflation occurring at $10^{9}$ GeV as
    calculated in Ref.~\cite{2007PhRvL..99v1301E}), and standard slow-roll inflation. \textit{Right:} Overlap reduction functions for various detector pairs considered here, normalized so that $\gamma(f) = 1$ for co-located and co-aligned L-shaped detectors.}
    \label{fig:stochastic}
\end{figure*}

\subsubsection{Stochastic backgrounds}
The sensitivity of a given XG network to the stochastic GW background of primordial origin quantifies its ability to probe early-universe physics. Typical stochastic background searches assume that the background is Gaussian, isotropic, stationary, and unpolarized, so the optimal search strategy is to look for excess correlated power between pairs of detectors~\cite{Allen:1997ad, Romano:2016dpx}. In this case, the sensitivity of the pair depends primarily on the detector PSDs and geometry, quantified via the overlap reduction function, $\gamma(f)$~\cite{Flanagan:1993ix}. Co-located and co-aligned L-shaped detectors have $\gamma(f)=1$, while for detectors separated by large distances and large relative angles, $\gamma(f)$ is an oscillatory function that asymptotes to zero at large frequencies, penalizing the sensitivity of that detector baseline. The overlap reduction functions for several detector baselines considered in this document are shown in the right panel of Fig.~\ref{fig:stochastic}.

The strength of the stochastic background is typically parameterized in terms of
\begin{align}
\Omega_{\mathrm{GW}}(f) &\equiv \frac{1}{\rho_{c}}\frac{d\rho_{\mathrm{GW}}}{d\ln{f}}= \Omega_{\alpha}\left(\frac{f}{f_{\mathrm{ref}}}\right)^{\alpha},
\end{align}
where $\rho_{\mathrm{GW}}$ is the energy density in GWs and $\rho_{c}$ is the critical energy density needed to close the universe. For a given value of the power-law index $\alpha$, the background amplitude that would be detectable with SNR $\rho$ and an observing time $T$ is given by
\begin{align}
\Omega_{\alpha} &= \frac{\rho}{\sqrt{2T}}\left[\int_{f_{\min}}^{f_{\max}} df \frac{(f/f_{\mathrm{ref}})^{2\alpha}}{\Omega_{\mathrm{eff}}^{2}(f)}\right]^{-1/2}\\
\Omega_{\mathrm{eff}} &= \frac{10\pi^{2}}{3H_{0}^{2}}f^{3}S_{\mathrm{eff}}(f).
\end{align}
The effective strain noise power spectral density is given by
\begin{align}
S_{\mathrm{eff}} = \left[ \sum_{I=1}^{M}\sum_{J>I}^{M}\frac{\gamma_{IJ}^{2}(f)\sin^{2}{\beta_I}\sin^{2}{\beta_J}}{P_{n,I}(f)P_{m,J}(f)}\right]^{-1/2},
\end{align}
where the indices $I,J$ indicate the interferometer, $\beta$ is the opening angle between the arms of each interferometer, and $P_n$ is the noise PSD.

The last row of Table~\ref{tab:network_science} gives the background amplitude that would be detectable with SNR$\,=\,3$ after one year of observing for each of the eight networks considered in this document at a reference frequency of 25~Hz for $\alpha=0$, which is the theoretical expectation for backgrounds produced by vanilla inflation~\cite{Christensen:2018iqi}. The left panel of Fig.~\ref{fig:stochastic} shows the power-law integrated (PI) curves~\cite{Thrane:2013oya} for each network, for which a stochastic background that crosses or lies tangent to the PI curve would be detected with SNR$\,=\,3$ after 1 yr of observing.

Because ET consists of three nearly co-located detectors and has the best projected sensitivity at low frequencies, the networks including ET are not penalized as strongly by the geometric $\gamma(f)$ factor and thus have the best projected sensitivity to the stochastic background. We neglect the effect of correlated noise, which may be significant for the co-located ET detectors~\cite[e.g.,][]{LIGOScientific:2014sej}. The exact sensitivity of the proposed XG networks will change due to the change in the overlap reduction function once the locations and orientations of the detectors are finalized, but the numbers quoted here are meant to be representative of XG detector capabilities.

It is worth noting that these values of $\Omega_{\mathrm{GW}}$ are calculated assuming that the primordial background can be perfectly separated from the foreground of merging compact binaries. New methods that exploit the statistical differences between the two signals are being developed to ensure this is possible by the time that XG data become available~\cite{2020PhRvD.102f3009S,2020PhRvL.125x1101B,2022Galax..10...34R,2022arXiv220901310Z,2023PhRvD.107f4048Z}.

\begin{figure*}
    \centering
    \includegraphics[width=0.99\textwidth]{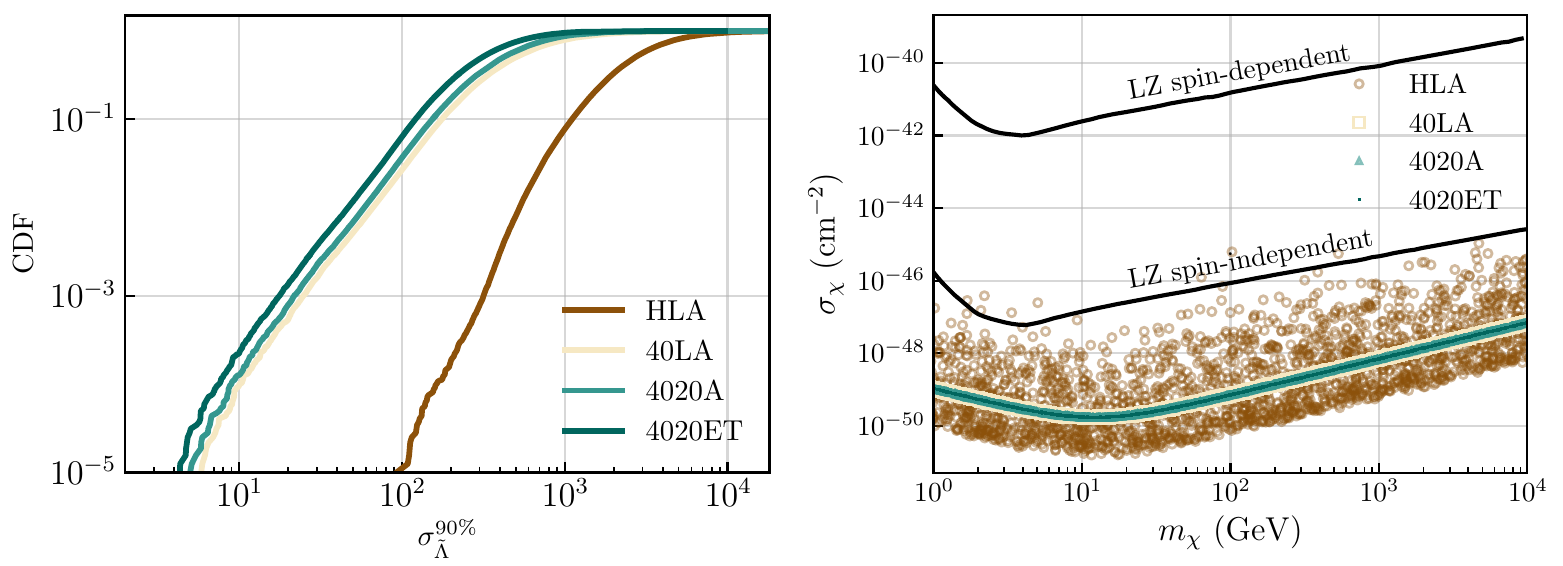}
    \caption{\textit{Left:} Cumulative distribution function (CDF) of the error in the measurement of effective tidal deformability in four representative detector networks with and without cosmic explorer. \textit{Right:} Comparison between the constraints obtained for $m_\chi \in [1, 10^4]$ GeV assuming an ambient DM density of $\rho_\chi = 1\ {\rm Gev/cm^3}$ for representative detector networks with the latest constraints from the direct detection experiment, LZ~\cite{LZ:2022ufs}.} 
    \label{fig:dark matter}
\end{figure*}

\subsubsection{Bosonic asymmetric dark matter}
The next generation of GW detectors are expected to observe $\sim 10^6$ BNS mergers per year with precise measurements of their effective tidal deformability. If neutron stars accumulate ambient dark matter, their EoS would deviate from that of pure neutron stars, in turn leading to a deviation in the effective tidal deformability of the binary system. The accumulation of dark matter particles in NS cores due to accretion over long timescales could potentially lead to the formation of a mini BH, destabilizing the NS and resulting in its implosion to form a BH without significantly increasing its mass. When this process occurs in neutron stars in coalescing binaries, one or both stars might be converted to a BH before they merge. Hence, the total rate of mergers of compact objects in the mass range $1-2\ M_\odot$ would have relative contributions majorly from BNS, BBH and significantly lower from NSBH systems.  The precise measurement of the effective tidal deformability parameter with the XG detectors, we would be able to distinguish between the sub-populations of compact mergers since the tidal deformability of BHs is zero, while NSs have non-zero values for the tidal deformability. This distinguishability using matter properties is essential in ascertaining the relative rates of BNS and BBH mergers. However, the rates are also informed by the collapse time of the NS to BH which in turn depends on the mass and scattering cross-section of the dark matter particles. Therefore, GW observations of BBHs and BNSs can potentially constrain particle properties of dark matter through the observed rates of different binary populations in the mass range of $1-2\ M_\odot$. 

In the left panel of Fig. \ref{fig:dark matter}, we show the capability of a subset of detector networks considered in the study to measure the effective tidal deformability of merging binaries. The addition of a single CE to the network drastically reduces the measured error - 0.001\% of events are detected with an error $\sigma_{\tilde\Lambda}^{90\%} \simeq 100$ in the HLA network, while the same fraction of events are detected with $\sigma_{\tilde\Lambda}^{90\%} < 10$ in networks with at least one CE.
The right panel in the same figure shows the effect of measured errors on the effective tidal deformability on the inferred dark matter constraints. For an assumed threshold of $\sigma_{\tilde\Lambda}^{90\%} = 50$, we derive the upper limits for the dark matter particle mass and scattering cross-section with baryons. We also present the upper limits for WIMPs reported by Lux-Zeplin experiment (LZ \cite{LZ:2018qzl,LZ:2019sgr,LZ:2022ufs}) to show how competitive these constraints can be with leading dark matter experiments.

\subsubsection{Ultra-light boson clouds around rotating black holes}

\begin{figure}
    \centering
    \includegraphics[width=\columnwidth]{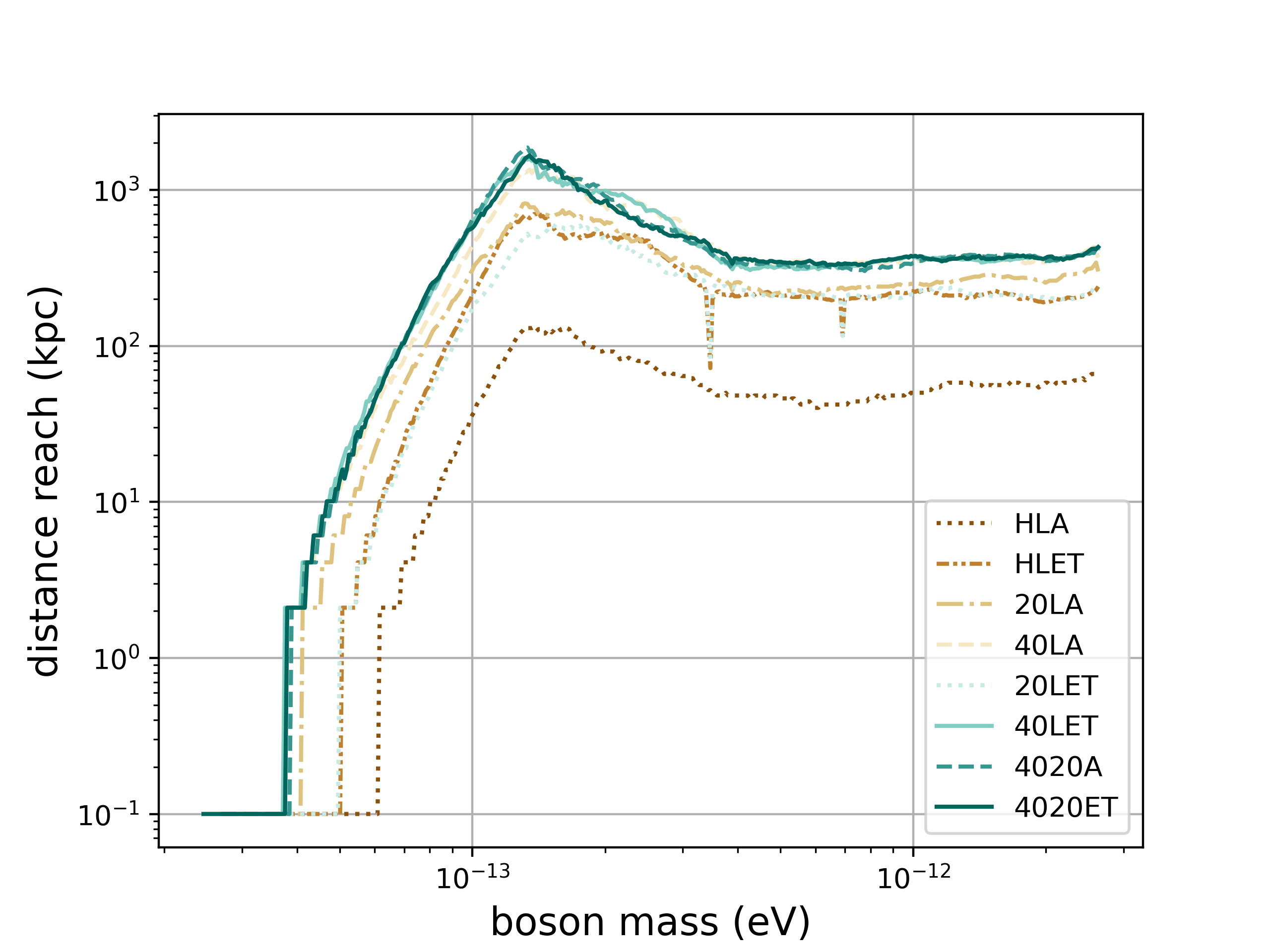}
    \caption{Astrophysical distance reach as a function of axion mass for ultralight boson clouds that could form around rotating black holes. Different colors correspond to different detectors.}
    \label{fig:cw-boson}
\end{figure}

Axions are ultralight bosons hypothesized to solve the strong-CP problem in QCD. If axions or other ultra-light scalar or vector ultralight bosons exist, they could appear spontaneously near rotating black holes and be bound to them if their Compton wavelength is comparable to the BH size \cite{Arvanitaki:2009fg, Arvanitaki:2014wva, Arvanitaki:2016qwi}. They could extract mass and energy from BHs over time, building up a macroscopic dark-matter ``cloud'' via a superradiance process \cite{Brito:2015oca, Arvanitaki:2016qwi} which might be easier to detect with vector (i.e., spin 1) bosons with near-term detectors \cite{Baryakhtar:2017ngi, Siemonsen:2019ebd, Jones:2023fzz}. The so-called ``gravitational atom'' could then emit quasi-monochromatic, persistent, GWs via boson-boson annihilation \cite{Brito:2017zvb}. Current detectors are able to detect the presence of boson cloud systems at the galactic center in the most recent observing run, for young spinning black holes (less than $10^5$ years) \cite{LIGOScientific:2021rnv}, but with the advent of XG observatories, those prospects will improve by a factor of 10-20. In Fig. \ref{fig:cw-boson}, we plot the distance reach, computed according to an analytic expression, Eq. (57) given in Ref.\,\cite{Astone:2014esa}, as a function of ultralight boson mass, in the small gravitational fine-structure constant $\alpha$ limit ($\alpha<0.1)$, assuming a uniform distribution of spins between [0.2,0.9], a log-uniform distribution of ages between [$10^3$,$10^7$] years, a coherence length of 10 days, and a Kroupa distribution \cite{Kroupa:2000iv} for black hole masses between [5, 100]$M_\odot$. This distance corresponds to detecting at least 5\% of black holes located at that distance away with a particular boson cloud. The improvements relative to the current detector era are immense, and are derived for a \emph{semi-coherent} all-sky search for boson cloud systems with fast-Fourier-transform length of $T_{\rm FFT}=10$ days and a threshold on our detection statistic, the critical ratio, of 3.4, as done in a similar analysis for ET design comparisons \cite{Branchesi:2023mws}. 

\section{Conclusions}
In this paper, we compared the relative performance of eight different GW observatory networks composed of upgraded LIGO, Virgo, KAGRA, LIGO-India, CE and ET. The spirit of this study was to understand which key science goals enumerated in the Cosmic Explorer White Paper \cite{Evans:2023euw} can be accomplished with networks that have no XG observatories and how adding one, two or three such observatories would strengthen the network. In particular, we have explored the role of upgraded LIGO-Livingston and LIGO-India operating at \asharp sensitivity in tandem with one or two next CE observatories and/or ET. As summarized in Table \ref{tab:network_science} there is great value in operating an upgraded LIGO when only one or two XG observatories are operating. Such a scenario could arise either because of scheduling or one or more of the XG observatories are in the commissioning or upgrade mode. A network composed of two CE observatories (one with 40\,km arms and one with 20\,km arms) and ET will deliver all the science goals identified in the White Paper \cite{Evans:2023euw}; such a network will be two orders-of-magnitude better with respect to almost every science metric considered in this study and will have an unprecedented discovery potential, observing BH binaries from epochs when the first stars were still being assembled and binary neutron stars from redshifts when the star formation was at its peak. 

\section*{Acknowledgements} \label{sec:acknowledgements} 
We thank the members of the Cosmic Explorer Project and the Scientific Advisory Committee for their input and feedback. 
IG, AD, BSS, RK, and DS were supported by NSF grant numbers  AST-2006384, AST-2307147, PHY-2012083, PHY-2207638, PHY-2308886, and PHYS-2309064 to B.S.S. SB acknowledges support from the Deutsche Forschungsgemeinschaft, DFG, Project MEMI number BE 6301/2-1.
K.G.A. acknowledges support from the Department of Science and Technology and Science and Engineering Research Board (SERB) of India via the following grants: Swarnajayanti Fellowship Grant DST/SJF/PSA-01/2017-18 and Core Research Grant CRG/2021/004565
K.G.A, E.B.  and B.S.S. acknowledge the support of the Indo-US Science and Technology Forum through the Indo-US Centre for Gravitational-Physics and Astronomy, grant IUSSTF/JC-142/2019. 
AB and KK acknowledge support from NSF award PHY-2207264.
E.B. is supported by NSF Grants No.~AST-2006538, PHY-2207502, PHY-090003 and PHY-20043, and by NASA Grants No. 20-LPS20-0011 and 21-ATP21-0010. E.B.~acknowledges support from the ITA-USA Science and Technology Cooperation program (CUP: D13C23000290001), supported by the Ministry of Foreign Affairs of Italy (MAECI).
MB acknowledges support from DOE under Award Number DE- SC0022348
GL acknowledges NSF award PHY-2208014, Nicholas and Lee Begovich, and the Dan Black Family Trust. A. Corsi acknowledges support from NSF grant PHYS-2011608. AHN acknowledges support from NSF grant PHY-2309240.
OAH and HP acknowledge support by grants from the Research Grants Council of Hong Kong (Project No. CUHK 2130822 and 4443086), and the Direct Grant for Research from the Research Committee of The Chinese University of Hong Kong.
TGFL acknowledge support by grants from the Research Foundation - Flanders (G086722N, I002123N) and KU Leuven (STG/21/061).
DEM, BJJS, and LS acknowledge the support of the Australian Research Council Centre of Excellence for Gravitational Wave Discovery (OzGrav), Project No. CE170100004.
C.P. acknowledges research support from the Italian Istituto Nazionale di Fisica Nucleare (INFN).
BJO, AP, and BR acknowledge support from NSF grants PHY-1912625, PHY-2309305, and AST-1907975.
 JDR acknowledges support from NSF grant PHY-2207270.
JRS and GL acknowledge support from NSF grant 2308985, Nicholas and Lee Begovich, and the Dan Black Family Trust.
SV acknowledges support from NSF award PHY-2207387.

\appendix

\section{Science metrics for other network configurations}
Other than the networks discussed in section \ref{sec:Networks}, we also examine the capabilities of eight other detector networks listed in Tab. \ref{tab:more_networks}. These networks contain detectors at A+ sensitivities as well. They were chosen to fulfill two objectives- first, to show the capabilities of \aplus detectors with CE observatories, and second, to assess the potential of a three-detector network contained in the US. For the latter, as CE-A and CE-B are very close to LHO and LLO, respectively, we introduce a new \textit{fiducial} location for CE, called CE-C, located in Lake Michigan, at the latitude of $45^\circ 10'3.67''$, the longitude of $-87^\circ 03'46.2''$, with the $x-$arm oriented $250.824^\circ$ counter-clockwise relative to the local east. Additionally, as the networks are taken to be located in the US (except 20L$^{\dag}$A$^{\dag}$), we do not consider Virgo and KAGRA detectors in our networks. However, it is important to note that a five-detector network including Virgo and KAGRA at A+ sensitivities will be extremely proficient in precise localization of binary mergers, playing a crucial role in fulfilling multimessenger objectives in the late 2020s \cite{Borhanian:2022czq,Gupta:2023evt}. The science capabilities of the eight new networks are listed in Tab. \ref{tab:network_science_ApAs}.
\begin{table}[h]
  \centering
  \begin{tabular}{l l l}
    \toprule
    \textsf{Number of XG } & \textsf{Network} & \textsf{Detectors in} \\ 
    \textsf{Observatories} & \textsf{Name}    & \textsf{the network} \\ \midrule
    \multirow{3}{*}{1 XG} 
    & 20L$^{\dag}$H$^{\dag}$ & CE-C 20 km, LLO (\aplus), LHO (\aplus) \\ 
    & 20L$^{\dag}$A$^{\dag}$ & CE-A 20 km, LLO (\aplus), LAO (\aplus) \\
    & 40L$^{\dag}$H$^{\dag}$ & CE-C 40 km, LLO (\aplus), LHO (\aplus) \\
    & 20LH & CE-C 20 km, LLO, LHO \\
    & 40LH & CE-A 40 km, LLO, LHO \\ \midrule
    \multirow{3}{*}{2 XG} 
    & 4020L$^{\dag}$ & CE-C 40 km, CE-A 20 km, LLO (A+) \\
    & 4020L & CE-C 40 km, CE-A 20 km, LLO \\
    & 4040L & CE-C 40 km, CE-A 40 km, LLO \\
    \bottomrule
\end{tabular}
\caption{We list the eight other network configurations, this time with detectors at \aplus sensitivities as well.\label{tab:more_networks}}
\end{table}

\begin{table*}
\caption{\label{tab:network_science_ApAs}Similar to Tab. \ref{tab:network_science}, but with networks at \aplus sensitivities as well. The eight networks used here are described in Tab. \ref{tab:more_networks}.}  
 \begin{tabular}{ p{.45\textwidth} | p{\colwidth} p{\colwidth} p{\colwidth} p{\colwidth} p{\colwidth} | p{\colwidth} p{\colwidth} p{\colwidth} } 
\hhline{=========}
& \multicolumn{8}{c}{\bf Network Performance}\\
\hhline{~--------}
\textbf{Science Goal Requirements} & \multicolumn{5}{c|}{\bf 1 XG} & \multicolumn{3}{c}{\bf 2 XG}  \\
\hhline{~--------}
& \textbf{20L$^{\dag}$H$^{\dag}$} & \textbf{20L$^{\dag}$A$^{\dag}$} & \textbf{40L$^{\dag}$H$^{\dag}$} & \textbf{20LH} & \textbf{40LH} & \textbf{4020L$^{\dag}$} & \textbf{4020L} & \textbf{4040L} \\ 
\hhline{---------}
{\bfseries BHs and NSs Throughout Cosmic Time }&  & & & & & & &  \\
Measure mass function, determine formation scenarios:&  & & & & & & &  \\
$N_{\rm BNS}$/yr, $z\ge1,$ $\delta z/z \le 0.2,$ $\delta m_{1}/m_{1}\le 0.3$ & 0 & 0 & 0 & 0 & 0 & 1 & 3 & 10\\
Detect the (injected) second Gaussian feature:& & & & & & & & \\
$N_{\rm BNS}$/yr, $m_1\ge1.5\msun,$ $\delta m_1/m_1 \le 0.1$ & 5 & 4 & 15 & 6 & 21 & 41 & 52 & 72\\
Unveiling the elusive population of IMBH:& & & & & & & & \\
$N_{\rm IMBBH}$/yr, $z\ge 3,$ $\delta z/z \le 0.2,$ $\delta m_{1}/m_{1}\le 0.2$ & 48 & 100 & 88 & 130 & 180 & 500 & 530 & 580\\
High-$z$ BBH formation channels and mass function: &  & & & & & & &\\
$N_{\rm BBH}$/yr $z\ge 10,$ $\delta z/z \le 0.2$ $\delta m_{1}/m_{1} \le 0.2$  & 3 & 0 & 24 & 5 & 30 & 79 & 85 & 129 \\
\hhline{---------}
{\bfseries MMA and Dynamics of Dense Matter} &  & & & & & & &   \\
GW170817-like golden sample: &  & & & & & & & \\
$N_{\rm BNS}$/yr $z\le0.06,$ $\Delta \Omega\le 0.1$\,deg$^2$ & 0 & 0 & 0 & 0 & 0 & 0 & 0 & 0 \\
r-process and kilonova-triggered follow up: &  & & & & & & &  \\
$N_{\rm BNS},$ $z\le0.1,$ $\Delta\Omega\le1$\,deg$^2$ & 0 & 0 & 0 & 0 & 0 & 0 & 1 & 4 \\
Jet afterglows, large-FOVs or small-FOV mosaicking: & & & & & & & & \\
$N_{\rm BNS}$/yr, $0.1<z\le2,$ $\Delta\Omega\le 10$\,deg$^2$
& 0  & 85 & 1 & 7 & 8 & 95 & 210 & 430 \\
Mapping GRBs to progenitors up to star-formation peak: & & & & & & & & \\
$N_{\rm BNS}$/yr, $z>2,$ $\Delta \Omega\le 100$\,deg$^2$ 
& 0 & 0 & 0 & 0 & 0  & 0 & 0 & 2\\
Pre-merger alerts 10 minutes before merger & & & & & & & & \\
$N_{\rm BNS}$/yr, $\Delta \Omega\le 100$\,deg$^2$ & 0 & 0 & 0 & 0 & 0 & 7 & 10 & 30  \\
NS EoS constraints:&  & &  & & & &   \\
$N_{\rm BNS}$/yr, SNR$\ge 100$ 
& 28 & 37 & 230 & 31 & 230 & 340 & 350 &  640\\
$N_{\rm BNS}$/yr, $\Delta R < 0.1$\,km & 4 & 0 & 36 & 20 & 64 & 140 & 180 & 270 \\
\hhline{---------}
{\bfseries New Probes of Extreme Astrophysics}&  & & & & & & &   \\
Pulsars with ellipticity $10^{-9}$ detectable in 1 year
& 3 & 3 & 5 & 3 & 5 & 9 & 9 & 20 \\
\hhline{---------}
{\bfseries Fundamental Physics and Precision Cosmology} &  & & & & & & &    \\
Constrain graviton mass:&  & & & & & &   \\
$N_{\rm BNS}$/yr, $z\ge5$  & 0 & 0 & 160 & 0 & 180 & 640 & 640 & 2600\\
$N_{\rm BBH}$/yr, $z\ge5$  & 2173 & 2169 & 3925 & 2232 & 3951 & 4431 & 4446 & 4962\\
Probing rare events:&  & & & & & & &   \\
$N_{\rm BBH}$/yr, SNR\,$> 100$  & 1200 & 1200 & 4800 & 1300 & 4900 & 6800 &  6800 & 10500 \\
$N_{\rm BBH}$/yr, SNR\,$> 1000$ & 1 & 1 & 1 & 1 & 1 & 5 & 5 & 9 \\
Precision tests of GR  (IMR and QNM): & & & & & & & & \\
BBH root sum square total SNR & 9700  & 9800 & 16,500  & 9900 & 17,000 & 19,000 & 19,000 & 23,000 \\
BBH root sum square post-inspiral SNR & 5200  & 5300  & 8200  & 5400  & 8300  & 9700  & 9700  & 12,000 \\
$N_{\rm BBH}$/yr, post-inspiral SNR\,$> 100$  & 243 & 251 & 1111 & 281 & 1158 & 1685 & 1706 & 2663 \\
Cosmology and tests of GR:&  & &   & & & & & \\
$N_{\rm BNS}$/yr, $z \le 0.5$, $\delta d_L/d_L \le 0.1$ and $\Delta \Omega\le 10$\,deg$^2$ & 0 & 4 & 0 & 5 & 7 & 18 & 60 & 95 \\
$N_{\rm BBH}$/yr, $\delta d_L/d_L \le 0.1,$ $\Delta \Omega\le 1\,\deg^2$ & 1 & 43 & 2 & 13 & 19 & 120 & 180 & 350 \\
Lensed BNS events/yr: & 62 & 61 & 160 & 64 & 170 & 210 & 210 & 290 \\
\hhline{---------}
{\bfseries Physics Beyond the Standard Model} & & & & & & & &  \\
Stochastic signal $\Omega_{\mathrm{GWBG}}$ in units of $10^{-10}$ & 0.6 & 1.1 & 0.35 & 0.3 & 0.18 & 0.01 & 0.01 & 0.005\\
Primordial black hole mergers: $N_{\rm pBBH}$/yr, $z>25$, $\Delta\,z/z < 0.2$: &  0 & 0 & 0 & 0 &  0 & 3 & 3 & 9 \\
Pop III black hole mergers $z>10$, $\Delta\,z/z < 0.1$: $N_{\rm PopIII}$/yr: & 0 & 0 & 0 & 0 & 1 & 32 & 39 & 98 \\
Max distance (Mpc) to detectable axion clouds in BHs & 0.9 & 0.9 & 1.4 & 1.0 & 1.6 & 1.6 & 1.8 & 1.5 \\
\hhline{=========}
\end{tabular}
\end{table*}

\section{Data release}
As mentioned in Sec. \ref{sec:pop}, we simulate one-year populations for BBH, BNS, NSBH, IMBBH, primordial BH and Pop-III BBH systems. We make the populations available on \href{https://doi.org/10.5281/zenodo.8087733}{Zenodo}~\cite{data}, along with an iPython notebook, called \texttt{intructions.ipy}, that demonstrates how one can use the available data. The data files contain the intrinsic and extrinsic parameters that describe the systems, as well as the SNR and the measurement errors on some of these parameters. 

\bibliography{references}
\end{document}